\pdfoutput=1 
\documentclass[12pt]{article}

\usepackage{inputenc}

\usepackage{amsmath,amssymb,bbm,color}
\usepackage{amsbsy}
\usepackage{fixmath}
\usepackage{euscript}
\usepackage{graphicx}
\usepackage{hyperref}
\usepackage{cite}
\numberwithin{equation}{section}

\newcommand{\msbar}{\overline{\text{MS}}}
\newcommand{\mc}{\text{MC}}
\newcommand{\lo}{\text{LO}}
\newcommand{\nlo}{\text{NLO}}
\newcommand{\order}[1]{\mathcal{O}\!\left(#1\right)}

\newcommand{\GeV}{\text{GeV}}
\newcommand{\DY}{\text{DY}}
\newcommand{\qbar}{{\bar q}}
\newcommand{\Dbar}{{\overline{D}}}

\newcommand{\shat}{{\hat s}}
\newcommand{\as}{{\alpha_s}}

\newcommand{\veps}{{\varepsilon}}
\newcommand{\talf}{{\tilde\alpha}}
\newcommand{\tbet}{{\tilde\beta}}
\newcommand{\tilp}{{\tilde{p}}}

\newcommand{\F}{{_{F}}}
\newcommand{\B}{{_{B}}}
\newcommand{\A}{{_{H}}}

\newcommand{\Peu}{\EuScript{P}}

\newcommand{\Weu}{\EuScript{W}}

\newcommand{\Kbbm}{\mathbbm{K}}
\newcommand{\ie}{{\it i.e. }}
\newcommand{\eg}{{\it e.g. }}
\newcommand{\cf}{{\it c.f. }}

\newcommand{\krknlo}{{\textsf{KrkNLO}}}

\newcommand{\herwig}[1]{\textsf{Herwig++#1}}
\newcommand{\sherpa}[1]{\textsf{Sherpa#1}}
\newcommand{\mcatnlo}[1]{\textsf{MC@NLO#1}}
\newcommand{\mcfm}[1]{\textsf{MCFM#1}}
\newcommand{\powheg}[1]{\textsf{POWHEG#1}}

\begin{document}

\begin{titlepage}


\begin{flushright}
\bf IFJPAN-IV-2015-1\\
    CERN-PH-TH-2015-061 \\
    MCnet-15-06
\end{flushright}

\vspace{5mm}
\begin{center}
    {\Large\bf Matching NLO QCD with parton shower \vspace{2mm}\\
               in Monte Carlo scheme -- the KrkNLO method$^{\star}$ }
\end{center}

\vskip 10mm
\begin{center}
{\large S.\ Jadach$^a$, W.\ P\l{}aczek$^b$, S.\ Sapeta$^c$, 
        A.\ Si\'odmok$^{a,c}$ and M.\ Skrzypek$^a$ }

\vskip 2mm
{\em $^a$Institute of Nuclear Physics, Polish Academy of Sciences,\\
  ul.\ Radzikowskiego 152, 31-342 Krak\'ow, Poland}\\
\vspace{1mm}
{\em $^b$Marian Smoluchowski Institute of Physics, Jagiellonian University,\\
ul.\ {\L}ojasiewicza 11, 30-348 Krak\'ow, Poland}\\
\vspace{1mm}
{\em $^c$CERN PH-TH, CH-1211, Geneva 23, Switzerland}
\end{center}

\vspace{5mm}
\begin{abstract}
\noindent
A new method of including the complete NLO QCD corrections 
to hard processes in the LO parton-shower Monte Carlo (PSMC) is presented.
This method, called \krknlo{}, requires the use of parton distribution functions
in a dedicated Monte Carlo (MC) factorization scheme, 
which is also discussed in this paper.
In the future, it may simplify introduction of the
NNLO corrections to hard processes and the NLO corrections to PSMC.
Details of the method and numerical examples of its practical implementation
as well as comparisons with other calculations, such as \mcfm{}, \mcatnlo{},
\powheg{}, for single $Z/\gamma^*$-boson production at the LHC are presented.
\end{abstract}

\vspace{2mm}
%

\vspace{15mm}
\begin{flushleft}
\bf IFJPAN-IV-2015-1\\
    CERN-PH-TH/2015-061\\
    MCnet-15-06 \\
\end{flushleft}

\vspace{10mm}
\footnoterule
\noindent
{\footnotesize
$^{\star}$This work is partly supported by 
 the Polish National Science Center grant DEC-2011/03/B/ST2/02632,
 the Polish National Science Centre grant UMO-2012/04/M/ST2/00240.
}

\end{titlepage}

\newpage
\tableofcontents

\newpage
\section{Introduction}

Higher-order perturbative corrections in Quantum Chromodynamics (QCD), important
for the LHC data analysis, are calculated order by order in the strong coupling,
$\as$, while some numerically important ones, related to soft and collinear
singularities, can be resumed to the infinite order.  The most valuable, albeit
technically difficult, way of QCD resummation is in form of a Monte Carlo (MC)
event generator~\cite{Buckley:2011ms}.  It is widely recognized that the most 
promising way of getting high precision QCD calculation for hadron collider 
data analysis is a common framework of fixed-order calculations combined with 
a parton shower Monte Carlo (PSMC).  The pioneering work, in which the complete 
first-order QCD corrections to the hard process of heavy boson production in 
hadron--hadron collision were combined with PSMC, was that of 
Ref.~\cite{Frixione:2002ik}.  Shortly later another interesting variant was 
proposed in Ref.~\cite{Nason:2004rx}. Presently both methods are available 
for many processes, see Ref.~\cite{Nason:2012pr}.

It is worth to mention that partial efforts in this direction
were done earlier in the course of development of the most popular PSMCs,
like for instance in Ref.~\cite{Seymour:1994we}
or \cite{Andre:1997vh} and in many other works.
In these earlier attempts, the distributions generated in PSMC
were improved using a tree-level exact matrix element (ME) of QCD,
while virtual corrections were neglected or taken in the leading-logarithmic
approximation~\cite{Catani:2001cc}.
It was also known for a long time, that an {\em ad hoc} approach, in which
PSMC differential distributions were corrected using the exact ME
and the overall normalization was corrected by hand to fixed-order
next-to-leading (NLO) integrated cross section, was providing distributions
in a quite good agreement with experimental data.
It may be therefore a little bit surprising that it took two decades to work
out a systematic method of combining the NLO-corrections to the hard process,
known from the early 1980's, see for instance 
Ref.~\cite{Altarelli:1979ub},
with the leading-order (LO) PSMC, also dating from the early 1980's.
Apart from the lack of interest in more precise QCD calculations
due to poor data quality, main reasons for
this much delayed development can be seen from problems addressed in
Refs.~\cite{Frixione:2002ik,Nason:2004rx}.
Namely, any such a method requires a very good
NLO-level analytic understanding/control
of distributions from PSMC and, 
either NLO-level complete phase space coverage for the hard process
or a practical methodology of correcting for the lack of it.
Luckily,  a new wave of developments of the LO parton showers, see 
Refs.~\cite{Bellm:2013lba,Sjostrand:2014zea,Gleisberg:2008ta,Nagy:2014mqa},
has lead to modernized PSMCs, better suited for merging/matching with the fixed-order 
QCD calculations, in particular with better or even complete coverage
of the hard process phase space.

It is now obvious that the next challenge on the way to even higher-precision 
perturbative QCD calculations
needed until the end of the LHC era two decades from now,
is to combine the fully exclusive NNLO corrections
to the hard process and the NLO parton shower.
The fixed-order NNLO corrections to many processes are already well established,
see for instance Refs.~\cite{Anastasiou:2003ds,Bozzi:2007pn}, 
but the NLO PSMC needed for such a progress is still not available,
except of feasibility studies summarized in Ref.~\cite{Jadach:2013dfd}.
Interesting partial solution of combining the NNLO-corrected hard process
with the LO parton shower can be found in%
\footnote{%
  In these methods only certain selected important
  distributions are upgraded to the NNLO level.}
refs.~\cite{Hamilton:2012rf,Hoeche:2014aia,Alioli:2013hqa}.
The present work is relevant for the above future developments it the sense
that it presents a simplified method of correcting the hard process
to the NLO level in combination with the LO parton shower (PS).
In other words, it offers a simpler alternative to
the {\sf \mcatnlo{}} and {\sf \powheg{}} methods of 
Refs.~\cite{Frixione:2002ik,Nason:2004rx},
which may hopefully pave the way to the NNLO hard process combined
with NLO PSMC.

The new method described here, nicknamed as \krknlo{},
was already proposed in  Ref.~\cite{Jadach:2011cr}, 
where its first numerical implementation was done 
on top of the dedicated toy model PSMC 
and was limited to the gluonstrahlung subset of the NLO corrections.
Later on it was tested numerically in a more detail
in Refs.~\cite{Jadach:2012vs,Jadach:2013dfd}.
In the present work the \krknlo{} method is implemented within
the standard PSMC 
\sherpa{ 2.0.0}~\cite{Gleisberg:2008ta}.
A pilot study of \krknlo{} implementation outside PS MC, using MC
event encoded in the event records produced by \herwig{ 2.7.0}~\cite{Bahr:2008pv,Bellm:2013lba,Gieseke:2012ft} 
and \sherpa{ 2.0.0} was also done.
Let us stress, however, that the overall simplifications
of the \krknlo{} method comes not completely for free, 
as it requires to use parton distribution functions (PDFs) 
in a special Monte Carlo (MC) factorization scheme
(obtained, however, easily from reprocessing the $\msbar$ PDFs),
and it is required that the basic LO PSMC provides for
the NLO-complete coverage of the hard process phase space
(this is also not a problem for all modern PSMCs).
Our method is simpler to implement in the case of PSMC with an ordering
based on the transverse momentum $k_T$ or a $q^2$ variable of Ref.~\cite{Platzer:2009jq},
inspired by the classic Catani--Seymour work \cite{Catani:1996vz}.
However, it can be also easily implemented on top
of PSMC that uses the angular ordering --
without the need of the so-called truncated showers required
in the {\sf \powheg{}} method, 
see Refs.~\cite{Alioli:2011nr,Jadach:2012vs} for more discussion on that.

The main advantage of the \krknlo{} method is
a simplification of the NLO corrections due to the use of PDFs
in the MC factorization scheme.
The implementation of the entire NLO corrections with the help of
a single multiplicative simple weight on top of the LO distribution
is a quite unique feature of the \krknlo{} method.

Numerical studies presented here will be extended in the future
publications to a wider range of distributions, energies, implementation variants,
including comparisons with experimental  data.

The paper is organized as follows. In Section 2 we introduce the kinematics and the phase space
parametrization for the considered process. 
In Section 3 we describe in detail the \krknlo\ method.
Sections 4 and 5 contain some numerical results of the \krknlo\ implementation: 
Section 4 from the fixed-order NLO cross-checks 
while Section 5 from the NLO$+$PSMC comparisons with 
the \mcfm, \mcatnlo\ and \powheg\ programs for the main $Z$-boson observables. 
Section 6 summarizes the paper. 
In Appendix A we add some details on the first gluon emission in the backward evolution in PSMC.

\section{Kinematics and phase space parametrization}
\label{sec:kinematics}

\begin{figure}[t]
  \centering
  \includegraphics[width=0.35\textwidth]{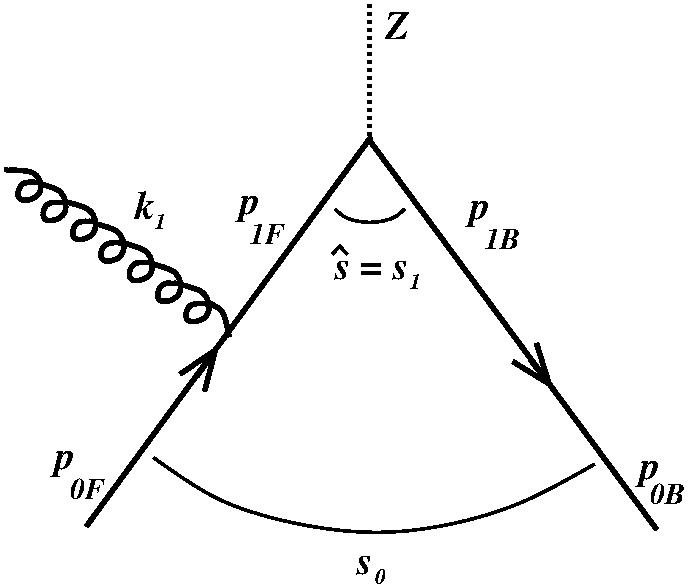}
  \caption{\sf
    Kinematics of $Z$-boson production in the $q\qbar$ channel.
  }
  \label{fig:DYdiag}
\end{figure}

In the present work we are going to concentrate on the Drell--Yan process, 
specifically production and decay of the heavy boson\footnote{For
brevity, in what follows, we shall often speak about the $Z$ boson only, but in
all cases we really mean $Z/\gamma^* \to \ell^+ \ell^-$.} $Z/\gamma^*$ in proton--proton
collisions. 
At the leading order (LO), 
$q\qbar \to Z$ is the only partonic subprocess that contributes. 
At the next-to-leading (NLO) level, 
the real correction $q\qbar \to Z g$ and
the virtual correction to $q\qbar \to Z$ contributes --
to be referred to collectively as $q\qbar$ channel.
The remaining NLO contributions, $q g\to Z q$ and  $\qbar g \to Z \qbar$,
are tree-level only --
to be called the $qg$ channel.

\subsection{Single emission}

Fig.~\ref{fig:DYdiag} illustrates part of the notation that will be used
throughout the paper.
The diagram shows the real correction to the $Z$-boson
production in the $q\qbar$ channel with the gluon four-momentum denoted by $k_1$.
The four-momenta of the incoming forward and backward partons, $p_{1F}$ and
$p_{1B}$ are related to
the four momenta of the incoming protons, $P_F$ and $P_B$, as follows
\begin{equation}
p_{0F} = x_F P_F\,, 
\qquad \qquad \quad
p_{0B} = x_B P_B\,,
\end{equation}
where
\begin{equation}
P_{F} = \frac{\sqrt{s}}{2}\left(1,0_T,1\right)\,, 
\qquad \qquad
P_{B} = \frac{\sqrt{s}}{2}\left(1,0_T,-1\right)\,.
\end{equation}

The invariant masses of the incoming parton pair prior to gluon emission
and that of the produced $Z$ boson 
are denoted by $s_0$ and $\shat=s_1$, respectively.
Their ratio%
\footnote{The subscript in $k_1$ and $s_1$ is kept to underline that, in the
context of PSMC, there is more parton emissions 
 further away from the hard process $q\qbar Z$ vertex.}
\begin{equation}
  z_1=\frac{s_1}{s_0}
    =\frac{\shat}{s_0}\,,
  \label{eq:zdef}
\end{equation}
can be related to light-cone variables of the emitted gluon,
as seen from the kinematics of the emitted gluon expressed
in terms of the \emph{light-cone Sudakov variables}
\begin{equation}
\label{eq:alpha-def}
\begin{split}
& \alpha_1  =  \frac{2 k_1 \cdot p_{0\B}} {s_0}\ =\ 
  \frac{2 k_1^+}{s_0},\quad
  \beta_1   =  \frac{2 k_1 \cdot p_{0\F}} {s_0}\ =\ 
  \frac{2 k_1^-}{s_0}\,,\quad
  \alpha_1 + \beta_1 \leq 1,\quad 
  \alpha_1, \beta_1 \geq 0,
\\&
  z_1      =  1 - \alpha_1 - \beta_1\,,  \quad
  k_{1T}^2  =  \alpha_1 \beta_1 s_0\,, \quad
  y_1      =   \frac{1}{2} \ln\frac{\alpha_1}{\beta_1}, \quad
  s_0      =  2p_{0\F} p_{0\B},
\end{split}
\end{equation}
spanned on the four-momenta $p_{1\F}$ and $p_{1\B}$ of the incoming partons,
prior to the gluon emission.
As seen in Eqs.~(\ref{eq:alpha-def}),
the variables $\alpha_1$, $\beta_1$ are simply the fractions of the light-cone
components of the gluon four-momentum to the centre-of-mass (CM) energy 
of the incoming partons (prior to the gluon emission).
The ratio $z_1$, the gluon transverse momentum $ k_{1T}$ and
the gluon rapidity $y_1$
are related to the $\alpha_1$, $\beta_1$ variables as well.

\begin{figure}[!t]
\centering
\includegraphics[width=90mm]{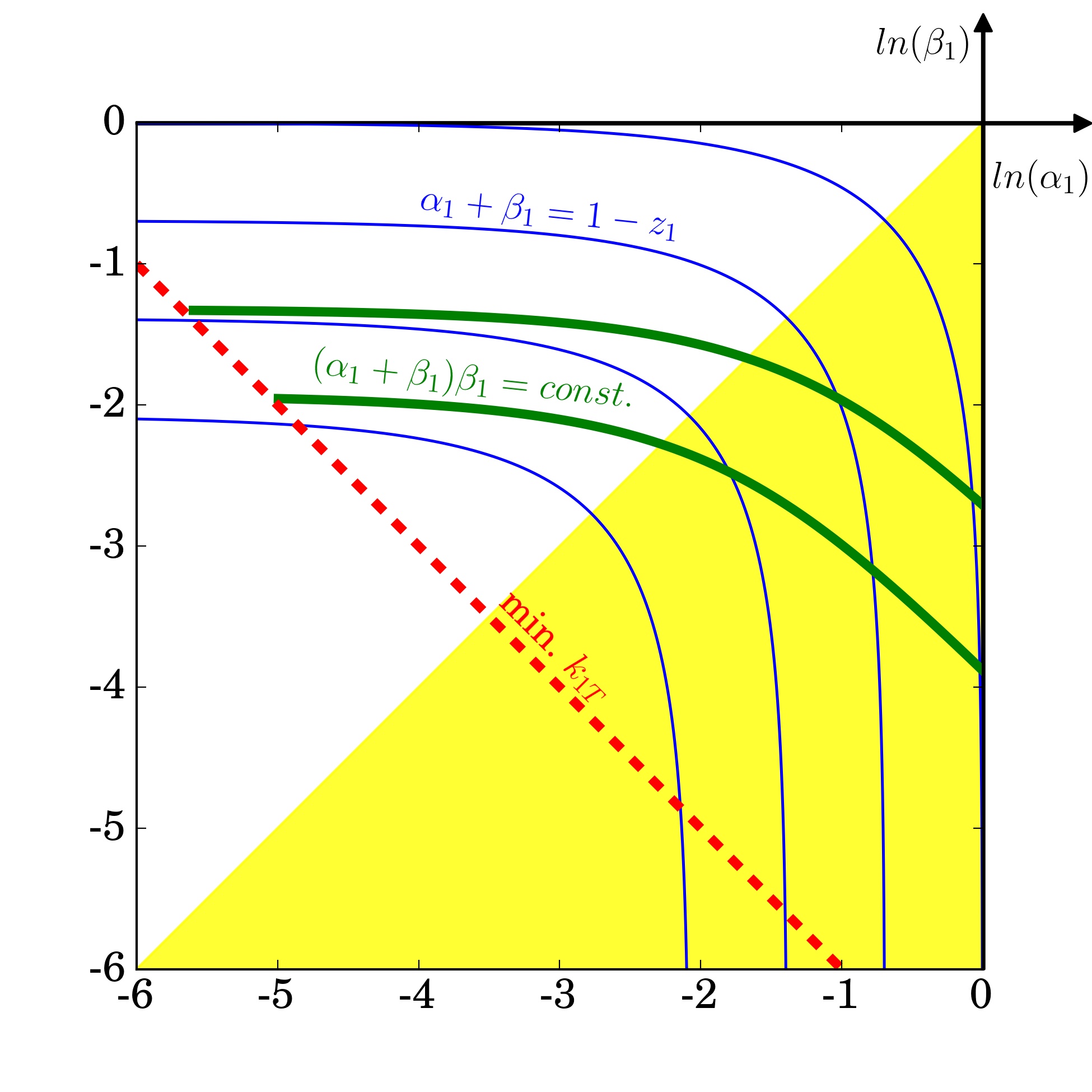}
\caption{\sf
The Sudakov logarithmic plane with lines marking
$q^2_{1\F}=\text{const.}$ (thick green)
and $z_1=\text{const.}$ (thin blue).  
The dashed (red) line marks IR cut-off on $k_{1T}=\text{const.}$.  
The shaded (yellow color) triangle marks the part of the phase space
where $\sim m_F$ dominates.}
\label{fig:sudakQ}
\end{figure}

In PSMC, the (eikonal) phase space measure of the emitted gluon
\begin{equation}
\frac{d^3 k_1}{2k_1^0}\;
\frac{1}{k_{1T}^2}
= \frac{\pi}{2}
\frac{d\phi_1}{2\pi}
\frac{d\alpha_1}{\alpha_1}
\frac{d\beta_1}{\beta_1}
\label{eq:single-eikonal-ps}
\end{equation}
is always split one way or another
into two parts belonging to the quark and antiquark emitters.
A sharp division along the $y_1=0$ angular boundary was used
in Ref.~\cite{Jadach:2011cr}, while in modern PSMCs,
see Refs.~\cite{Bellm:2013lba,Gleisberg:2008ta,Nagy:2014mqa},
a more gentle division, 
introduced in Refs.~\cite{Nagy:2008eq,Nagy:2008ns}
and inspired by the Catani--Seymour work \cite{Catani:1996vz},
is exploited:
\begin{equation}
\frac{d\alpha_1}{\alpha_1}
\frac{d\beta_1}{\beta_1}
(m_\F+m_\B)
= \frac{ d\alpha_1 d\beta_1}{\beta_1(\alpha_1+\beta_1)}
+ \frac{ d\alpha_1 d\beta_1}{\alpha_1(\alpha_1+\beta_1)},\quad
m_\F=\frac{\alpha_1}{\alpha_1+\beta_1},\;
m_\B=\frac{\beta_1}{\alpha_1+\beta_1},
\label{eq:CS-soft-partition-functions}
\end{equation}
where $m_{\F,\B}$ are the
so-called {\em soft partition functions}.
What is important for our methodology in the following is that the sum
of the two parts attributed to two emitters/showers reproduces
the original 1-particle phase space without any gaps and 
(or well-controlled) overlaps. 
Clearly the $m_F+m_B=1$ property of the functions used in
Eq.~(\ref{eq:CS-soft-partition-functions}) takes care of that.

Once the above (overlapping) separation of the emission phase space
into two part is applied,
a different {\em evolution variable} of PSMC in each of them
is defined
\begin{equation}
  q^2_{1\F}  = s_0 (\alpha_1+\beta_1)\beta_1, \qquad  \qquad
  q^2_{1\B}  = s_0 (\alpha_1+\beta_1)\alpha_1\,,
  \label{eq:q2vardef}
\end{equation}
instead of a common one, like $k_T^2=\alpha_1\beta_1 s_0$ or rapidity.

\begin{figure}[!t]
\centering
  \includegraphics[width=0.70\textwidth]{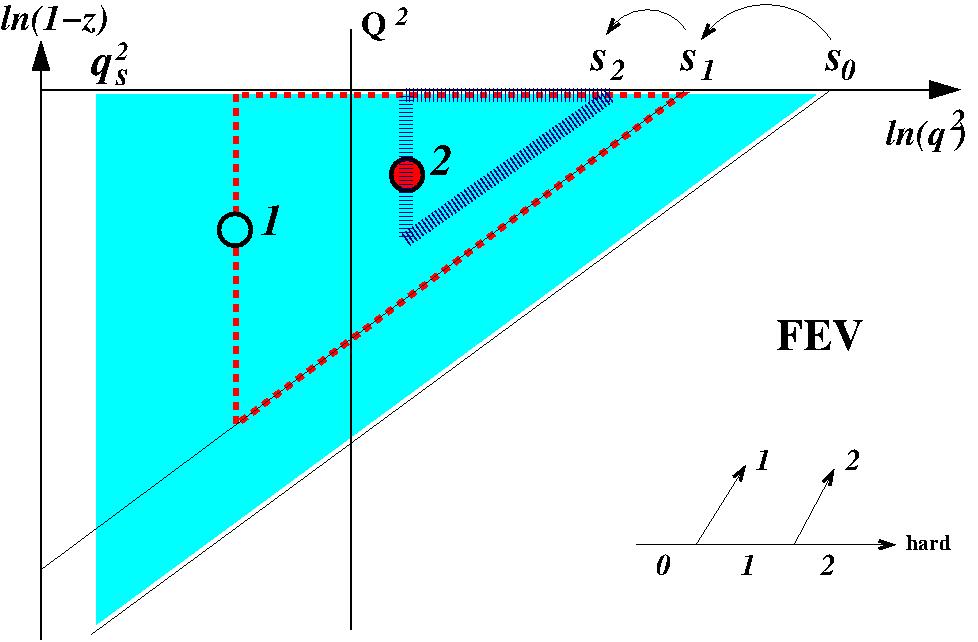}
\caption{\sf
The illustration of the kinematic boundaries in the forward evolution
(FEV)  algorithm with $n=2$ particles.
The parton no.~1 is emitted as a first one, 
within the biggest shaded triangular (blue) area
defined by $(1-z_1)^2> q_1^2/s_0$ and $q_1^2>q_s^2$.
The parton no.~2 is generated as a smaller triangular
area marked with dashed (red) line according to
$(1-z_2)^2> q_2^2/s_1$ and $q_2^2>q_1^2$.
Third parton is not generated, but its would-be-allowed space is
marked by the smallest triangle below $s_2=\hat{s}.$
The vertical line marking the upper phase-space boundary $q^2<Q^2$
in $\Dbar(Q^2,x)$ of Eq.~(\protect\ref{eq:PSMCfev2scale}) is also marked.
}
\label{fig:FEVkinema}
\end{figure}

Using the above evolution variable and $z_1$, the single-emission
(eikonal) phase space~(\ref{eq:single-eikonal-ps}) is easily re-parametrized
\begin{equation}
\frac{d\alpha_1 d\beta_1 }{\alpha_1\beta_1 }
= \frac{dq_{1\F}^2}{q_{1\F}^2} \frac{dz_1}{1-z_1}
 +\frac{dq_{1\B}^2}{q_{1\B}^2} \frac{dz_1}{1-z_1}.
\end{equation}
The relation between the old and new variables are illustrated
graphically in Fig.~\ref{fig:sudakQ}.
The transformation back to the Sudakov variables is different for each part:
\begin{equation}
\begin{split}
&\beta_1=\frac{(q^2_{1\F}/s_0)}{1-z_1},\qquad  \qquad
 \alpha_1=1-z_1-\frac{(q^2_{1\F}/s_0) }{1-z_1},
\\&
 \alpha_1=\frac{(q^2_{1\B}/s_0)}{1-z_1},\qquad  \qquad
 \beta_1=1-z_1-\frac{(q^2_{1\B}/s_0) }{1-z_1}.
\end{split}
\end{equation}
The upper phase-space limit $\alpha_1+\beta_1\leq 1$ transforms into 
\begin{equation}
\label{eq:q2zlimits}
z_1\geq 0 \qquad \qquad \text{and}\qquad \qquad q^2_{1\F,1\B}\leq s_0,
\end{equation}
while the positivity conditions, $\alpha_1>0$ and $\beta_1>0$,
enforce the IR-boundary cut-offs
\begin{equation}
(1-z_1)^2 > \frac{q^2_{1\F}}{s_0}
\qquad \qquad \text{or} 
\qquad \qquad (1-z_1)^2 > \frac{q^2_{1\B}}{s_0}\,,
\end{equation}
for the two parts, correspondingly.
Also, in most of the phase space region populated
according to the $\sim m_{\F}$ factor we may approximate  
$q^2_{1\F} \simeq k_{1T}^2$.
Similarly $q^2_{1\B} \simeq k_{1T}^2$ 
in the $\sim m_{\B}$ phase-space sector.
The above kinematical limits are also shown
on the logarithmic Sudakov plane in Fig.~\ref{fig:sudakQ} 
for the $\sim m_{F}$ sector.
NB. The IR cut-off $k_{1T}^2>k_{T \min}^2$ marked in this figure
translates into a slightly stronger
cut-off on $1-z_1$, easily calculable.
The same kinematical limits for one emission are also illustrated
directly in terms of the $1-z_1$ and $q_{1\F}^2$ variables
in Fig.~\ref{fig:FEVkinema}, 
including also the second emission for the purpose of the following discussion.

The essential ingredient of the 1-particle phase space reorganization
towards PSMC is introduction of the on-shell ``effective beams''
$p_{1\F}$ and $p_{1\B}$, such that $p_{1\F}+p_{1\B}=p_{0\F}+p_{0\B}-k_1$.
Their definition is not unique.
For example, for the $\sim m_F$ branch one may choose
\begin{equation}
\label{eq:p0FB}
p_{1\B}=(1-\epsilon) p_{0\B}, \quad
p_{1\F}=p_{0\F}-k_1 + \epsilon p_{0\B},\quad
\epsilon =\beta_1/(1-\alpha_1).
\end{equation}
The same for the $\sim m_B$ branch modulo the obvious variables interchange.
All that in the rest frame of $P_0=p_{0\F}+p_{0\B}$.
However, in practice of a typical PSMC with the backward evolution,
all four-momenta are reconstructed
starting from $q^2_{1\F,1\B}$ and $z_1$ variables in 
the rest frame of the effective beams $P_1=p_{1\F}+p_{1\B}$,
which are constructed in the first place.

Technical details of the construction of effective beams
are not so important for our analysis.
Just as an illustration let us define explicitly one possible construction,
which was introduced in Ref.~\cite{Platzer:2009jq}.
In the rest frame of the hard process and the effective beams 
$P_1=p_{1\F}+p_{1\B}=(\sqrt{\hat{s}},0,0,0)$
one may construct all four-momenta -- starting from the $(q_{1\F},z_1,\phi_1)$ set,
then translating it into $(\alpha_1,\beta_1)$ and using (for $\sim m_F$):
\begin{equation}
\label{eq:q2zalfbet}
\begin{split}
&k_1^\mu = \alpha^*_1 p_{1\F}^\mu  + \beta^*_1 p_{1\B}^\mu +q_{T,1}^\mu,
\quad q_{T,1}^2= \alpha^*_1 \beta^*_1  \hat{s},
\\&
p_{0\B}^\mu = \frac{\beta_1+z_1}{z_1} p_{1\B}^\mu ,\quad
p_{0\F}^\mu = P_1^\mu +k_1^\mu - p_{0\B}^\mu,
\\&
\alpha^*_1=\frac{1-\beta_1-z_1}{\beta_1+z_1}\; ,\quad
\beta^*_1=\frac{1}{\beta_1+z_1}\; \frac{\beta_1}{z_1}.
\end{split}
\end{equation}
Finally, knowing $P_0=P_1-k_1$ one may transform all newly constructed four-momenta 
$k_1,\; p_{0\F},\; p_{0\B} $ to the rest frame of $P_0$. 

Altogether, the complete reorganization of the 1-real emission phase space
from a simple form based on the Sudakov variables to an equivalent
parametrization using the variables of PSMC, based on the backward evolution
algorithm (applying the Catani--Seymour soft-partition factor $m_F$),
keeping track of the kinematical limits,
and defining 
$\hat{x}=x_1=x_0z_1$, $\hat{s}=s_1=sx_1$,
looks as follows:
\begin{equation}
\label{eq:kinema_single}
\begin{split}
d\sigma_{1\F} &\simeq
   \int\limits_0^{1}\!\! dx_1\;  D(\mu_F^2,x_1)
   \int\limits_{\alpha_1+\beta_1\leq 1} 
       \frac{d\beta_1 d\alpha_1}{\beta_1 (\alpha_1+\beta_1)}\;
   \bar{P}(z_1)\;
   d\sigma_0(sx_0z_1)
\\&
 =\int\limits_0^1 dx_1
  \int\limits_{q^2_{\min}}^{s_0} \frac{dq^2_1}{q^2_1}
  \int\limits_0^{1-\sqrt{q_1^2/s_0}}\!\!\! dz_1\; 
   \frac{\bar{P}(z_1)}{1-z_1}\;
   d\sigma_0(sx_1) \;
   D(\mu_F^2,x_1)
 \\&
 = \int\limits_0^{1}\!\! d\hat{x}
   \int\limits_{q^2_{\min}}^{s} \frac{dq^2_1}{q^2_1}
   \int\limits_{x_0}^{1} dz_{1}\;
   \theta_{ (1-z_1)^2 s\hat{x}/z_1 \geq q_1^2  } \;
   P_{q\qbar}(z_1)\;
   d\sigma_0(s\hat{x})\; 
   D\Big( \mu_F^2, \frac{\hat{x}}{z_1}\Big),
 \end{split}
\end{equation}
where $q_1^2=q_{1\F}^2$,
$\bar{P}(z)\equiv (1-z)P_{q\qbar}(z)$
and $P_{q\qbar}(z)$ is the DGLAP~\cite{DGLAP} splitting function. 
It is important to see that the full phase coverage requires
integration over $q_\F^2$ to extend above the effective mass squared
$\hat{s}=s_2$ of the LO hard process.
Since  $q_\F^2 \simeq k_{1T}^2$, it means that the transverse momentum
above $\hat{s}$ is included in the above phase space.

In Section~\ref{sec:NLOwith-PS} presenting numerical results,
it will be commented more on what kind of evolution variable
is chosen in the parton shower generation of \sherpa{} and \herwig.

\subsection{Multiple emissions}
\label{eq:MultiEmissionKinema}

Although the above 1-emission
kinematics is enough  for most of our prescription
for the NLO-correcting of the hard process,
for better understanding of the role of the
underlying multi-emission PSMC, it is useful
to extend the 1-emission treatment of the
kinematics to two and more emissions
in the initial-state parton showers.

\begin{figure}[ht]
\centering
\includegraphics[width=168mm]{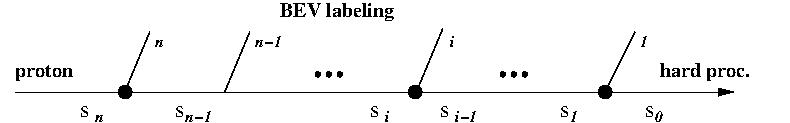}
\caption{\sf Labelling in the reconstruction of the four-momenta.
}
\label{fig:BEVlabeling}
\end{figure}
The important technical point is the choice of the numbering (labelling)
of the emitted particles.
In Fig.~\ref{fig:FEVkinema} we use the numbering 
in the emission chain (ladder) starting at the incoming
hadron and ending next to the hard process,
that is the labelling of the forward evolution (FEV) algorithm.
From now on we switch to the backward evolution (BEV) labelling which
starts next to the hard process and ends at the incoming hadron,
see Fig.~\ref{fig:BEVlabeling} for illustration.

Let us summarize briefly on 
the \emph{effective beam technique}, for simplicity
limiting its description to a single tree $F$ of emissions (shower).
The sequence of the effective emitter beams
$(\tilp_{i\F},\tilp_{i\B})$ is defined starting from the hard process, with
the four-momentum $P_i$, such that
\[
 P_i =\tilp_{i\F}+\tilp_{i\B},\quad
 P_{i+1}=P_{i}+k_{i},\quad 
 P_i^2=\hat{s}/\Big(\prod_{j=1}^i z_i \Big),
\]
and they are used to span the four-momentum of the emitted gluon
\[
 k_i=\talf_i \tilp_{i\F} + \tbet_i \tilp_{i\B} +k_{Ti},\quad
 \talf_i = \frac{ k_i\tilp_{i\B} }{\tilp_{i\F}\tilp_{i\B} },\quad
 \tbet_i =\frac{ k_i\tilp_{i\F} }{\tilp_{i\F}\tilp_{i\B} },
\]
introducing the Sudakov variables $\talf_i,\;\tbet_i $.
These Sudakov variables are related to the evolution variable $q_{i\F}^2$ 
and the light-cone variable $z_i$ of PSMC
as follows:
\begin{equation}
\tbet_i=\frac{q^2_i\; Z_i}{\hat{s} (1-z_i)},\quad
\talf_i=1-z_i-\tbet_i,\quad Z_i= \prod_{j=1}^i z_j,\quad
\hat{s}=s\hat{x}=sx_0.
\end{equation}

\begin{figure}[!t]
\centering
  \includegraphics[width=0.70\textwidth]{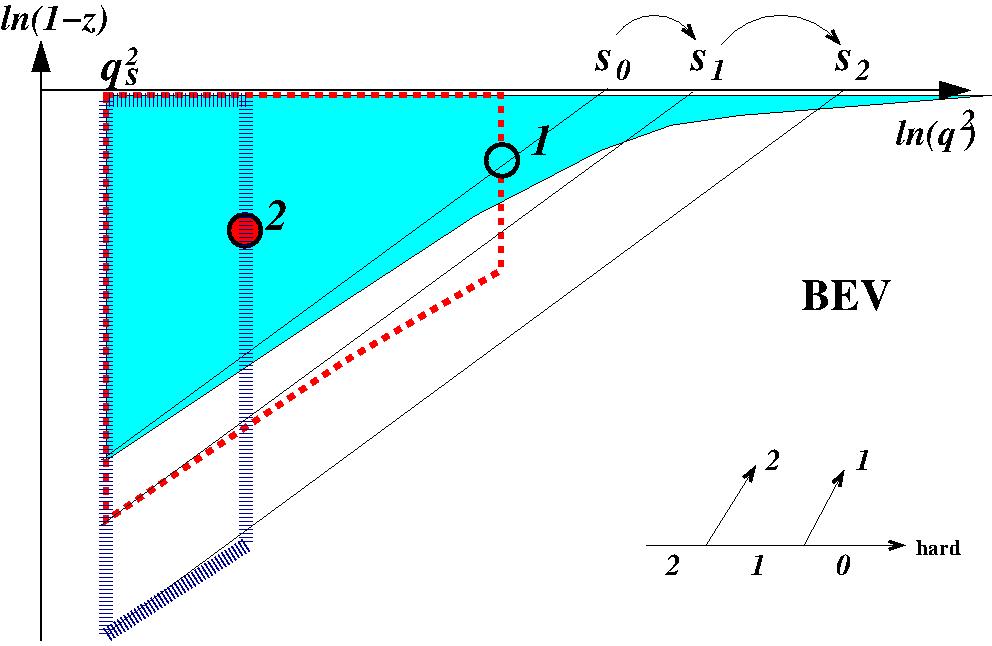}
\caption{\sf
The illustration of the kinematic boundaries in the BEV algorithm
with $n=2$ particles.
The parton no.~1 is emitted 
within the semi-triangular shaded (blue) area
defined by $(1-z_1)^2> q_1^2/(s_0/z_1)$ and $s>q_1^2>q_s^2$.
The parton no.~2 is generated within the second
semi-trapezoid area marked by the dashed (red) line according to
$(1-z_2)^2> q_2^2/(s_1/z_2)$ and $q_1^2>q_2^2>q_s^2$.
The third parton is not generated, but its would-be-allowed space
is marked by the leftmost almost-trapezoid line.
}
\label{fig:BEVkinema1}
\end{figure}

Finally, in the recursive backward reconstruction of the four-momenta
starting from the hard process,
one employs the Sudakov-like decomposition of the emitted parton
in terms of the emitters {\em after} the emission%
\footnote{Generalizing Eq.~(\ref{eq:q2zalfbet}).}:
\begin{equation}
\begin{split}
&
 k_{i}^\mu = \alpha^*_i \tilp_{(i-1)\F}^\mu  
           +\beta^*_i \tilp_{(i-1)\B}^\mu +q_{T,i}^\mu,\quad
\\&
 \alpha^*_i=\frac{\talf_i}{1-\talf_i}\; ,\quad
 \beta^*_i =\frac{1}{1-\talf_i}\; \frac{\tbet_i}{z_i}.
\end{split}
\end{equation}
All exact kinematical limits (including ordering in the evolution variable)
are represented by the following inequalities:
\begin{equation}
\label{eq:orderBEV}
 (1-z_i)^2 \geq q_i^2 Z_i /\hat{s} \qquad \text{and} \qquad
 q^2_{i+1} \leq q^2_{i} \leq \hat{s} (1-z_i)^2/Z_i.
\end{equation}
In particular, the kinematical limits for the first emission in the 
backward-evolution (BEV) labelling are
\[
q^2_{\max,1} = \hat{s} (1-z_1)^2/z_1,\quad
z^{\max}_1   \simeq  1-q_0/\sqrt{\hat{s}}.
\]
The above kinematical limits in terms of the BEV variables $q_i^2$ and $z_i$
look more complicated than in the FEV scenario,
although they represent exactly the same phase space region,
and are illustrated graphically in Fig.~\ref{fig:BEVkinema1}.

\begin{figure}[!t]
\centering
  \includegraphics[width=0.70\textwidth]{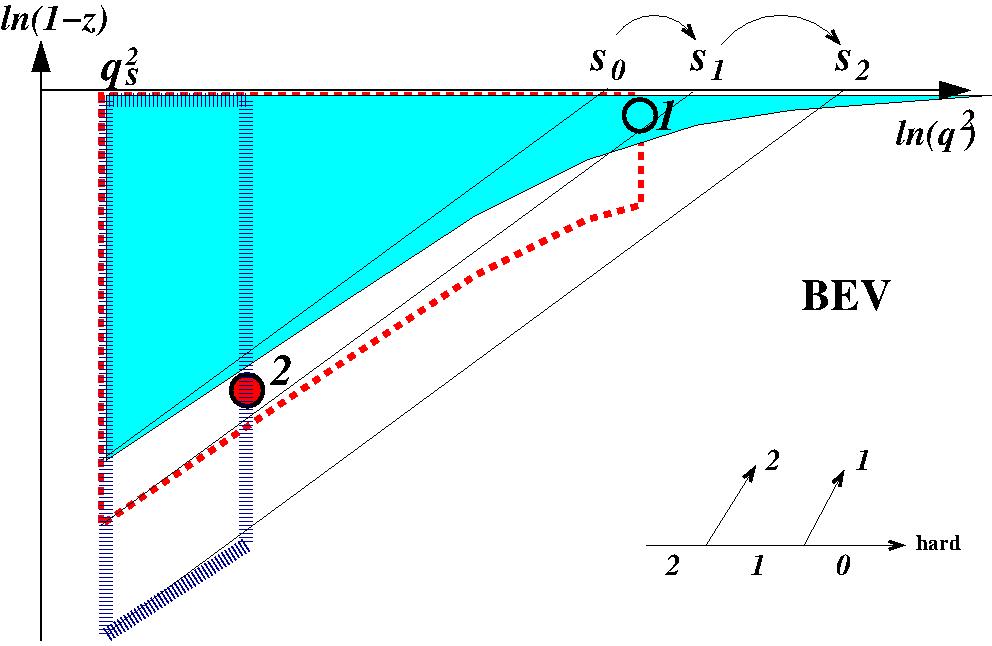}
\caption{\sf
The illustration of the kinematic of the BEV algorithm
with $n=2$ particles in the extreme case when
the parton no.~1 is emitted with $q_1^2>s_0=\hat{s}$ and
the parton no.~2 is generated within the area not accessible for
the parton no.~1 due to a higher IR boundary.
}
\label{fig:BEVkinema2}
\end{figure}
As already underlined, in terms of the BEV variables,
the phase space gets {\em apparently widened} after each emission,
for instance $q^2>\hat{s}$ is already available for the first emission
and, due to lowering of the IR boundary by the $1/z_1$ factor,
more phase space is available for the second emission.
This phenomenon, important for the full coverage of the phase space,
is illustrated graphically in Fig.~\ref{fig:BEVkinema2}.
It is, of course, an artefact of the BEV phase-space parametrization,
which in the FEV world corresponds to the phase-space {\em reduction} 
due to energy conservation.

We omit here the discussion of the ``kinematical cross-talk'' between two parton showers,
which means that for emissions 
with the common $q^2$-ordering in two initial showers
(as in any realistic PSMC),
the emission in one shower reduces the available phase space
in the other shower.
This effect is easily incorporated in the kinematical construction (mappings)
of PSMC.
The only thing one has to watch out is  
the correctness of the soft-emission limit in the case of two and more emissions,
see for instance the discussion in Ref.~\cite{Platzer:2009jq}.
This subject will be covered in a more detail in our future publications.

\section{KrkNLO methodology}
\label{sec:krknlo-meth}
Very briefly, the essence of the \krknlo{} prescription
defined in Ref.~\cite{Jadach:2011cr} is that NLO corrections are introduced
by a multiplicative weight on top of distributions from the LO PS
acting, either inside PSMC or outside it, on a MC event record.
This NLO weight is sensitive to the parton four-momentum with the highest $q^2$
(or maximum $k_T^2$  in the case of $k_T$-ordering),
although in Ref.~\cite{Jadach:2011cr} it was demonstrated that such a
NLO multiplicative weight works also in the case of the angular ordering,
provided that summation over all emitted partons is performed.

In \mcatnlo{}~\cite{Frixione:2002ik}, the correcting weight is essentially additive and the 
NLO$\,-\,$LO correction to PSMC is provided from outside 
in form of additional MC events, with a non-positive-definite weight.
In parts of the phase space which are not covered by the LO PS, 
extra events provide the entire NLO distributions (positive weights), 
otherwise extra events, 
correcting the LO distributions to the NLO level,
have typically (inconvenient) negative weights.

In \powheg{}~\cite{Nason:2004rx}, the entire NLO correction to LO PS  
is provided by an external MC generator 
-- the highest $k_T^2$ emission is isolated/subtracted from PSMC,
following the double-logarithmic Sudakov exponential factor,
and generated according to the NLO distribution outside PSMC,
while trailing emissions with lower $k_T^2$ 
(suppressed by the Sudakov exponent)
are left for generating within LO PSMC.

Both \krknlo{} and \mcatnlo{} require good analytical control of the LO PS distributions,
while for \powheg{} it is less important.
In addition, \krknlo{} requires that LO PS fills in the entire NLO phase space
with some non-zero distribution.
The Sudakov suppression is also exploited in \krknlo{}, 
but in a different way than in \powheg{}.

In the standard NLO corrections to a hard process with $\msbar$ PDFs,
part of the NLO corrections feature degenerate 
1-dimensional longitudinal phase space with $k_T^2=0$ exactly.
These corrections enforce in practice certain convolution on the $\msbar$ PDFs, 
which in the case of \powheg{} and \mcatnlo{} is done in-flight inside MC.
In the \krknlo{} prescription, the implementation of these corrections 
is moved outside MC (simplifying it).
This leads to the use  in \krknlo{} of the modified LO PDFs, 
in the so-called MC factorization scheme.
This reorganization is mandatory
because one of the main aims of the \krknlo{} method,
that the NLO corrections are implemented with a well-behaved multiplicative 
positive weight, is not compatible with such a degenerate collinear phase-space
contribution to the NLO MC distributions.

Last but not least, any scheme for correcting the hard process of LO PSMC
to the NLO level requires a formal proof that the resulting distributions
are indeed of the NLO class (no double-counting, no NLO leak).
Such a proof in an algebraic form is not trivial, 
not only because the NLO total cross section has to be verified, 
but also any NLO-class observable (experimental event selections) 
has to be properly reproduced.
In other words, it has to be done using functional space of 
{\em all}\ the NLO observables.
In the case of \krknlo{} such a proof was done in Ref.~\cite{Jadach:2011cr},
both algebraically and numerically,
for a toy-model PSMC with the angular ordering.
Here, we shall provide such an algebraic proof starting from
the NLO-corrected multiparton distributions for
realistic PSMC of the kind implemented in \sherpa{} and \herwig{}
using the BEV algorithm.

In the following, we are going to collect building blocks for the NLO weight,
then we shall elaborate on the multiparton distributions of LO PSMC
without and with the NLO weight of \krknlo{}.
We shall pay particular attention to the question of the completeness
of the phase space in PSMC and to the equivalence
between the backward and forward evolution algorithms in PSMC.
Finally, we shall show that for an arbitrary NLO-class observable, 
\krknlo{} gives the same result as simpler NLO calculations 
with the collinear PDFs, instead of PSMC, 
such as for instance \mcfm{}~\cite{MCFM}.

\subsection{NLO-correcting weight}
\label{sec:sigNLO}
Let us collect the ingredients for construction of the NLO corrections
to the hard process of the $Z$-boson production and decay
in the proton--proton collisions.

The fully differential NLO cross section of the production 
and decay of the $Z$ boson in the quark--antiquark annihilation process,
with the simultaneous emission of a single real gluon, can be cast 
in a well-known compact form, see Ref.~\cite{Jadach:2011cr}:
\begin{equation}
d^5\sigma^{NLO}_{q\qbar}(\alpha,\beta, \Omega)=
\frac{C_F \alpha_s}{\pi}\;
\frac{d\alpha d\beta}{\alpha\beta}\;
\frac{d\varphi}{2\pi}\;
d\Omega
\Bigg[
 \frac{d\sigma_0(\hat{s},\theta_\F)}{d\Omega}
 \frac{(1-\beta)^2}{2}
+\frac{d\sigma_0(\hat{s},\theta_\B)}{d\Omega}
 \frac{(1-\alpha)^2}{2}
\Bigg],
\label{eq:nlo-gen-formula}
\end{equation}
where the Sudakov variables\footnote{For better readability of the formulae
given in this section, we shall suppress ``1'' in the lower index of the Sudakov
variables as well as other kinematical variables,  such as $s$ 
and $q^2_{F,B}$.} ($\alpha,\beta$) are spanned on
momenta of the effective beams of $q$ and $\qbar$ prior to the gluon emission, see
Eq.~(\ref{eq:alpha-def}).
The Born differential cross section
$\frac{d\sigma_0(\hat{s},\theta_\B)}{d\Omega}$ 
for $Z$-boson production and decay is well known
(see for instance \cite{Alioli:2011nr} for the exact expression).
The solid angle $\Omega=(\theta,\phi)$ is the direction of the lepton
from the decaying $Z$ boson in its rest frame
and $\hat{s} = (1-\alpha-\beta)s_1$ (see Fig.~\ref{fig:BEVlabeling}).
The angles $\theta_\F$ and $\theta_\B$ depend on $\alpha$ and $\beta$ as well --
their precise definition was given%
\footnote{
 In fact they coincide with the polar angles with respect to the effective
 beams defined in Eq.~(\ref{eq:p0FB}).
}
in Ref.~\cite{Berends:1980jk}.
The integration over luminosities
of $q$ and $\qbar$ is not yet included.

How does  the above compare with the distributions from PSMC
also restricted to the single gluon emission?
The differential cross section for
the gluon emission from the quark emitter (\ie the $\sim m_F$ part in
Eq.~(\ref{eq:CS-soft-partition-functions})) in PSMC reads
\begin{equation}
\label{eq:dsig5LOF}
\begin{split}
d^5\sigma^{\F}_{q\qbar}(\alpha,\beta, \Omega)
&=\frac{C_F \alpha_s}{\pi}\;
  \frac{dq_\F^2}{q_\F^2}\;
  \frac{d\varphi}{2\pi}
   P_{q\qbar}(z) dz\;
\frac{d\sigma_{0}}{d\Omega}\big(\hat{s},\hat\theta \big) d\Omega
\\&
=\frac{C_F \alpha_s}{\pi}\;
 \frac{d\alpha d\beta}{(\alpha+\beta)\beta}\;
 \frac{d\varphi}{2\pi} d\Omega\;
 \frac{1+(1-\alpha-\beta)^2}{2}
 \frac{d\sigma_{0}}{d\Omega}\big(\hat{s},\hat\theta \big),
\end{split}
\end{equation}
where $q_F$ is the evolution variable defined in Eq.~(\ref{eq:q2vardef})
and $\hat\theta$ is another effective angle in $Z$ decay specific to LO PSMC,
for instance the so-called Collins--Soper angle \cite{Collins:1977iv}.
Adding the gluon emission from $\qbar$ simply amounts to
the $\alpha\leftrightarrow\beta$ symmetrization, resulting in
\begin{equation}
\label{eq:dsig5LO}
d^5\sigma^{\lo}_{q\qbar}(\alpha,\beta, \Omega)
=d^5\sigma^\F_{q\qbar} + d^5\sigma^\B_{q\qbar}
=\frac{C_F \alpha_s}{\pi}\;
 \frac{d\alpha d\beta}{\alpha\beta}\;
 \frac{d\varphi}{2\pi} d\Omega\;
 \frac{1+(1-\alpha-\beta)^2}{2}
 \frac{d\sigma_{0}}{d\Omega}\big(\hat{s},\hat\theta \big),
\end{equation}
where
\begin{equation}
d^5\sigma^\F_{q\qbar}  = \frac{\alpha}{\alpha+\beta} d^5\sigma^{\lo}_{q\qbar}
              = m_\F d^5\sigma^{\lo}_{q\qbar},\quad
\qquad
d^5\sigma^\B_{q\qbar}  = \frac{\beta}{\alpha+\beta}  d^5\sigma^{\lo}_{q\qbar}
              = m_\B d^5\sigma^{\lo}_{q\qbar}.
\end{equation}
The integration limits are not explicit, but they are
the same as in Eq.~(\ref{eq:alpha-def}).

In our discussion, we shall often use the following objects:
the additive NLO correction
\begin{equation}
\label{eq:d5beta1}
 d^5\bar\beta_{q\qbar} (\alpha,\beta, \Omega)
= d^5\sigma^{\nlo}_{q\qbar}(\alpha,\beta, \Omega)
- d^5\sigma^{\lo}_{q\qbar}(\alpha,\beta, \Omega)
\end{equation}
and the NLO multiplicative weight for the $q\qbar$ channel
\begin{equation}
\label{eq:W1nlo}
W^{(1)}_{q\qbar}(\alpha,\beta)
   = 1+\frac{d^5\bar\beta_{q\qbar}}{d^5\sigma^{\lo}_{q\qbar}}
   = \frac{d^5\sigma^{\nlo}_{q\qbar}}{d^5\sigma^{\lo}_{q\qbar}}.
\end{equation}
The above weight is especially simple in the case of averaging over
the angles in $Z$ decay:
\begin{equation}
\langle W^{(1)}_{q\qbar} \rangle_{\Omega}=
W^{q\bar q}_R =  1- \frac{2\alpha\beta}{1+(1-\alpha-\beta)^2}\,,
\label{eq:wt-real}
\end{equation}
and it can be used in approximate implementations of the NLO corrections.

The analogous weight for the $qg$ channel is
\begin{equation}
\langle W^{(1)}_{qg} \rangle_{\Omega}=
W^{qg}_R  = 1 + \frac{\beta(\beta + 2z)}{(1-z)^2 + z^2}\,.
\label{eq:wt-real-qg}
\end{equation}

After summing up the contributions from the two emitters,
$d^5\sigma^{\lo}_{q\qbar}$
is obtained, which is exactly the same%
\footnote{
 In spite of the differences of the LO PSMC distributions
 for each emitter separately.}
as in Ref.~\cite{Jadach:2011cr}.
The important consequence of the above is that
many details of the matching between the $\msbar$ NLO
corrections and the LO PSMC in the present \krknlo{} implementation
remains the same as in Ref.~\cite{Jadach:2011cr}.
In particular, the virtual plus soft-real correction, 
when the PDFs in the MC factorization scheme are used,
is the same as in Ref.~\cite{Jadach:2011cr} and reads\footnote{Note that ${\bar
B}^{ij}_\mc$, though conceptually similar, is not identical to the well known
$\bar B$ used in the context of \powheg{}~\cite{Frixione:2007vw}, as the quantity
from Eq.~(\ref{eq:wt-ovs}) is defined in the MC scheme.
}
\begin{equation}
{\bar B}^{ij}_\mc = 1+\Delta_{VS}^{ij}, 
\label{eq:wt-ovs}
\end{equation}
where 
\begin{align}
\Delta_{VS}^{q\bar q}  &=  
      \frac{\alpha_s}{2\pi} C_F \left[\frac{4}{3}\pi^2 -\frac{5}{2}\right]\,,
&
\Delta_{VS}^{qg}  &=   0.
\label{eq:wt-vs}
\end{align}
As one can see, this virtual+soft-real correction is constant, \ie kinematics-independent.

\subsection{MC factorization scheme}
\label{sec:mc-scheme}

In this subsection we extend the definition of the MC factorization scheme, 
introduced in Ref.~\cite{Jadach:2011cr} for the quark--antiquark channel, 
to the NLO DY process with the quark--gluon initial state.  For the completeness and convenience
of the reader we first provide the main formulae for the $q\bar{q}$ initial state.

The NLO $q\bar{q}$-channel  coefficient function for the DY process in the $\overline{\rm MS}$
factorization scheme is given by \cite{Jadach:2011cr}
\begin{equation}
C_{2q}^{\overline{\rm MS}}(z) = \frac{\alpha_s}{2\pi}\, C_F \left\{ \delta(1-z) \left(\frac{4}{3}\pi^2 - \frac{7}{2}\right) + \left[2\,\frac{1 + z^2}{1-z} \ln\frac{(1-z)^2}{z} \right]_+ \right\}.
\label{eq:C2q-MSbar}
\end{equation}
The corresponding coefficient function in the MC factorization scheme, 
defined in Ref.~\cite{Jadach:2011cr}, reads
\begin{equation}
C_{2q}^{\rm MC}(z) = \frac{\alpha_s}{2\pi}\, C_F \left[\delta(1-z) \left(\frac{4}{3}\pi^2 - \frac{7}{2}\right) - 2(1-z)_+\right] =  \delta(1-z) \Delta_{VS} - \frac{\alpha_s}{\pi}\, C_F (1-z),
\label{eq:C2q-MC}
\end{equation}
where $\Delta_{VS} \equiv \Delta_{VS}^{q\bar q}$ is the virtual plus soft-real gluonstrahlung
correction given in Eq.~(\ref{eq:wt-vs}).

From the above equations, following Ref.~\cite{Jadach:2011cr}, 
we can obtain a $q\bar{q}$ contribution
to the relation between the  MC-scheme and  $\overline{\rm MS}$-scheme quark (antiquark) PDFs:
\begin{equation}
\Delta C_{2q}(z) = \frac{1}{2}\left[ C_{2q}^{\overline{\rm MS}}(z)\; -\; C_{2q}^{\rm MC}(z) \right] = \frac{\alpha_s}{2\pi}\, C_F \left[ \frac{1+z^2}{1-z}\ln\frac{(1-z)^2}{z} + 1-z \right]_+ .
\label{eq:DeltaC2q}
\end{equation}

Similarly, for the NLO $qg$-channel contribution to the DY process we have:
\begin{eqnarray}
C_{2g}^{\overline{\rm MS}}(z) &=& \frac{\alpha_s}{2\pi}\, 
T_R \left\{ \left[z^2+(1-z)^2\right] \ln\frac{(1-z)^2}{z} 
   - \frac{7}{2}z^2 + 3z + \frac{1}{2}\right\},
\label{eq:C2g-MSbar} \\
C_{2g}^{\rm MC}(z) &=& \frac{\alpha_s}{2\pi}\, T_R \,\frac{1}{2}(1-z)(1+3z),
\label{eq:C2g-MC} \\
\Delta C_{2g}(z) &=& C_{2g}^{\overline{\rm MS}}(z)\; -\; C_{2g}^{\rm MC}(z)= \nonumber
\\ 
     & = & \frac{\alpha_s}{2\pi}\, T_R 
        \left\{ \left[z^2+(1-z)^2\right] \ln\frac{(1-z)^2}{z} + 2z(1-z)\right\}.
\label{eq:DeltaC2g}
\end{eqnarray}

Using Eqs.~(\ref{eq:DeltaC2q}) and (\ref{eq:DeltaC2g}), we can relate the MC-scheme quark (antiquark)
PDFs to the corresponding $\overline{\rm MS}$-scheme PDFs in the following way
\begin{equation}
f_{q(\bar{q})}^{\rm MC}(x,Q^2) = f_{q(\bar{q})}^{\overline{\rm MS}}(x,Q^2)\; + \int_x^1 \frac{dz}{z}\,
f_{q(\bar{q})}^{\overline{\rm MS}}\left(\frac{x}{z},Q^2\right) \Delta C_{2q}(z) \; + \int_x^1
\frac{dz}{z}\, f_g^{\overline{\rm MS}}\left(\frac{x}{z},Q^2\right) \Delta C_{2g}(z)\,. 
\label{eq:PDFs-MC-quark}
\end{equation}
The above relation is universal, \ie\ process-independent, at the NLO level.
It is simply because it is defined uniquely with respect to the $\msbar$  scheme.
 
The gluon PDF is equal between the MC and $\msbar$ schemes up to the
$\order{\as^2}$ corrections for processes with no gluons at the Born
level, such as the Drell--Yan process considered in this work. 
Hence for the DY process we may use
\begin{equation}
f_g^{\rm MC}(x,Q^2) = f_g^{\overline{\rm MS}}(x,Q^2)\,.
\label{eq:PDFs-MC-gluon}
\end{equation}

As one can see, in the MC factorization scheme 
the NLO coefficient functions in both the $q\bar{q}$
and $qg$ channels are substantially simpler 
than the corresponding ones in the $\overline{\rm MS}$ scheme;
in particular they are free of logarithmic singular terms. 
Since the latter terms are absorbed into the MC PDFs,
i.e. they are included in a resummed way, 
one may also expect that the higher-order QCD corrections  
in the MC scheme are smaller than in the standard $\overline{\rm MS}$ scheme.

Let us finally add a comment
on the universality (process-independence) of the MC factorization scheme.
This issue was discussed quite extensively in Ref.~\cite{Jadach:2011cr}
for the  DIS and DY processes,
albeit restricting the problem to QED-like gluonstrahlung diagrams only.
The main point is that the MC factorization scheme 
is in reality defined operationally as a modification of the $\msbar$ scheme, 
consequently it inherits automatically the universality from the latter%
\footnote{
In particular, the MC scheme is neither the DIS nor the DY scheme,
as can be seen from the fact that the coefficient functions
for any of these processes in the MC scheme are not equal to $\delta(1-z)$.
}.
On the other hand, it is true
that the procedure of defining extra ${\cal O}(\epsilon^{0})$ terms, 
added to the $\msbar$ collinear counter-terms to define the PDFs in the MC 
scheme, is clearly guided by the inspection of a number
of simplest physical processes, like DIS, DY and/or Higgs production.
However, once this is done, these extra terms are frozen 
and the resulting (modified) counter-terms are ready to apply for any other process.
They also define completely the PDFs in the MC scheme.
What is new in the present work with respect to Ref.~\cite{Jadach:2011cr}
is the inclusion of the quark--gluon transitions.
Generally, the transformation of the PDFs from the $\msbar$
to the MC scheme is a matrix in the flavor space.
The DY process at the NLO level, discussed in the present work, 
fixes only a subset of terms in this matrix,
while the remaining ones will be fixed by inspecting the NLO corrections
to the process of gluon--gluon fusion into the Higgs boson%
\footnote{
They are set temporarily to zero in the present work,
as for the DY process they become relevant only at NNLO.}.
At the next step, after including Higgs production in the game,
the MC scheme will be fully defined and will be applicable to any process,
including also more color particles in the final state.

\subsection{Multiparton reality of PSMC}
\label{sec:PSreality}
In the above we have restricted ourselves to a single emission 
and hence omitted all the multi-emission reality of PSMC.
Let us elaborate on that in a more detail now, 
because the NLO correcting weights are put not on top of single-emission distributions
but on top of multiparton distributions of LO PSMC, so they have to
be known and controlled for the hard process within the NLO precision,
preferably in a closed algebraic form.

Let us start from LO PSMC in the forward-evolution (FEV) formulation.
The equivalent backward-evolution (BEV) formulation will be presented later on.
Restricting ourselves temporarily to the pure gluonstrahlung case,
the FEV differential cross sections of gluons
emitted  from the $q$ and $\qbar$ emitters%
\footnote{We adopt a convention in which $\sum_{n=1}^0 d_n =1 $.}
reads as follows
\begin{equation}
\label{eq:PSMCfev4}
\begin{split}
&\sigma^{\lo}_\mc
=\int d x_{\F} dx_{\B} d\Omega\;
\sum_{n_\F=0}^{\infty}\; \sum_{n_\B=0}^{\infty}
\int d\sigma^{\lo}_{n_\F n_\B},
\\&
d\sigma^{\lo}_{n_\F n_\B}=
\prod_{i=1}^{n_\F} 
\prod_{j=1}^{n_\B} 
\Big\{
  \int d^3\rho^\F_i \theta_{q^2_{i-1} > q^2_{i} > q_s^2}\;
  e^{ -S_\F(q_{i-1}^2,q_i^2) }
\Big\}
\Big\{ 
  \int d^3\rho^\B_j  \theta_{q^2_{j-1} > q^2_{j} > q_s^2}\;
  e^{ -S_\B(q_{j-1}^2,q_j^2) }
\Big\}
\\&~~~\times
e^{ -S_\F(q_{n\F}^2,q_s^2) }
e^{ -S_\B(q_{n\B}^2,q_s^2) }
\frac{d\sigma}{d\Omega}(s x_\F x_\B,\hat\theta)
\frac{1}{Z_{n\F}}
D^\F_\mc \Big( q_s^2, \frac{x_\F}{Z^\F_{n\F} } \Big)\;
\frac{1}{Z_{n\B}}
D^\B_\mc \Big( q_s^2, \frac{x_\B}{Z^\B_{n\B} } \Big)\;,
\end{split}
\end{equation}
where $Z_i=\prod_{l=1}^i z_l,\; Z_0=1,\; q_0^2\equiv s$ and 
$\bar{\Peu}(z)=(1-z) \Peu(z)$,
$\Peu(z)\equiv \frac{C_F \alpha_s}{\pi} P_{q\bar{q}}(z) $.
The principal evolution variable $q_i^2$
was introduced in Section~\ref{sec:kinematics}, Eq.~(\ref{eq:q2vardef}),
and the labelling of the emissions starts from the hard process%
\footnote{This is unnatural for the present FEV scenario, 
  but better suited for the BEV algorithm in the following.},
as in Fig.~\ref{fig:BEVlabeling}.
The emission distributions for the ladder labelled with $\F$  are the following:
\begin{equation}
\begin{split}
&d^3\rho^\F_i=d^3\rho^\F_i(s_{ij})=
   \frac{d\tbet_i d\talf_i}{\tbet_i (\talf_i+\tbet_i)}
   \frac{d\phi_i}{2\pi}\;\;
   \bar{\Peu}(1-\talf_i-\tbet_i)\;
   \theta_{\talf_i>0}\; \theta_{\talf_i+\tbet_i<1}
\\&~~~~~~
=\frac{dq_i^2 dz_i}{q_i^2}
   \frac{d\phi_i}{2\pi}
   \theta_{(1-z_i)^2s_{ij}>q_i^2}
   \Peu(z_i)
  =\frac{dq_i^2}{q_i^2} 
   \frac{d\phi_i}{2\pi} dz_i
   \frac{ \bar\Peu(z_i)}{1-z_i}
   \theta_{(1-z_i)^2s_{ij}>q_i^2},
\end{split}
\end{equation}
where the Sudakov function reads
\begin{equation}
S_\F(q_b^2,q_a^2) = S_\F(s_{ij}|q_b^2,q_a^2) 
\equiv \int\limits_{q_a^2<q_i^2<q_b^2} d^3\rho^\F_i(s_{ij}),
\end{equation}
and for the ladder $\B$ they look the same, 
except for the $\talf_i \leftrightarrow \tbet_i$ swap.

The important variable $s_{ij}$ entering $d^3\rho^{\F,\B}_i$ and $S^{\F,\B}$
for the single shower/ladder
was already defined  in Section~\ref{sec:kinematics} as
$s_{i}=\hat{s}/Z_i$, with $\hat{s}=sx_\F x_\B$.
For two showers, in any realistic PSMC, the emissions are generated
(and the four-momenta are reconstructed)
simultaneously in both showers using the competition algorithm,
in which a common $q^2$-ordering in both showers is emerging in a natural way%
\footnote{This method leads to forward-backward
  symmetric distributions, contrary to generating first the emissions
  from $q$ and later on all the remaining emissions from $\qbar$.
}.
Within such a common ordering for the two showers,
the variable $s_{ij}=\hat{s}/(Z_i^\F Z_j^\B )$ includes 
all $z$'s from the emissions in both the ladders,
starting from the hard process%
\footnote{In the PSMC jargon this is referred to
               as a recoil mechanism.}.

Strictly speaking, 
the above implicit ``kinematical coupling'' of the two showers
through $s_{ij}$ variable prevents us from rewriting the distributions
of Eq.~(\ref{eq:PSMCfev4}), without any approximation,
into the traditional convolution of two LO PDFs and the hard process,
as it was possible in the toy PS MC in Refs.~\cite{Jadach:2011cr,Jadach:2012vs}.
However, a slight modification of the kinematical coupling 
(modulo N$^3$LO corrections) allows us to get from  Eq.~(\ref{eq:PSMCfev4})
the following equivalent factorized inclusive formula
\begin{equation}
\label{eq:PSMCfev2D}
\begin{split}
&\sigma^{\lo}_\mc 
=\int d x_{\F} dx_{\B} d\Omega\;
D^\F_\mc(\hat{s}, x_{\F}) D^\B_\mc(\hat{s}, x_{\B}) 
\frac{d\sigma}{d\Omega}(s x_\F x_\B,\hat\theta),
\end{split}
\end{equation}
where $d\sigma/d\Omega$ is the hard cross section from Eq.~(\ref{eq:nlo-gen-formula}) 
and the LO PDF $D^{F}_\mc$ is
resulting from the FEV algorithm run {\em separately} for each single shower,
written in form of the following time-ordered (T.O.) exponential%
\footnote{See Ref.~\cite{GolecBiernat:2007xv} for the precise formal derivation
 of the T.O. exponent from the Markovian FEV algorithm.}
\begin{equation}
\label{eq:PSMCfevSingle}
\begin{split}
&D_\mc^\F(\hat{s}, x_{\F})=
D^\F_\mc ( q_s^2, x_{\F} )
e^{ -S_\F(\hat{s} | s,q_s^2  ) }\,+
\\&
+\sum_{n=1}^{\infty}\;
\int
e^{ -S_\F( \hat{s}| s,q_1^2    ) } 
d^3\rho^\F_1(s_1)\; e^{  -S_\F(  s_1|q_1^2,  q_2^2) } 
d^3\rho^\F_2(s_2)\; e^{  -S_\F(  s_2|q_2^2,  q_3^2) } 
\dots
\\&~~~\times
e^{ -S_\F(  s_{n-1}|q_{n-1}^2,q_n^2 ) } d^3\rho^\F_n(s_n)\;
e^{ -S_\F(     s_n| q_{n}^2, q_s^2 ) }
\prod_{i=1}^{n}
\theta_{s > q^2_{i-1} > q^2_{i} >  q_s^2}
D^\F_\mc ( q_s^2, x_{s} )\;
\delta_{x_F= x_{s}  \prod_{j=1}^n z_j },
\end{split}
\end{equation}
with $s_i=\hat{s}/Z_i$,  $s_0=\hat{s}$,
$q_0^2\equiv s$.
The other PDF, $D^\B_\mc(\hat{s}, x_{\B})$, is defined analogously.
 
It is important to note that the objects $D^{F,B}_\mc$ appearing in
Eq.~(\ref{eq:PSMCfev2D}) are not just scalar functions but they have
non-trivial and well-defined internal
structure, as explicitly seen in Eq.~(\ref{eq:PSMCfevSingle}).
In particular, the MC PDFs, $D^{F,B}_\mc$, result from the Markovian process and
thanks to kinematical mappings they respect the phase-space constraints exactly. 
Therefore, they are not equal to
the standard DGLAP parton distributions functions,
in particular they integrate
emissions up the absolute phase-space limit, 
\cf Eqs.~(\ref{eq:alpha-def}) and (\ref{eq:q2zlimits}), 
rather than stopping at some arbitrary scale $Q^2=\mu_F^2$.

The reader may check, analyzing one and two emissions in a detail,
that the effect of the above ``kinematical coupling'' of the two showers
through the variable $s_{ij}$ is
conveniently absorbed by the construction of the four-momenta defined 
in Section~\ref{eq:MultiEmissionKinema},
hence Eq.~(\ref{eq:PSMCfev2D}) is equivalent to Eq.~(\ref{eq:PSMCfev4})
up to the N$^2$LO level, \ie\ neglecting the N$^3$LO and higher corrections.
The above equivalence could also be tested numerically,
similarly as was done in Ref.~\cite{Jadach:2011cr}.

Why do we insist on the FEV representation of PSMC knowing that
any typical PSMC is built using the BEV algorithm?
The important reason is that
in any methodology of combining fixed-order perturbative
corrections with PSMC one has to make an algebraic contact
with the standard diagrammatic perturbative calculations, 
including factorization theorems, 
resummation techniques, etc.,
which are all defined within the standard Lorentz-invariant phase space (LIPS).
The FEV parton shower works directly within the LIPS%
\footnote{
 Modulo kinematical mappings, effective beam technique, recoils, etc.
},
while in the BEV algorithm, the relation between LIPS and the PSMC distributions
is obscured due to the presence
of ratios of the PDFs in the BEV distributions.
The way out is to define a framework in which the BEV and FEV distributions
are by construction {\em exactly identical}.
This requires certain non-trivial extra care and effort.
We shall follow this path in the following.

The PDF of Eq.~(\ref{eq:PSMCfevSingle}) looks pretty standard,
for instance, it is directly implementable in form of 
the FEV Markovian MC.
Its peculiarity is already quite obvious for a single emission $n=1$,
where the upper phase-space boundary limit is not just regular $q_1^2<\hat{s}$,
but more complicated $q_1^2<(1-z_1)^2 \hat{s}/z_1$, reflecting the realistic
phase-space boundary of the hard process, 
while the lower boundary is just regular $q_1^2>q_s^2$,
i.e.\ the same as from the ordering in PSMC.
This peculiarity influences mainly the first emissions, closest to the hard process%
\footnote{
  Already for the second emission, the limit $q_2^2<q_1^2$ is more important
  than  $q_2^2<(1-z_2)^2 \hat{s}/(z_1z_2)$, which can be neglected modulo NNLO.
}, as discussed in Section~\ref{sec:kinematics}.
The consequence of the above peculiarity
is that for our aim of the rigorous correspondence of the multiparton distributions
between the FEV to BEV algorithms
we need an auxiliary PDF with two competing 
factorization scales, $Q^2$ and $\hat{s}$:
\begin{equation}
\label{eq:PSMCfev2scale}
\begin{split}
&
\Dbar_\mc^\F( \hat{s}|Q^2, x_{\F})=
\Dbar^\F ( \hat{s}|q_s^2, x_{\F} )e^{ -S_\F( \hat{s}| Q^2,q_s^2) }
+\sum_{n=1}^{\infty}\; \int
 e^{ -S_\F( \hat{s}| Q^2,q_1^2 )}
\\&~~~~~~~~~~\times
 \prod_{i=1}^n
 \Big\{
   d^3\rho^\F_i(s_i)\;
   e^{ -S_\F( s_{i}| q_{i}^2,q_{i+1}^2  )}
   \theta_{Q^2 > q^2_{i-1} > q^2_{i} >  q_s^2}
 \Big\}
 \Dbar^\F ( \hat{s}| q_s^2, x_{s} )\;
 \delta_{x_F= x_{s}  \prod_{j=1}^n z_j },
\end{split}
\end{equation}
where $q_{n+1}\equiv q_s$
and it coincides with the PDF of Eq.~(\ref{eq:PSMCfev4}) 
for $Q^2=s=\hat{s}/x_F$, i.e.
\begin{equation}
\label{eq:Dbar2D}
 D_\mc^\F(\hat{s}, x_{\F}) 
 = \Dbar_\mc^\F(\hat{s}|\hat{s}/x_F, x_{\F})
 =\Dbar_\mc^\F( \hat{s}|\infty, x_{\F}).
\end{equation}

The main difference between this new PDF and the one introduced earlier in
Eq.~(\ref{eq:PSMCfevSingle}) is the $\theta_{Q^2 > q^2_{i-1} > q^2_{i} > q_s^2}$
function, see Fig.~\ref{fig:FEVkinema}, that restrict emissions only to the
region below the scale $Q^2$, which in principle is an arbitrary parameter, as
long as $Q^2 \ll \hat{s}$. In that sense the new PDFs, $\Dbar_\mc^{F,B}$,
are closer to the standard DGLAP parton distribution functions.

With the above definitions we may work out the BEV algorithm
providing the LO distributions, 
which by construction are {\em exactly the same} as from the FEV algorithm%
\footnote{
 Modulo the kinematical coupling mentioned above.
}:
\begin{equation}
\label{eq:LOdiff}
\begin{split}
&d\sigma^{\lo}_{n_\F n_\B} =
\frac{d\sigma}{d\Omega}(s x_\F x_\B,\hat\theta) 
\prod_{i=1}^{n_\F}
\Big\{
   d^3\omega^\F_i\; \theta_{q^2_{i-1} > q^2_{i}}
\Big\}
\prod_{j=1}^{n_\B}
\Big\{ 
   d^3\omega^\B_j\; \theta_{q^2_{j-1} > q^2_{j}}
\Big\}
\\&~\times
e^{ -\Delta_\F( x^\F_{n\F}|q_{n\F}^2,q_s^2 ) }
e^{ -\Delta_\B( x^\B_{n\B}|q_{n\B}^2,q_s^2 ) }
D^\F( \hat{s}, x_\F)\;
D^\B( \hat{s}, x_\B)\;
d x_{\F} dx_{\B} d\Omega,
\\&
x^\F_i=x_\F/Z^\F_i,\; x^\B_j=x_\B/Z^\B_j,\;
\hat{s}= s  x_\F x_\B,
\end{split}
\end{equation}
where $x^\F_i=\prod_{m=1}^i z_m$ includes $z_i$ from  $d^3\omega^\F_i$
and similarly for  $x^\B_j$ from  $d^3\omega^\B_i$.
The single-emission distributions and form factors are defined as follows:
\begin{equation}
\begin{split}
&
d^3\omega^\A_i
= \frac{dq_i^2 dz_i}{q_i^2} \frac{d\phi_i}{2\pi}
  \Kbbm_\mc(x_{i-1}|z_i,q_i^2)\;
  e^{-\Delta_\mc( x_{i-1}|q_i^2,q_{i-1}^2)},
\\&
\Kbbm_\mc(x^*|z,q^2) \equiv
   \Peu(z_i)\;
   \theta_{(1-z)^2 sx^*/z >q^2}\;
   \frac{\Dbar_\mc(  sx^*/z |q^2,x^*/z)}%
        {\Dbar_\mc(  sx^*   |q^2,x^*)    },
\\&
\Delta_\mc( x^*|q_{j-1}^2,q_j^2) \equiv
   \int_{q_j^2}^{q_{j-1}^2} \frac{dq^2}{q^2}
   \int_{ x^*}^1 \frac{dz}{z}
   \Kbbm_\mc( x^*|z,q^2),
\end{split}
\end{equation}
and the shower label $\A=\F,\B$ was temporarily omitted 
wherever it was unambiguous to do so.

Although it may not look obvious, the integrated cross section
\[  
\sigma^{\lo}_\mc=
  \sum_{n_\F, n_\B=0}^\infty \int d\sigma^{\lo}_{n_\F n_\B}
\]
and all the multiparton distributions are {\em exactly} the same as in 
Eqs.~(\ref{eq:PSMCfev2D}) and (\ref{eq:PSMCfevSingle}).
Let us present a sketchy proof of that%
\footnote{ The result of the detailed proof is the same, of course.}.
Directly from the BEV algorithm
for the PDF of Eq.~(\ref{eq:PSMCfevSingle}) of the showering quark we get
\begin{equation}
\label{eq:PDFbev2}
\begin{split}
& D^\F_\mc(\hat{s},x_\F) \Big|_{BEV}=
\Dbar_\mc( s\hat{x}|s,\hat{x}  )
\Big\{
  e^{-\Delta_\mc(s,q_s^2|x_\F)} \,+
\\&~
+\sum_{n=1}^\infty
 \bigg[
 \prod_{i=1}^n
   \int_{q_s^2}^{q_{i-1}^2}\frac{dq_i^2}{q_i^2} 
   \int_{x_{i-1}}^1\frac{dz_i}{z_i}\;
   e^{-\Delta_\mc( x_{i-1}|q_i^2,q_{i-1}^2)}
   \Kbbm_\mc(x_{i-1}|z_i,q_i^2)
 \bigg]
   e^{-\Delta_\mc( x_{n}|s,q_{n}^2)}
\Big\}.
\end{split}
\end{equation}
In order to prove that the above PDF is the same as from the FEV algorithm one
exploit the fact that in the T.O. exponential representation
of the double-scale PDF of Eq.~(\ref{eq:PSMCfev2scale}) 
we may detach (factorize off) the first emission, obtaining
the following integral equation:
\begin{equation}
\label{eq:Detach}
\begin{split}
&\Dbar_\mc^\F( \hat{s}|Q^2, x)=
\Dbar^\F ( \hat{s}| q_s^2, x )e^{ -S_\F(  \hat{s}| Q^2,q_s^2) }+
\\&
+\int_{q_{s}^2}^{Q^2} \frac{dq_1^2}{q_1^2}
 \int_{0}^1 dx_1
 \int_{0}^1 dz_1\;
 \Peu(z_1) \theta_{(1-z_1)^2 \hat{s}/z_1>q_1^2}\;
 e^{ -S_\F( \hat{s}| Q^2,q_1^2) } 
\Dbar_\mc(\hat{s}/z_1 |q_1^2,x_{1} )
\delta_{x=z_1 x_1}.
\end{split}
\end{equation}
Differentiating both sides by $\partial/\partial(\ln q_s^2)$
and then integrating $\int_{\ln q_a^2}^{\ln q_b^2} d(\ln q_s^2)$,
one obtains 
\begin{equation}
\label{eq:DeltaBEV2MC}
e^{-S_\mc(\hat{s}|q_b^2,q_a^2)}=
e^{-\Delta_\mc( x|q_b^2,q_a^2)}
\frac{\Dbar_\mc(\hat{s}|q_b^2, x)}%
     {\Dbar_\mc(\hat{s}|q_a^2, x)}.\quad
\end{equation}
The above identity can be used many times in order
to eliminate the ratios of the PDFs in the BEV formula
and transform without any approximation the BEV distributions
of Eq.(\ref{eq:PDFbev2}) into 
the FEV distributions of Eq.~(\ref{eq:PSMCfevSingle}).

\subsection{NLO weight in parton shower and its algebraic validation}
\label{sec:validation}

Having defined the framework in which the FEV and BEV
distributions of LO PSMC are identical, 
we may define a MC weight correcting the hard process 
in the (notoriously) inefficient FEV LO PSMC,
but having advantage of a straightforward connection to the perturbative expansion,
and apply it within LO PSMC implemented using the efficient BEV algorithm.
The NLO completeness of such a scheme 
can be proven analytically within FEV 
without any troubles, due to the absence of the `obscure' 
ratios of the PDFs of BEV.
The only extra cost is that in order to
gain a perfect FEV$\,\leftrightarrow\,$BEV compatibility,
we had to introduce the special LO PDFs $\Dbar(\hat{s}|Q^2,x)$
with two factorization scales, at least for the purpose of the
discussion of the NLO completeness of the \krknlo{} scheme,
while in practice
we shall be able to obtain $\Dbar$ from the standard $\msbar$ PDFs

The differential cross section in which the hard process is NLO-corrected
for the $q\qbar$ channel in the \krknlo{} method reads as follows%
\footnote{ We adopt a convention that $\sum_{n=1}^2 d_n =0 $.
}:
\begin{equation}
\label{eq:WTnloSum}
 d\sigma^{\nlo}_{n_\F n_\B}=
 \Big(1+\Delta_{VS} 
       +\sum_{i=1}^{n_\F} W^{[1]}_{q\qbar}(\talf^\F_i,\tbet^\F_i)
       +\sum_{j=1}^{n_\B} W^{[1]}_{q\qbar}(\talf^\B_j,\tbet^\B_j)
 \Big)
  d\sigma^{\lo}_{n_\F n_\B},
\end{equation}
where 
\begin{equation}
\label{eq:WTnloBeta}
  W^{[1]}_{q\qbar} = \frac{d^5\bar\beta_{q\qbar}}{d^5\sigma^{\lo}_{q\qbar}}
   = \frac{d^5\sigma^{\nlo}_{q\qbar} -d^5\sigma^{\lo}_{q\qbar} }{d^5\sigma^{\lo}_{q\qbar}}
   =W^{(1)}_{q\qbar}-1,
\end{equation}
see Eqs.~(\ref{eq:W1nlo}) and ~(\ref{eq:wt-vs}).
The structure of this weight, 
with the summation over the emitted gluons,
is a straightforward generalization of the weight used in many
MC programs with multiphoton exponentiated corrections,
see for instance Ref.~\cite{Jadach:1999vf}.

As pointed out in Ref.~\cite{Jadach:2012vs}, in order to get the complete NLO
corrections to the hard process, it is enough to retain
in the sum in Eq.~(\ref{eq:WTnloSum}) only a single term,
the one with the maximum $k_T^2$.
Since our $q^2$-variable is practically identical to $k_T^2$
we may retain only one term for a gluon
with the maximum $q_{i_\F}^2$ or  $q_{j_\B}^2$,
from one of the two showers.
In the case of the BEV algorithm with competition, 
this gluon is just the one which was generated first.

Similarly, as we have detached the first emission from the T.O. exponential
representation of the PDF in Eq.~(\ref{eq:Detach}),
we now factorize off explicitly the gluon emissions
with the maximum $q_\F^2$ and  $q_\B^2$
from the product of two PDFs in Eq.~(\ref{eq:PSMCfev2D}):
\begin{equation}
\label{eq:detach2PDF}
\begin{split}
&D^\F_\mc(\hat{s},\hat{x}_\F) D^\B_\mc(\hat{s},\hat{x}_\B)
=e^{-S_\F(\hat{s}|s_{1\F},q_s^2)} e^{-S_\B(\hat{s}|s_{1\B},q_s^2)}
 \Dbar^\F_\mc(\hat{s}|q_s^2,\hat{x}_\F) 
 \Dbar^\B_\mc(\hat{s}|q_s^2,\hat{x}_\B)
\\&
+\!\! \int\limits_{q_s^2<q_{1\F}^2<s_{1\F} } 
 \!\!\!\!\!\!\!\!\!\!d^3\rho_{1}^\F(\hat{s}/z_{1\F})
 e^{-S_\F(\hat{s}|s_{1\F},q_{1\F}^2)} e^{-S_\B(\hat{s}|s_{1\B},q_{1\F}^2)}
 \frac{1}{z_{1\F}}
 \Dbar^\F_\mc(\hat{s}|q_{1\F}^2, \hat{x}_\F/z_{1\F}) 
 \Dbar^\B_\mc(\hat{s}|q_{1\F}^2, \hat{x}_\B)
\\&
+\!\! \int\limits_{q_s^2<q_{1\B}^2<s_{1\B} } 
 \!\!\!\!\!\!\!\!\!\!d^3\rho_{1}^\B(\hat{s}/z_{1\B})
 e^{-S_\F(\hat{s}|s_{1\F},q_{1\B}^2)} e^{-S_\B(\hat{s}|s_{1\B},q_{1\B}^2)}
 \Dbar^\F_\mc(\hat{s}|q_{1\B}^2, \hat{x}_{\F}) 
 \frac{1}{z_{1\B}}
 \Dbar^\B_\mc(\hat{s}|q_{1\B}^2, \hat{x}_\B/z_{1\B}),
\end{split}
\end{equation}
where
$ s_{1\F} = \hat{s}/\hat{x}_{\F}$ and
$s_{1\B} = \hat{s}/\hat{x}_{\B}$.

In order to improve readability, we are going to omit temporarily
(until Eq.~(\ref{eq:EtVoila5}))
the argument $\hat{s}$ in the $\Dbar^{F,B}_\mc(\hat{s}|\dots)$ 
and the $S_{\F,\B}(\hat{s}|\dots)$ functions.
Moreover, the ``kinematical coupling'' of the two showers is will be restored,
that is we are going to replace in the following
$s_{1\F},s_{1\F} \to s=\hat{s}/(x_{\F}x_{\B})$.
The above decomposition is easily exploited in calculating the NLO
cross section with the MC weight
of Eq.~(\ref{eq:WTnloSum}) truncated to a single term $W^{[1]}_{q\qbar}$
with the maximum $q^2$:
\begin{equation}
\label{eq:EtVoila2}
\begin{split}
&\sigma^{\nlo}[J] = \sum_{n_\F,n_\B=1}^{\infty} \int d\sigma^{\nlo}_{n_\F n_\B}
= \int d x_\F d x_\B d\Omega \;
  \frac{d\sigma}{d\Omega}(s x_\F x_\B,\hat\theta)
 (1+\Delta_{VS})
 e^{-S_\F(s,q_s^2)} e^{-S_\B(s,q_s^2)}
\\&~~~~~~~~~~~~~~~~~~~~~~~~~~~~~~~~~~~~~~~~~~~~~~~~~~~~~~\times
 \Dbar^\F_\mc(q_s^2,x_\F) \Dbar^\B_\mc(q_s^2,x_\B)
 J_{\lo}(x_\F,x_\B)
\\&
+\int d x_\F d x_\B d\Omega \;
 \Bigg\{
 \int\limits_{q_s^2<q_{1\F}^2<s } 
 \!\!\!\!\!\!\!\!d^3\rho_{1}^\F( s x_\F x_\B )
 (1+\Delta_{VS}+ W^{[1]}_{q\qbar}(k_1))
 \frac{d\sigma}{d\Omega}(s x_\F x_\B z_{1\F},\hat\theta)
\\&~~~~~~~~~~~~~~\times
 e^{-S_\F(s,q_{1\F}^2)} e^{-S_\B(s,q_{1\F}^2)}
 \Dbar^\F_\mc(q_{1\F}^2, x_\F) 
 \Dbar^\B_\mc(q_{1\F}^2, x_\B)
 J_{\nlo}(x_\F,x_\B,z_{1\F},k^2_{1T})
\\&~~~~~~~~~~~~~~~~~~~~~
+\!\! \int\limits_{q_s^2<q_{1\B}^2<s } 
 \!\!\!\!\!\!\!\!d^3\rho_{1}^\B( s x_\F x_\B )
 (1+\Delta_{VS}+ W^{[1]}_{q\qbar}(k_1))
 \frac{d\sigma}{d\Omega}(s x_\F x_\B z_{1\B},\hat\theta)
\\&~~~~~~~~~~~~~~\times
 e^{-S_\F(s,q_{1\B}^2)} e^{-S_\B(s,q_{1\B}^2)}
 \Dbar^\F_\mc(q_{1\B}^2, x_{\F})
 \Dbar^\B_\mc(q_{1\B}^2, x_{\B})
 J_{\nlo}(x_\F,x_\B,z_{1\B},k^2_{1T})
\Bigg\},
\end{split}
\end{equation}
where we have introduced $x_\F$ and $x_\B$ {\em before} the emission, as in a
typical fixed-order NLO calculation.  They are related to $\hat x_F$ and $\hat
x_B$ from Eq.~(\ref{eq:detach2PDF}) in such a way that if the emission happens
on the leg $F$, then $x_F=\hat{x}_\F/z_{1\F}$ and $x_\B=\hat{x}_\B$. And
symmetrically if emission occurs on the leg $B$.
The same expression we can get directly from distributions of the BEV algorithm,
as shown explicitly in Appendix~\ref{sec:AppedA}.

For the purpose of the proof of compatibility of the above formula with the
fixed-order calculation (\mcfm{}) for the entire class of the LO and NLO
observables, we have introduced in the above the {\em jet functions},
$J_{\lo}(x_\F,x_\B)$ and  $J_{\nlo}(x_\F,x_\B,z_1,k^2_{1T})$, which satisfy the
following properties:
\begin{equation}
\label{eq:jetfun}
 J_{\nlo}(x_\F,x_\B,z_1,k^2_{1T})\ \to\ J_{\lo}(x_\F,x_\B)
  \quad \text{for }  \quad k_{1T}^2 \to 0,
  \quad \text{or }   \quad z_1 \to 1,
\end{equation}
often referred to as an infra-red safety requirement.
In the experimental practice, the function $J_{\nlo}$ represents 
the most general 2-dimensional
histogramming in the transverse momentum of the gluon (or the heavy boson)
and the $z_1$ variable, related to the rapidity difference between the heavy boson
and the emitted gluon 
(or the ratio of the effective mass of the heavy boson and the heavy boson plus gluon system).
The meaning of the above limit $k_{1T}^2 \to 0$ is that
for the first bin in $k_{1T}^2$, the one including the $k_{1T}^2 = 0$ point,
we are not allowed to do any binning in $z_1$
(we have to sum up inclusively over all $z_1$,
similarly as for $J_{\lo}$ we sum up inclusively over all  $k_{1T}^2$).

In order to recover the fixed-order NLO formula of \mcfm{},
all terms ${\cal O}(\alpha_s^2)$
have to be carefully eliminated.
It is easy to do it for the $\sim d^3\rho\; W^{[1]}_{q\qbar}$ parts,
which are formally ${\cal O}(\alpha_s^1)$.
We replace in them
$\Dbar^\F_\mc(\hat{s}|q_{1\F,1\B}^2,  x_{\F}) 
 \to \Dbar^\F_\mc(\hat{s}|s,  x_{\F})$
and also eliminate safely the Sudakov exponents%
\footnote{ The collinear singularity in $q^2\to 0$ is killed 
  by $W^{[1]}_{q\qbar}$,
  so $\int dq^2/q^2 e^{-S(q_2...)}$ cannot give a $\sim 1/\alpha_s$ contribution.
}:
\begin{equation}
\label{eq:EtVoila3}
\begin{split}
&\sigma^{\nlo}[J]
= \int d x_\F d x_\B d\Omega \;
 \Big\{
 e^{-S_\F(s,q_s^2)} e^{-S_\B(s,q_s^2)}
 \Dbar^\F_\mc(q_s^2,x_\F) \Dbar^\B_\mc(q_s^2,x_\B)
 \Big\}
\\&~~~~~~~~~~~~~~~~~~~~~~~~~~~\times
 (1+\Delta_{VS})
 \frac{d\sigma}{d\Omega}(s_1,\hat\theta)
 J_{\lo}(x_\F,x_\B)
\\&
+\int d x_\F d x_\B d\Omega \;
 \Bigg\{
 \int\limits_{q_s^2<q_{1\F}^2<s } 
 \!\!\!\!\!\!\!\!d^3\rho_{1}^\F( s_1)
 e^{-S_\F(s,q_{1\F}^2)} e^{-S_\B(s,q_{1\F}^2)}
 \Dbar^\F_\mc(q_{1\F}^2, x_\F) 
 \Dbar^\B_\mc(q_{1\F}^2, x_\B)
\\&~~~~~~~~~~~~~~~~~~~~~
  + \int\limits_{q_s^2<q_{1\B}^2<s } 
 \!\!\!\!\!\!\!\!d^3\rho_{1}^\B( s_1 )
 e^{-S_\F(s,q_{1\B}^2)} e^{-S_\B(s,q_{1\B}^2)}
 \Dbar^\F_\mc(q_{1\B}^2, x_{\F})
 \Dbar^\B_\mc(q_{1\B}^2, x_{\B})
\Bigg\}
\\&~~~~~~~~~~~~~~~~~~~~~~~~~~~~\times
 (1+\Delta_{VS})
 \frac{d\sigma}{d\Omega}(s x_\F x_\B z_{1},\hat\theta)
 J_{\nlo}(x_\F,x_\B,z_1,k^2_{1T})
\\&
+\int d x_\F d x_\B d\Omega \;
 \!\!\!\!\!\!\!\! \int\limits_{0< q_{1\F}^2,q_{1\B}^2 <s } 
 \!\!\!\!\!\!\!\! \big( d^3\rho_{1}^\F( s_1 ) +d^3\rho_{1}^\B( s_1 )\big)
 W^{[1]}_{q\qbar}(k_1)
 \frac{d\sigma}{d\Omega}(s x_\F x_\B z_{1},\hat\theta)
\\&~~~~~~~~~~~~~~~~~~~~~~~~~~~~~\times
 \Dbar^\F_\mc(s, x_{\F})
 \Dbar^\B_\mc(s, x_{\B})
 J_{\nlo}(x_\F,x_\B,z_1,k^2_{1T}),
\end{split}
\end{equation}
where we were also allowed to replace $z_{1\F},z_{1\B}\to z_1$,
because in terms of the Sudakov variables before the emission it is the same variable.
We have also introduced $s_1=s x_\F x_\B$.

The remaining parts $ \sim (1+\Delta_{VS})$ are less trivial.
In order to use again the identity of Eq.~(\ref{eq:detach2PDF})
for folding in three integrals within the $\{ \dots \}$ braces
back into a simple product of two PDFs, we have to do something with
the $z_1$-dependence in $d\sigma/d\Omega$ 
and the $k_{1T}^2$-dependence of the jet function $J_{\nlo}$ in the second part.
The solution is to add and subtract a term proportional
$ \frac{d\sigma}{d\Omega}(s_1,\hat\theta) J_{\lo}(x_\F,x_\B) $ 
in the second part,  regroup and use again Eq.~(\ref{eq:detach2PDF}):
\begin{equation}
\label{eq:EtVoila4}
\begin{split}
&\sigma^{\nlo}[J]
= \int d x_\F d x_\B d\Omega \;
 \Dbar^\F_\mc(s,x_\F) \Dbar^\B_\mc(s,x_\B)
 (1+\Delta_{VS})
 \frac{d\sigma}{d\Omega}(s_1,\hat\theta)
 J_{\lo}(x_\F,x_\B)
\\&
+\int d x_\F d x_\B d\Omega \;
 \Bigg\{
 \int d^3\rho_{1}^\F( s_1)
 \Dbar^\F_\mc(q_{1\F}^2, x_\F) 
 \Dbar^\B_\mc(q_{1\F}^2, x_\B)
\\&~~~~~~~~~~~~~~~~~~~~~
  + \int\limits d^3\rho_{1}^\B( s_1 )
 \Dbar^\F_\mc(q_{1\B}^2, x_{\F})
 \Dbar^\B_\mc(q_{1\B}^2, x_{\B})
\Bigg\}
\\&~~~~~~~~~~~~~~~~~~~~\times
\Big[
 \frac{d\sigma}{d\Omega}(s x_\F x_\B z_{1},\hat\theta)
 J_{\nlo}(x_\F,x_\B,z_1,k^2_{1T})
-\frac{d\sigma}{d\Omega}(s_1,\hat\theta) J_{\lo}(x_\F,x_\B)
\Big]
\\&
+\int d x_\F d x_\B d\Omega \;
 \int d^5\rho^{LO}_{q\qbar}\;
 W^{[1]}_{q\qbar}(k_1)
 \Dbar^\F_\mc(s, x_{\F})
 \Dbar^\B_\mc(s, x_{\B})
 J_{\nlo}(x_\F,x_\B,z_1,k^2_{1T}),
\end{split}
\end{equation}
where we have profited from finiteness of the integrals,
in order to omit $(1+\Delta_{VS})$ and the Sudakov exponents 
wherever possible
and to set the lower integration limits of $q^2$ to zero.

The final clean-up of ${\cal O}(\alpha_s^2) $ terms involves the change 
$q_{1\F}^2\to s$ and $q_{1\B}^2\to s$ in the PDFs in the second
integral and recombining it with the third one:
\begin{equation}
\label{eq:EtVoila5}
\begin{split}
&\sigma^{\nlo}[J]
= \int d x_\F d x_\B d\Omega \;
 (1+\Delta_{VS})
 \frac{d\sigma}{d\Omega}(s_1,\hat\theta)
 J_{\lo}(x_\F,x_\B)
 \Dbar^\F_\mc(\hat{s}|s,x_\F)
 \Dbar^\B_\mc(\hat{s}|s,x_\B)
\\&
+\int d x_\F d x_\B d\Omega \; \int
\Big\{
 d^5\rho^{\lo}_{q\qbar}\;
 \big[1+ W^{[1]}_{q\qbar}(k_1) \big]
 J_{\nlo}(x_\F,x_\B,z_1,k^2_{1T})
\\&~~~~~~~~~~~~~~~~
-(d^3\rho_{1}^\F+d^3\rho_{1}^\B)
 \frac{d\sigma}{d\Omega}(s_1,\hat\theta) J_{\lo}(x_\F,x_\B)
\Big\}
 \Dbar^\F_\mc(\hat{s}|s, x_{\F})
 \Dbar^\B_\mc(\hat{s}|s, x_{\B}).
\end{split}
\end{equation}
We may finally eliminate the MC weight of  Eq.~(\ref{eq:WTnloBeta})
going back to the NLO distributions:
\begin{equation}
\label{eq:MCsemiAN}
\begin{split}
&\sigma^{\nlo}[J]
= \int d x_\F d x_\B d\Omega \;
 (1+\Delta_{VS})
 \frac{d\sigma}{d\Omega}(s_1,\hat\theta)
 J_{\lo}(x_\F,x_\B)
 D^\F_\mc(\hat{s},x_\F)
 D^\B_\mc(\hat{s},x_\B)
\\&
+\int d x_\F d x_\B d\Omega
\Big\{
 d^5\rho^{\nlo}_{q\qbar}\;
 J_{\nlo}(x_\F,x_\B,z_1,k^2_{1T})
- d^5\rho^{\lo}_{q\qbar}\; J_{\lo}(x_\F,x_\B)\Big\}
\\&~~~~~~~~~~~~~~~~~~~~~~~~~~~~~~~~~~~~~~~~~~~~~~~~~~~~~~~~~~\times
 D^\F_\mc(\hat{s}, x_{\F})
 D^\B_\mc(\hat{s}, x_{\B}),
\end{split}
\end{equation}
where $D_\mc^{\F,\B}(\hat{s}, x)$ of Eq.~(\ref{eq:Dbar2D}) was also recovered.

The above looks like an example of the fixed-order NLO calculation formula
employing the technique of soft-collinear counter-terms
following the Catani--Seymour (CS) work~\cite{Catani:1996vz},
with the explicit definition of an arbitrary NLO observable
using the $J$-function.
For the sake of completeness, such a formula in the standard $\msbar$
scheme using the CS method is shown explicitly
in Eq.~(\ref{eq:DYinCS}) in Appendix~\ref{append:B}
for the same $q\qbar$ channel.

However, there are two important differences between
the formulas in Eq.~(\ref{eq:DYinCS}) and in the above
Eq.~(\ref{eq:MCsemiAN}):
missing the $\sim \delta(k_{1T}^2)\Sigma $ term and non-$\msbar$ PDFs
in Eq.~(\ref{eq:MCsemiAN}).
The real emission integral with subtraction, 
in spite of slightly different notation, is identical.
These differences are, of course, due to the differences between 
the MC and $\msbar$
factorization schemes, and are well understood, see the discussion
in the following Section~\ref{sec:nlo-bench} and in Appendix \ref{append:B}.
The important bonus of the comparison of Eqs.~(\ref{eq:DYinCS})
and (\ref{eq:MCsemiAN}) is that we can establish
the relation between the quark PDF used in the parton shower MC
and the corresponding PDF of the $\msbar$ scheme, 
modulo NNLO,
in a solid and unambiguous way:
\begin{equation}
f_q^\mc(Q^2,x)=D_\mc(Q^2,x)=\Dbar_\mc(Q^2|Q^2/x,x),
\label{eq:fq-D-rel}
\end{equation}
where $f_q^{\mc}$ is the quark PDF in the MC scheme  defined
in Eq.~(\ref{eq:PDFs-MC-quark}).

In this way we have proven algebraically that 
the \krknlo{} scheme {\em is equivalent to the fixed-order NLO calculation} 
in the entire functional space of the NLO-class experimental observables%
\footnote{
 It seems that this kind of an explicit rigorous, albeit tedious,
 algebraic proof is not available for the other
 methods of combining the NLO corrections with the LO parton shower.
}.

\subsection{Summarizing the KrkNLO method}

In the following, we summarize and comment on the key elements of the \krknlo{}
method, starting from the $q\qbar$ channel only and then adding the $qg$ channel.

\begin{enumerate}
\item
The first element is the reorganization of the NLO corrections, 
the same as Ref.~\cite{Jadach:2011cr}.
Starting from the standard $\msbar$ calculation,
a strictly collinear part of the NLO corrections is removed thanks
to the clever redefinition of the LO PDFs from the $\msbar$ scheme to the MC scheme,
see Eq.~(\ref{eq:PDFs-MC-quark}).
In a nutshell, in the standard $\msbar$ procedure, one subtracts from
the diagrammatic real$\,+\,$virtual results a pure $\sim P(z)/\epsilon$ pole,
while in the case of \krknlo{} one subtracts the $\Gamma_\mc(z,\epsilon)$
function, which contains extra non-pole terms.
As discussed in Ref.~\cite{Jadach:2011cr},
see also Eq.~(\ref{eq:GammaMC}), this $\Gamma_\mc$ function
is just the integral over the LO distribution 
of the single-gluon emission in the parton shower MC%
\footnote{ It is the so-called soft-collinear counter-term.
 Although it is defined with the help of PSMC for the DY process, 
 its definition is universal.},
written in $4+\epsilon$ dimension, together with the Sudakov form factor,
obeying typical parton shower sum rules.
This element of the \krknlo{} prescription remains the same 
as in Ref.~\cite{Jadach:2011cr},
provided the LO distribution is the same as in the present study.
This assumption is valid because the LO distribution of Eq.~(\ref{eq:dsig5LO})
is indeed the same in Ref.~\cite{Jadach:2011cr} and in the Catani--Seymour inspired
implementations of the single-gluon emissions in \sherpa{} and \herwig{}.
What remains to be checked is whether the upper phase-space limit,
$\alpha+\beta\leq 1$, is not spoiled in a given PSMC
by the use of the backward evolution algorithm.
We are going to come back to this point shortly.

\item
The second element of the \krknlo{} prescription is the construction,
implementation, and algebraic validation of the multiplicative weight
introducing the NLO corrections in the multiparton environment of PSMC.
The weight of Eq.~(\ref{eq:W1nlo}) is implemented following
Eq.~(\ref{eq:WTnloSum}).
The summation over gluons is necessary only for the angular ordering,
while for the $q^2$-ordering or $k_T$-ordering it is enough
to keep just one term for the gluon next to the hard process.
Validity of above method can be proven rigorously,
see Section~\ref{sec:validation}, in the case when the initial-state PSMC
is realized using the forward Markovian evolution (FEV) algorithm,
in the presence of any observable-defining function $J$, 
see Eq.~(\ref{eq:jetfun}).
Such a FEV algorithm would be terribly inefficient for the resonant $Z$-boson
production, but it is mathematically perfectly well defined.
(In such a {\em gedanken} FEV scenario with the complete phase-space coverage,
the output PDF from the MC at the hard process scale will be 
automatically in the MC scheme,
following Eq.~(\ref{eq:PDFs-MC-quark}),
provided that the input PDF at low $q_s^2$ is in the $\msbar$ scheme%
\footnote{ In other words, such a FEV MC performs not only the LO DGLAP evolution
    but also the transition from
    one to another factorization scheme.}.)

\item
The above validation proof of the NLO weight has to be extended to the (efficient)
backward evolution (BEV) scenario of the typical PSMC, such as \sherpa{} or
\herwig{}. 
It was shown in Section~\ref{sec:PSreality} that this is feasible, 
by means of formulating the twin FEV and BEV algorithms, using the same PDFs,
which produce {\em exactly the same} multiparton exclusive distributions.
What is highly non-trivial is that the full phase-space coverage is not lost.
The price for this was that we had to introduce the auxiliary PDF $\Dbar(\hat{s}|Q^2,x)$
of Eq.~(\ref{eq:PSMCfev2scale})
with two competing factorization scales,
in addition to the single-scale PDF of Eq.~(\ref{eq:PSMCfevSingle}).
Luckily, at the end this auxiliary PDF can be finally eliminated from the BEV algorithm.
Its main role is to provide theoretical control for the FEV$\,\to\,$BEV transition,
desired in view of the previous point.

\item
The elimination of the auxiliary  parton distribution function $\Dbar(\hat{s}|Q^2,x)$ can be seen best
with the analysis presented in Appendix~\ref{sec:AppedA},
where the transition  FEV$\,\to\,$BEV
is analyzed in a fine detail for $n=0,1$ emissions.
In Eq.~(\ref{eq:NOglu}) for $n=0$, we see that we may replace
$\Dbar(\hat{s}|s,x)\to D(\hat{s},x)$ using Eq.~(\ref{eq:Dbar2D}).
For $n=1$ in Eq.~(\ref{eq:BEV2dipole}), one may show that,
modulo ${\cal O}(\alpha_s^2)$ terms, the ratio of $\Dbar$-functions can be
replaced 
with any kind of the LO PDFs and the exponents, $\exp(-\Delta)$, can be neglected.
What cannot be approximated there, 
it is the exact implementation of the evolution kernel,
keeping the correct  phase-space limits due to the $\theta$-functions.
This is, however, not the problem, 
because any standard BEV implementation with the veto algorithm does that correctly%
\footnote{More detailed analysis of this kind was done but is left beyond
   the scope of the present paper.}.
We conclude that, in the BEV scenario, all $\Dbar$-functions go away and
only the single-scale PDFs in the MC factorization scheme
seen in  Eq.~(\ref{eq:NOglu}) are left finally in the game%
\footnote{%
 One can show algebraically that the effective PDF generated by the above 
 BEV at low $q_s^2$ will be in the $\msbar$ scheme --
 the BEV algorithm is not only undoing the LO evolution,
 but also the transition from the $\msbar$ to MC scheme.
}.
\end{enumerate}

In the above analysis we have omitted $qg$ channel.
For the $Z/\gamma^*$-boson production process however, is not difficult to include 
it following the same steps described above%
\footnote{%
  All ingredient distributions are well known in the literature.}
In our implementation of the \krknlo{} method in PSMC, 
whenever in the first step of the BEV PS 
algorithm the transition from the quark (antiquark) to the gluon is generated, 
we associate with such an event
the weight computed according to Eqs.~(\ref{eq:wt-real-qg}) and (\ref{eq:wt-ovs}).
This weight corresponds to observables averaged over the $Z/\gamma^*$-boson decay angles
and is sufficient for the current paper,
as we are going to concentrate on the $Z/\gamma^*$-boson transverse momentum 
and rapidity distributions.
The discussion of the $Z$-decay leptonic observables, 
as well as more details on the implementation of the $qg$ channel,
is reserved for a separate publication.

\section{Fixed-order NLO benchmarks}
\label{sec:nlo-bench}

For the numerical evaluation of the cross sections\footnote{Unless stated otherwise in the text.} at the LHC 
for the proton--proton collision energy of $\sqrt{s}=8$~TeV
we have chosen the following set of Standard Model (SM) input parameters:
\begin{eqnarray}\label{eq:pars}
M_Z = 91.1876 \; {\rm GeV}, & \quad & \Gamma_Z =  2.4952  \; {\rm GeV},
\nonumber  \\
M_W = 80.4030 \; {\rm GeV}, & \quad & \Gamma_W = 2.1240 \; {\rm GeV},
\\
G_{\mu} = 1.16637\times 10^{-5} \; {\rm GeV}^{-2}, 
& \quad & m_t = 173.2  \; {\rm GeV},
\nonumber  \\
\alpha_s(M_Z^2)=0.13938690,
\nonumber  
\end{eqnarray}
and the $G_{\mu}$-scheme~\cite{LHC-YR} for the electroweak sector of the Standard Model.  To
compute the hadronic cross section we also use the {\tt MSTW2008} LO set of parton
distribution functions~\cite{Martin:2009iq}, and take the renormalization and
factorization scales to be $\mu_R^2=\mu_{F}^2=M_Z^2$. 
The only detector acceptance cuts are imposed 
on the invariant mass of the final-state lepton pair ($Z/\gamma^{\star}$-boson):
\begin{equation}
50~{\rm GeV}<M_{l\bar{l}}<150~{\rm GeV}.
\label{eq:Mllcut}
\end{equation}

In order to check that our settings are identical in all used programs, we began
with the comparisons at the Born level. The results presented  in Table.~\ref{tab:lo-settings-val} show
a very good agreement (within statistical errors) between different programs.
\begin{table}[t]
\centering
\begin{tabular}{|c||c|c|c|}
  \hline
                           & \mcfm{}  & \sherpa{} & \herwig{}  \\
  \hline \hline                        
  $\sigma_\text{tot}$ [pb] &  $936.9 \pm 0.1$     &  $937.2 \pm 0.2$      &  $937.0 \pm	0.6$   \\ 
  \hline
\end{tabular}
\caption{\sf
Values of the total cross section with statistical errors at the Born level for 
the Drell--Yan process in the $\msbar$ scheme. 
}
\label{tab:lo-settings-val}
\end{table}

\subsection{PDFs in MC scheme}
\label{sec:PDFSinMCscheme}

In Section~\ref{sec:mc-scheme}, we have introduced the MC factorization
scheme and explained why this scheme is better suited for matching the NLO
results with the parton shower than the standard $\msbar$ scheme.

The MC factorization scheme comes with a new set of parton distribution
functions that can be obtained from the standard $\msbar$ PDFs via the
relations given in Eqs.~(\ref{eq:PDFs-MC-quark}) and~(\ref{eq:PDFs-MC-gluon}).
Note that the convolution terms in Eq.~(\ref{eq:PDFs-MC-quark})
introduce the dependence on the renormalization scale via the strong
coupling multiplying the expressions for $\Delta C_{2q}$ and $\Delta C_{2g}$ in
Eqs.~(\ref{eq:DeltaC2q}) and (\ref{eq:DeltaC2g}). In what follows, we shall
always choose that scale to be equal to the scale $Q^2=\mu_F^2$ of the $\msbar$ PDF.

For clarity of the discussion, before presenting the complete NLO corrections to
the Drell--Yan process, we shall be starting in what follows from results limited to the pure
$q\qbar$ channel. This case corresponds to setting the gluon coefficient
functions of Eqs.~(\ref{eq:C2g-MSbar}) and (\ref{eq:C2g-MC}) to zero. That
in turn implies that, for the two schemes to be consistent at $\order{\as}$, the
last term in the relation between the quark PDFs in the $\msbar$ and MC schemes from
Eq.~(\ref{eq:PDFs-MC-quark}) should be omitted.

\begin{figure}[!t]
\centering
\includegraphics[width=0.48\textwidth]{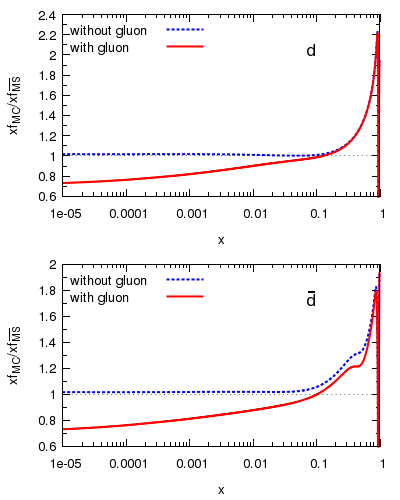}
\includegraphics[width=0.48\textwidth]{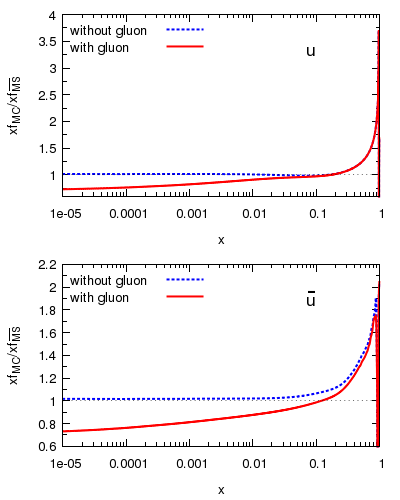}
\caption{\sf 
Ratios of MC PDFs to the standard $\msbar$ PDFs at $Q^2 = 100\, \GeV$.
The two curves in each plot correspond to the case with (red-solid line) and
without (dotted-blue line) the convolution with the gluon distribution in
Eq~(\ref{eq:PDFs-MC-quark}).}
\label{fig:mc-pdfs}
\end{figure}

The MC PDFs for light quarks and light antiquarks, obtained from the
{\tt MSTW2008} LO set \cite{Martin:2009iq} are shown in
Fig.~\ref{fig:mc-pdfs} through ratios to the $\msbar$ PDFs.  The curves correspond
to the choice of $Q^2 = 100\, \GeV$. The same scale was used as the argument of
$\as$. Each panel shown in Fig.~\ref{fig:mc-pdfs} contains two curves. The solid
line anticipates including both the $q\qbar$ and $qg$ channels, hence it was
obtained exactly with the formula (\ref{eq:PDFs-MC-quark}). On the other
hand, the dotted line corresponds to the case of the pure $q\qbar$ channel, hence
only the convolution with $\Delta C_{2q}$ was used, as explained earlier.

As seen in Fig.~\ref{fig:mc-pdfs}, the MC PDFs without the gluon, \eg without
the last term in Eq.~(\ref{eq:PDFs-MC-quark}), are very similar to the $\msbar$
quarks at low and moderate $x$. The MC PDFs in this region are only about 2\%
higher. At large $x$ the ratio increases, reaching the value $2$--$3$ near $x=1$.
Note that the MC PDFs are always higher than the standard $\msbar$ quark PDFs, which
results from the fact that the second term on the RHS of
Eq.~(\ref{eq:PDFs-MC-quark}) is always positive.
Adding the gluon component to the MC quark makes a little difference at large
$x$, whereas the small and moderate $x$ regions are affected significantly. We see that
the convolution with  $\Delta C_{2g}$ is negative and, at small $x$, it can
reduce the $\msbar$ quark PDFs even by 20\%.
We have found a very similar picture for other quark flavors as well for other $Q^2$
values.

\begin{figure}[!t]
\centering
\includegraphics[width=0.49\textwidth]{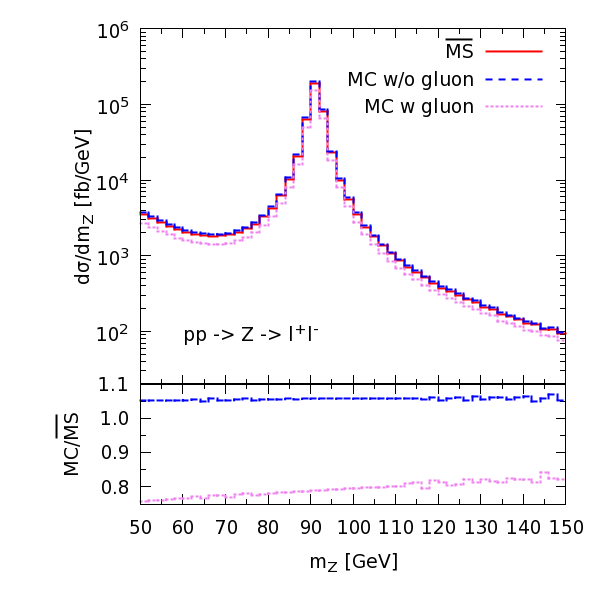}
\includegraphics[width=0.49\textwidth]{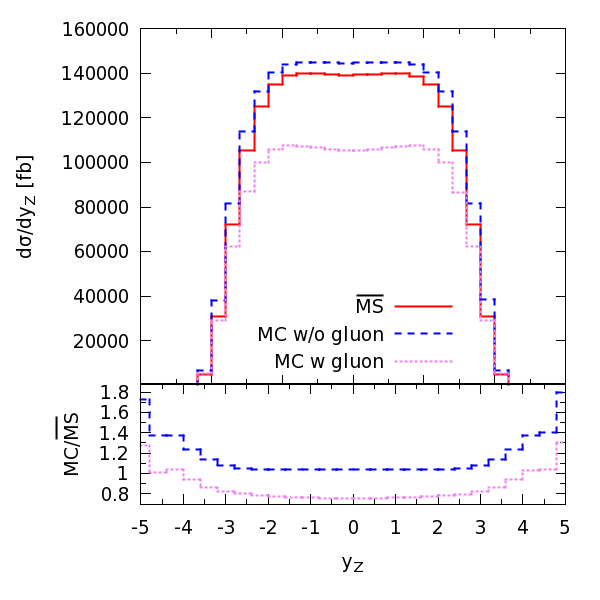}
\caption{\sf
Distributions of the transverse momentum and rapidity from the
LO calculation from the \mcfm{} program with the standard $\msbar$ MSTW2008LO PDFS
and with PDFs in the MC factorization scale with and without including the gluon
convolution in Eq~(\ref{eq:PDFs-MC-quark}).
}
\label{fig:dy-dist-lo} 
\end{figure}

Before moving to NLO, it is interesting to check how the LO
result changes when switching from  the $\msbar$ to MC PDFs.
From the LO point of view, both PDFs are equivalent, as
the differences are formally of the ${\cal O}(\as^2)$ order and higher. 
We can see, however, some numerical differences in the LO distributions between
those cases.
And, indeed, as shown in Fig.~\ref{fig:dy-dist-lo}, the differential
distributions of the dilepton mass and the dilepton rapidity differ at LO
depending on whether the $\msbar$ or the MC PDFs are used. 
In the case of the
MC PDFs without the gluon convolution, \cf Eq.~(\ref{eq:PDFs-MC-quark}), the
mass spectrum is $\sim5\%$ higher than that with the $\msbar$ PDFs
over a broad range around the peak.  
On the other hand, when we include the gluon convolution, the result with MC
PDFs gets up to $\sim20\%$ below that of $\msbar$.
The dilepton rapidity distribution with the
MC PDFs with (without) the gluon is smaller (larger) by a similar amount for central 
rapidities and grows above
$100\%$ in the forward and backward regions. This can be related to the large
$x$ behavior seen in Fig.~\ref{fig:mc-pdfs} via the formula $x_{1,2} =
\displaystyle \frac{m_Z}{\sqrt{s}}\exp\left(\pm y_{Z}\right)$ 
valid for $2\to 1$ kinematics. Here, $x_1$ and $x_2$ are the usual hadron's energy
fractions of the incoming partons, while $y_Z$ is the rapidity of the produced
$Z$-boson.
The large rapidities in the forward or backward direction
correspond to one of the $x$'es being large and that is the region where the
differences between the $\msbar$ PDFs and the MC PDFs are substantial.

\subsection{MC and $\msbar$ schemes at NLO}
\label{sec:mc-msbar-atNLO}

The cross section for the Drell--Yan process at NLO in the $\msbar$ scheme can be
schematically written as
\begin{equation}
\sigma_\text{\DY}^{\msbar}   = 
\sum_{\substack{i,j = q,\qbar\\  i \neq j}}
\!\!\!
f_i(x_1) \otimes \left[\delta(1-z) + \as \widetilde C_q^{\, \msbar} \right] \otimes f_j(x_2)
\ + \!
\sum_{\substack{i = q,\qbar\,; j = g\\ i = g\,; j = q,\qbar}}
\!\!\!\!\!\!\!
f_i(x_1) \otimes  \as \widetilde C_g^{\, \msbar} \otimes f_j(x_2)\,,
\label{eq:msbar-dy}
\end{equation}
where $f_i(x_{1,2})$ are the standard $\msbar$ parton distribution
functions, whose scale dependence is understood,
and $\otimes$ denotes the convolution via integration over the $z$ variable.
The first term in the square brackets
is just the Born contribution, which is followed by the NLO correction in the
$q\qbar$ channel. The second sum corresponds to the NLO contribution from the
$qg$ channel. 
In both cases, the functions $\widetilde C_{q,g}^{\, \msbar}$ have non-trivial
$z$-dependence (suppressed here for clarity of notation) and they are related to
the coefficient functions from Eqs.~(\ref{eq:C2q-MSbar}) and~(\ref{eq:C2g-MSbar})
as $\widetilde C_{q,g}^{\, \msbar} = 
\frac{1}{\displaystyle \as}\, C_{2q,2g}^{\msbar} (z)$. 
Hence, the dependence on $\as$ is explicit in Eq.~(\ref{eq:msbar-dy}) and all
the following formulae of this section. 

When going to the MC scheme, the $\msbar$ coefficient functions in
Eq.~(\ref{eq:msbar-dy}) need to be replaced by their MC scheme counterparts
and the PDFs need to be transferred to the MC scheme as well according to
Eq.~(\ref{eq:PDFs-MC-quark}). This leads to the formula
\begin{eqnarray}
\sigma_\text{\DY}^{\mc}   & = &
\sum_{\substack{i,j = q,\qbar\\  i \neq j}}
\!
\Big(f_i(x_1)  + f_i(x_1)\otimes \as \Delta \widetilde C_q + f_g(x_1) 
     \otimes \as \Delta \widetilde  C_g\Big)
\nonumber \\[-20pt]
& & \hspace{35pt}
\otimes \left[\delta(1-z) + \as \widetilde  C_q^{\mc} \right] \otimes 
\Big(f_j(x_2)  + f_j(x_2)\otimes \as \Delta \widetilde  C_q + f_g(x_2) 
     \otimes \as \Delta \widetilde  C_g\Big)
\nonumber \\
& &
+\
\sum_{i = q,\qbar}
\Big(f_i(x_1)  + f_i(x_1)\otimes \as \Delta \widetilde  C_q + f_g(x_1) 
     \otimes \as \Delta \widetilde  C_g\Big)
\otimes  \as \widetilde  C_g^{\mc} \otimes f_g(x_2)
\nonumber \\
& &
+\
\sum_{j = q,\qbar}
f_g(x_1)
\otimes  \as \widetilde  C_g^{\mc} \otimes 
\Big(f_j(x_2)  + f_j(x_2)\otimes \as \Delta \widetilde  C_q + f_g(x_2) 
     \otimes \as \Delta \widetilde  C_g\Big)\,,
\label{eq:mc-dy}
\end{eqnarray}
where, again
$\widetilde C_{q,g}^{\, \mc} = 
\frac{1}{\displaystyle \as}\, C_{2q,2g}^{\mc} (z)$, 
with the MC coefficient functions defined in
Eqs.~(\ref{eq:C2q-MC}) and~ (\ref{eq:C2g-MC}), and similarly for
$\Delta \widetilde C_{q,g} = 
\frac{1}{\displaystyle \as}\,\Delta  C_{2q,2g} (z)$, with the latter given in
Eqs.~(\ref{eq:DeltaC2q}) and~(\ref{eq:DeltaC2g}).
 
By construction (\cf section~\ref{sec:mc-scheme}), terms proportional to a given
partonic luminosity, $f_i f_j$, are identical in both factorization schemes up
to the order $\order{\as}$. The results in Eqs.~(\ref{eq:msbar-dy}) and
(\ref{eq:mc-dy}) differ however at $\order{\as^2}$, which is beyond NLO and
therefore such a difference is allowed.

In order to validate our transformation from the $\msbar$ to the MC scheme, we
have performed an explicit check of the equivalence of the two schemes up to
$\order{\as}$, as well studied numerical importance of the higher order terms.
The calculations were performed using our standard setup defined at the
beginning of Section~\ref{sec:nlo-bench}.

\begin{table}
\centering
\begin{tabular}{|c|c||c|c|}
\hline
\multicolumn{4}{|c|}{LO}                                       \\ \hline
\multicolumn{2}{|c||}{$\sigma^{(0)}_\text{DY}$\,[pb] $q\bar q$ channel} &
\multicolumn{2}{|c|}{$\sigma^{(0)}_\text{DY}$\,[pb] both channels} \\[3pt] 
\hline
$\msbar$                              &  $936.79\pm 0.30$ 
&$\msbar$ & $988.9 \pm 0.30$ \\ \hline
MC                                    &  $989.18 \pm 0.32$ & 
MC   & $778.8 \pm 0.20$ \\
{\small \it in which:} &  & 
{\small \it in which:}  &                      \\ \hline
\multicolumn{4}{|c|}{$\order{\alpha_s}$ }          \\ \hline
$f_q f_\qbar$ \hspace{-1.5em}           & $25.79 \pm 0.04$ & 
$ f_q f_{\bar q} + f_\qbar f_q$ & ~~~~$54.8   \pm 0.4$ \\
$ f_\qbar f_{q} $\hspace{-1.5em}   & $25.79 \pm 0.02$ & 
$f_q f_g + f_\qbar f_g$ & $-271.4 \pm 0.4$ \\ \hline
\multicolumn{4}{|c|}{$\order{\alpha_s^2}$ }          \\ \hline
  & $0.64 \pm 0.01$ & 
 & $6.70  \pm 0.20$ \\ \hline
\end{tabular}
\caption{\sf
Values of the total cross section with statistical errors for the the Drell-Yan process at LO in $\msbar$ and
MC factorization schemes. The results were obtained with \mcfm{} 6.6~\cite{MCFM}.
The $\order{\as}$ admixture in the MC results is split into contributions
proportional to various terms of Eq.~(\ref{eq:mc-dy}).
}
\label{tab:mcvsmsbarmcfm-lo}
\end{table}

Let us start from comparing the cross sections at LO. This corresponds to
setting the coefficient functions $\widetilde C_{q,g}=0$ in
Eqs.~(\ref{eq:msbar-dy}) and~(\ref{eq:mc-dy}).
As shown in the previous section, the LO cross sections will differ between the
two schemes because of the PDFs.
To check the extent to which this happens,
we performed an explicit computation with \mcfm{}~\cite{MCFM} in both factorization schemes for
either the pure $q\qbar$ channel or both channels.
 
Indeed, as demonstrated in Table~\ref{tab:mcvsmsbarmcfm-lo}, 
the LO $\msbar$ and MC cross sections
are not identical, both in the case of the pure $q\qbar$ channel, where the LO
result with MC PDFs is $\sim$5\% higher, and in the case where both channels are
included, where the MC result is ~$\sim$20\% lower. 
The beyond-LO, $\order{\as}$ and $\order{\as^2}$, terms
are also given in the table. We see that most of the difference comes from
$\order{\as}$.  The $\order{\as^2}$ terms are in fact very small, below 1\% in
both cases.
It is interesting to note that the difference between $\sigma_\DY^{(0)}$ 
in the two schemes, for the case with both channels, 
comes primarily from the term proportional to 
$\as f_q \otimes \Delta C_g \otimes f_g$, hence it originates from
the large gluon luminosity.

\begin{table}
\centering
\begin{tabular}{|c|c||c|c|}
\hline
\multicolumn{4}{|c|}{NLO}                                       \\ \hline
\multicolumn{2}{|c||}{$\sigma^{(1)}_\text{DY}$\,[pb] $q\bar q$ channel} &
\multicolumn{2}{|c|}{$\sigma^{(1)}_\text{DY}$\,[pb] both channels} \\[3pt] 
\hline
$\msbar$  & $336.36 \pm 0.09$ 
&$\msbar$ & $157.9  \pm 0.10$ \\ \hline
MC        & $352.96 \pm 0.09$ & 
MC        & $305.8  \pm 0.10$ \\
{\small \it in which:} &  & 
{\small \it in which:}  &                      \\  \hline
\multicolumn{4}{|c|}{$\order{\alpha_s}$ }          \\ \hline
$f_q \Delta \widetilde C_q f_\qbar$            & 25.79 $\pm$ 0.04 
& $f_q \Delta \widetilde C_q f_\qbar + f_\qbar \Delta \widetilde C f_q $ & ~~~$355.00 \pm 0.29$  \\[0.3em]
$ f_\qbar \Delta \widetilde C_q f_q $  & $25.79 \pm 0.02$ 
& $f_q \widetilde C^\mc_q f_\qbar + f_\qbar \widetilde C^\mc_q f_q$ & ~$-17.10 \pm 0.14$  \\[0.3em]
$f_{q,\qbar} \widetilde C_q^{\mc} f_{\qbar, q}$  & $284.77 \pm 0.08$ 
& $f_q \widetilde C^\mc_g f_g + f_\qbar \widetilde C^\mc_g f_g$ & $-180.10 \pm 0.10$\\[0.3em]
 sum & $336.35 \pm 0.09$ & sum & ~~~$157.80 \pm 0.34$ \\ \hline
\multicolumn{4}{|c|}{$\order{\alpha_s^2} + \order{\alpha_s^3}$ } \\ \hline
MC  & $16.61 \pm 0.02$ & MC &  $147.9 \pm 0.20$\\ \hline
\end{tabular}
\caption{\sf
Values of the NLO contribution to the total cross section with statistical errors for the the Drell--Yan process in the $\msbar$
and MC factorization schemes. 
The results were obtained with \mcfm{} 6.6~\cite{MCFM} and its version adjusted to
the MC scheme, \mcfm{}$^*$.
The $\order{\as}$ corrections in the MC results is split into contributions
proportional to various terms of Eq.~(\ref{eq:mc-dy}).
}
\label{tab:mcvsmsbarmcfm-nlo}
\end{table}

We turn now to a similar comparison at NLO.
Table~\ref{tab:mcvsmsbarmcfm-nlo} 
gives the NLO-only results in the two  
schemes for the cases of $q\qbar$ and both channels. 
At NLO, we need to be careful what we take for the coefficient functions. In the
$\msbar$ scheme, we just use the original \mcfm{}~6.6~\cite{MCFM} implementation.
However, in order to carry out consistent NLO calculations in the MC scheme,
we had to change the coefficient functions from $C_{2q,2g}^{\msbar} (z)$ to 
$C_{2q,2g}^{\mc} (z)$. We dubbed this modified version of the program \mcfm{}$^*$
and used it together with the MC PDFs to compute our NLO predictions in that scheme.

We see that also the
subleading corrections to the Drell--Yan process are not the same in the MC and $\msbar$
schemes (numbers in the two first lines of Table~\ref{tab:mcvsmsbarmcfm-nlo}). This is allowed provided that all the
difference comes from terms of the order $\as^2$ and beyond. As shown
in Table~\ref{tab:mcvsmsbarmcfm-nlo}, by extracting only the $\order{\as}$ terms
of the MC result and summing them up, we recover the $\msbar$ cross
section exactly.
There, we also give subleading correction introduced via the MC PDFs and we see
that they contribute at most $5\%$ to the NLO correction in the MC scheme in the
case of the pure $q\qbar$ channel but can be quite sizable when both channels are
considered.

Finally, it would be interesting to fit the quark and gluon PDFs in the MC factorization
scheme directly to experimental data, in order to minimize higher-order effects.
For that we would need to define the gluon PDF in the MC factorization scheme using
a process in which the gluon PDF enters already at the LO level, 
\eg\ the Higgs-boson production process.
This will be done in our future publication.

Let us conclude that we have successfully validated the MC factorization
scheme by explicitly showing that all numerical differences w.r.t. the $\msbar$
results are of the order of $\as^2$ and $\as^3$. 
We emphasize that the validation of the new MC scheme presented in this section
provides a highly non-trivial check as various components, given schematically
in Eq.~(\ref{eq:PDFs-MC-quark}), come from different parts of the
calculation (PDFs, coefficient functions) and the agreement up to $\as$ is much
more sophisticated than a purely algebraic relation.

\section{Results for NLO with parton shower}
\label{sec:NLOwith-PS}

Numerical implementation of \krknlo{} is presently done mainly using the  \sherpa{}
PSMC of Ref.~\cite{Gleisberg:2008ta} version 
2.0.0\footnote{\tt https://sherpa.hepforge.org/doc/SHERPA-MC-2.0.0.html } 
with the dipole organization of  the parton shower distributions 
inspired by the Catani--Seymour work~\cite{Catani:1996vz},
see Ref.~\cite{Schumann:2007mg}.
However, the evolution variable in Ref.~\cite{Schumann:2007mg} is chosen to be
the transverse momentum distribution, while in the actual \sherpa{} 2.0.0 implementation
it is the $q^2\sim\alpha(\alpha+\beta)$ variable 
of Section~\ref{sec:kinematics}, see Ref.~\cite{Hoeche:2009rj}.
The dipole shower implemented in \herwig{}, see Ref.~\cite{Platzer:2009jq},
is quite similar.
Actually, the choice of the evolution variable is not critical
in the \krknlo{} method, as long as the full coverage of the phase space for the emission
closest to the hard process
is assured, and the PSMC distribution is under the perfect control.
In fact, for any choices of the dipole PS evolution variable in \sherpa{} or \herwig{}
the same distribution of the single-gluon emission is obtained, 
provided that the contributions for the quark and antiquark emitters are added.
For the \sherpa{} MC program, we have checked numerically the completeness of the coverage
of the phase space of the first emission from the backward evolution
by examining the gluon distribution on the $\alpha,\beta$ plane.

In the following, the \krknlo-matched results will be compared mostly with the
results of the \mcatnlo{} technique, implemented on top of the same parton shower
within \sherpa{}\footnote{Unfortunately, the \powheg{} method is not available in
\sherpa{ 2.0.0} framework.}. Using the same parton shower in both cases ensures that all
differences in the matched results come purely from the differences in the
methods themselves rather than details of the shower
implementation. 
Nevertheless, the difference between the dipole showers currently implemented in
\sherpa{} and \herwig{} turns out to be small. 
From the point of view of our study, it amounts mainly to using the $q$ variable in
the former and $k_T$ in the latter.  These two evolution variables are, however,
very close to each other in practice. Hence, in our final discussion, we shall also compare
the \krknlo{} results with those of \powheg{}, as implemented in \herwig{}\footnote{For 
our comparisons we used the version of \powheg{} without restrictions of the phase space, which is
similar to the original \powheg{} method described in Ref.~\cite{Nason:2004rx}.} 
using an automated setup based on the \textsf{Matchbox} 
framework~\cite{Platzer:2011bc} with adaptive sampling~\cite{Platzer:2011dr}.

In the numerical validation of the \krknlo{} approach, we shall first 
compare it to the fixed-order NLO calculation, as implemented in the \mcfm{} program.
Later on, the comparison with the \mcatnlo{} implementation of \sherpa{} will
be discussed.
In order to minimize the influence of various higher-order effects
on the difference between  \krknlo{} and \mcfm{}, we shall initially limit
our numerical exercises to a simplified version, in which only the $q\qbar$ channel is kept.
Another initial limitation, with the same aim in mind, will be to
profit from the option in \sherpa{} to stop the backward evolution in the parton shower
just after the first emission (starting from the hard process).
This will help us to eliminate the influence of the differences in the so-called
recoil schemes~\cite{Hoeche:2009xc}, which start to play a role from the second emission onwards. 
The running $\alpha_s$ will be also introduced gradually.
Since we study the differences in the matching methods, 
we have switched off in \sherpa{} non-perturbative 
effects including the intrinsic $k_T$, multiparton interactions and hadronization. 
However, we would like to stress that there is no problem with 
switching them on for the \krknlo{} method. 

In all the following results for the \krknlo{} method the PDFs in the MC factorization
scheme, discussed in Sec.~\ref{sec:PDFSinMCscheme}, will be used.
 The factorization scale will be set to $\mu_F= M_Z$.
As for the choice of the renormalization scale
in the matched NLO$\,+\,$PS results, we note that there is always some level of
arbitrariness.  In the pure fixed-order NLO calculation, in the case of DY, one usually sets
$\mu_R=\mu_F = M_Z$. On the other hand, the LO parton shower uses $\mu_R^2 = q^2$,
with the latter being 
closely related to the evolution variable of the shower.
The parton shower is unitary, hence the choice of the argument of $\as(\mu_R^2)$
does not influence the total cross section, 
it changes, however, shapes of some distributions, \eg of $p_{T,Z}$. 
The differences between $\as(q^2)$ and $\as(M_Z^2)$ are
formally of the higher order, that is beyond NLO and, from that perspective, 
they are equivalent in the context of NLO$\,+\,$PS matching.
They can, however, be numerically relevant.
 
In what follows, for the \krknlo{} method we adopt the following procedure. 
The argument of $\as$ in the
virtual correction is always set to $M_Z^2$. For the real correction, all
emissions except for the first one (in BEV) are set by the parton shower according to its
current $q^2$ value. For the first emissions, we consider two choices: $q^2$ and
$M_Z^2$, both in the real correction and in the Sudakov form factor, to keep the shower
unitary. The difference between the results corresponding to these two choices
will be indicative of the size of beyond-NLO terms.

\subsection[Initial results for $q\qbar$ channel only]
           {Initial results for $\mathbold{q\qbar}$ channel only}

\begin{table}[!t]
\centering
  \begin{tabular}{|l|c|}
 \hline 
           &      $\sigma_\text{tot}^{q\bar q}$ [pb]  \\ \hline \hline
\mcfm{}       &      $1273.4  \pm  0.1$        \\
\mcatnlo{}     &      $1273.4  \pm  0.1$        \\
\powheg{}     &      $ 1272.1  \pm  0.7$          \\ \hline 
\krknlo{} $\alpha_s(q^2)$    &      $1282.6  \pm  0.2$        \\ 
\krknlo{} $\alpha_s(M_Z^2)$     &     $1285.3  \pm  0.2$        \\ 
\hline 
  \end{tabular}
  \caption{\sf
   Values of the total cross section with statistical errors for the Drell--Yan process (the $q\qbar$ channel only) from the \krknlo{} method
    compared to the fixed-order result of \mcfm{} and the results of \mcatnlo{} and \powheg{}.
  }
  \label{tab:krknlo-qqbar-xsectio}
\end{table}

\begin{figure}[!t]
\centering
  \includegraphics[width=0.50\textwidth]{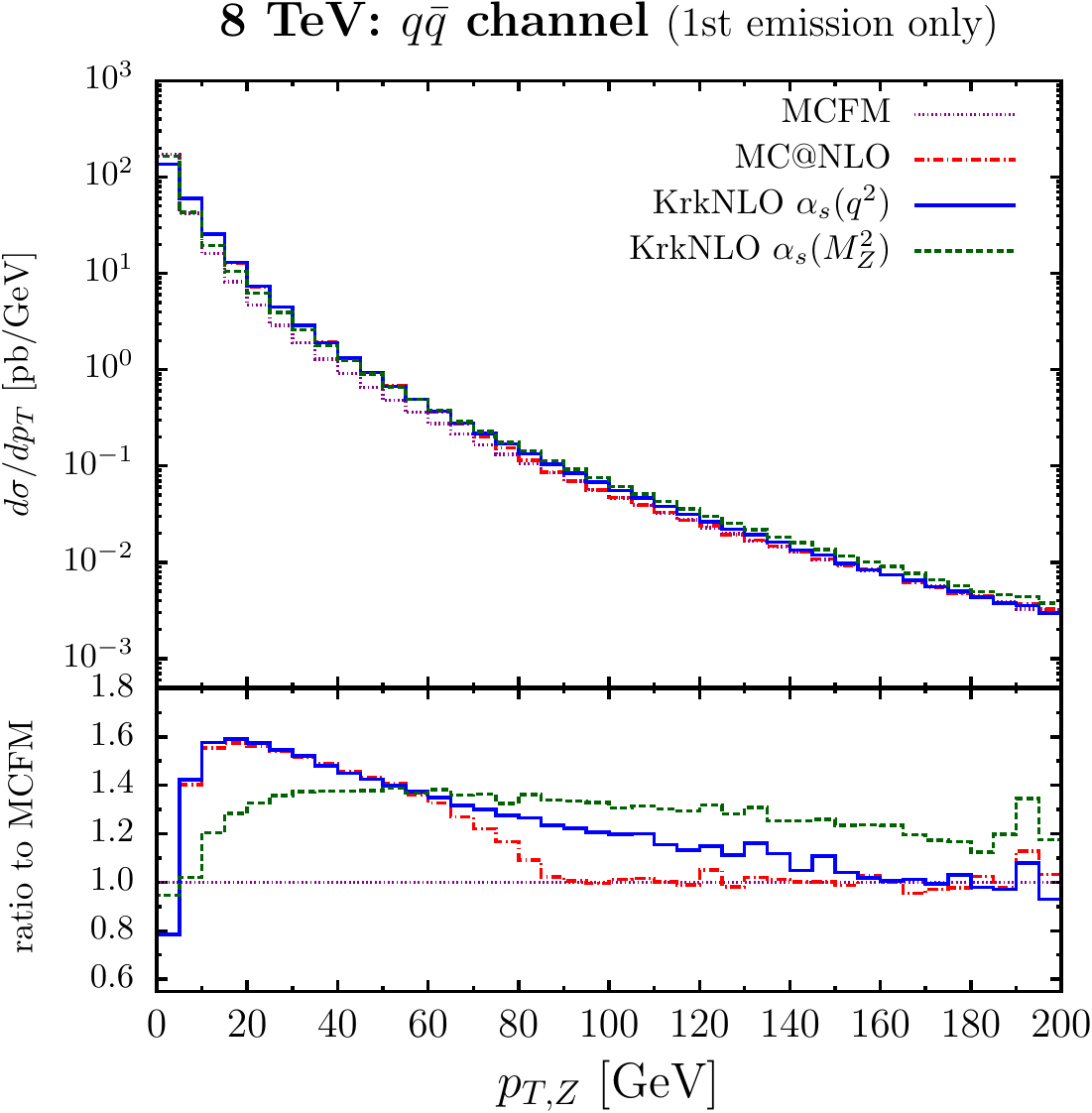}
  \includegraphics[width=0.48\textwidth]{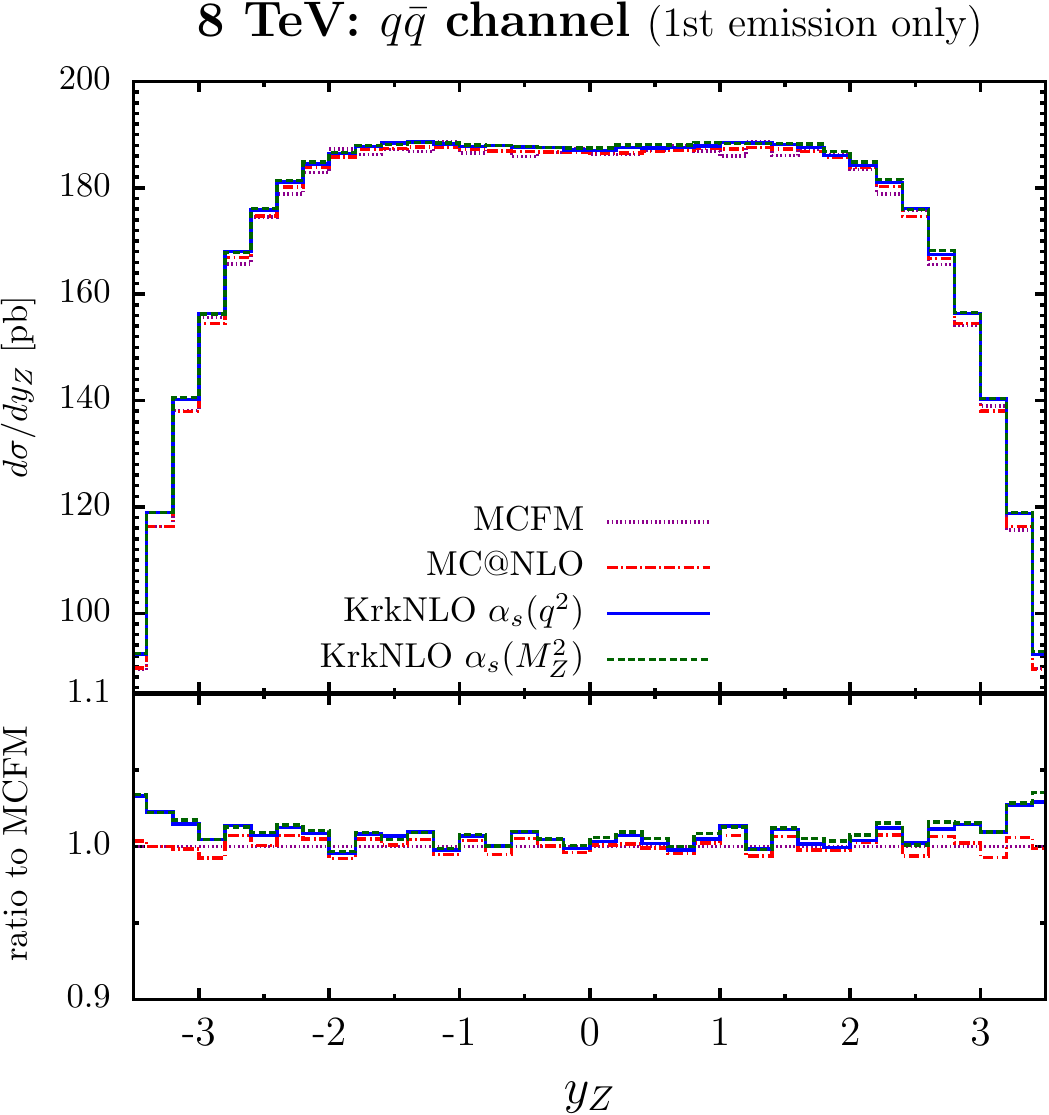}
\caption{\sf
Comparisons of the transverse momentum and rapidity distributions from
\mcfm{}, \mcatnlo{} and two versions of \krknlo{,} one with fixed $\alpha_s(M_z^2)$
(also in \mcfm{})
and another one with running  $\alpha_s$ depending on the transverse momentum squared.
\krknlo{}~method is implemented within \sherpa{} parton shower MC.
The exercise is restricted to the $q\qbar$ channel only and the parton shower 
backward evolution is stopped after the first emission, the closest to the hard process.
\label{fig:krk-qqbar-1em}
}
\end{figure}  

The first round of numerical tests of \krknlo{} focuses on comparisons with \mcfm{}
and \mcatnlo{} for the $q\qbar$ channel only and stopping the parton shower after the first
emission.
The corresponding results for the total cross section
are presented in Table~\ref{tab:krknlo-qqbar-xsectio}
and for the transverse momentum and rapidity distributions
in  Fig.~\ref{fig:krk-qqbar-1em}.
As we can see, the total cross section and the rapidity distributions from \krknlo{}
agree very well with those of \mcfm{} and \mcatnlo{}.
The difference between the \krknlo{} results and the pure fixed-order ones is below $1\%$ and
comes from the $\order{\as^2}$ contamination due to the MC PDFs, \cf
Eq.(\ref{eq:mc-dy}). Also, the differences between the $\as(q^2)$ and
$\as(M_Z^2)$ are at the per-mille level.

On the other hand, there are remarkable differences in the $p_{T,Z}$ distributions,
especially at the lower end. They compensate in the total cross section
between the first and the following bins.
However, the choice of the running or non-running $\alpha_s$ is numerically
important for the $p_T$ distribution at small and moderate values.
The perfect agreement of \mcatnlo{} and \mcfm{} at higher $p_T$ is of course
enforced by construction.  This is not the case in \krknlo{}, 
where some part of higher-order effects, 
beyond NLO, is included
(which is also the case in \powheg{}).
We see, however, that both \mcatnlo{} and \krknlo{} coincide below $\sim 80\,$GeV
in the case with $\as(q^2)$.

\begin{figure}[!t]
\centering
  \includegraphics[width=0.50\textwidth]{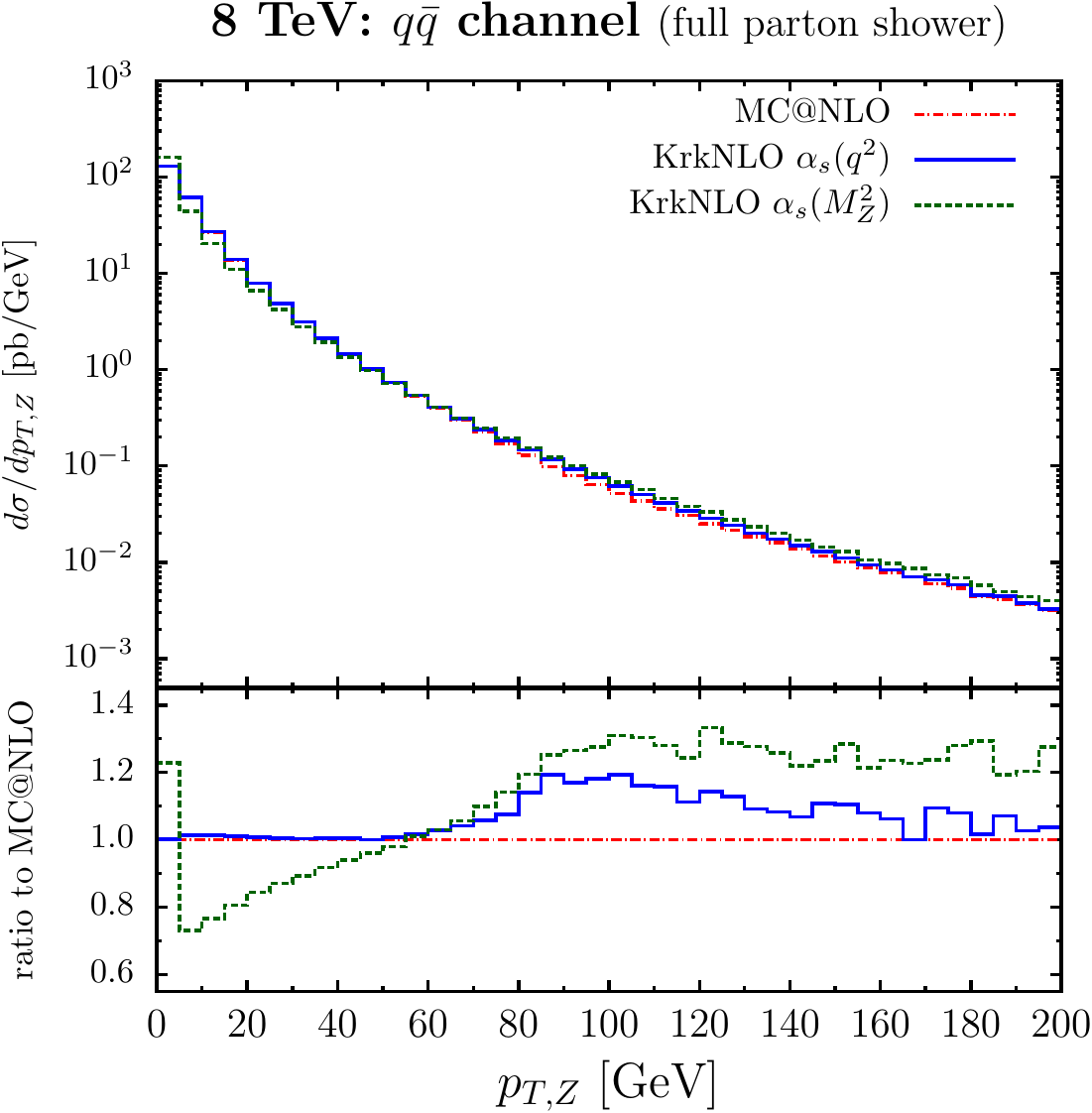}
  \includegraphics[width=0.48\textwidth]{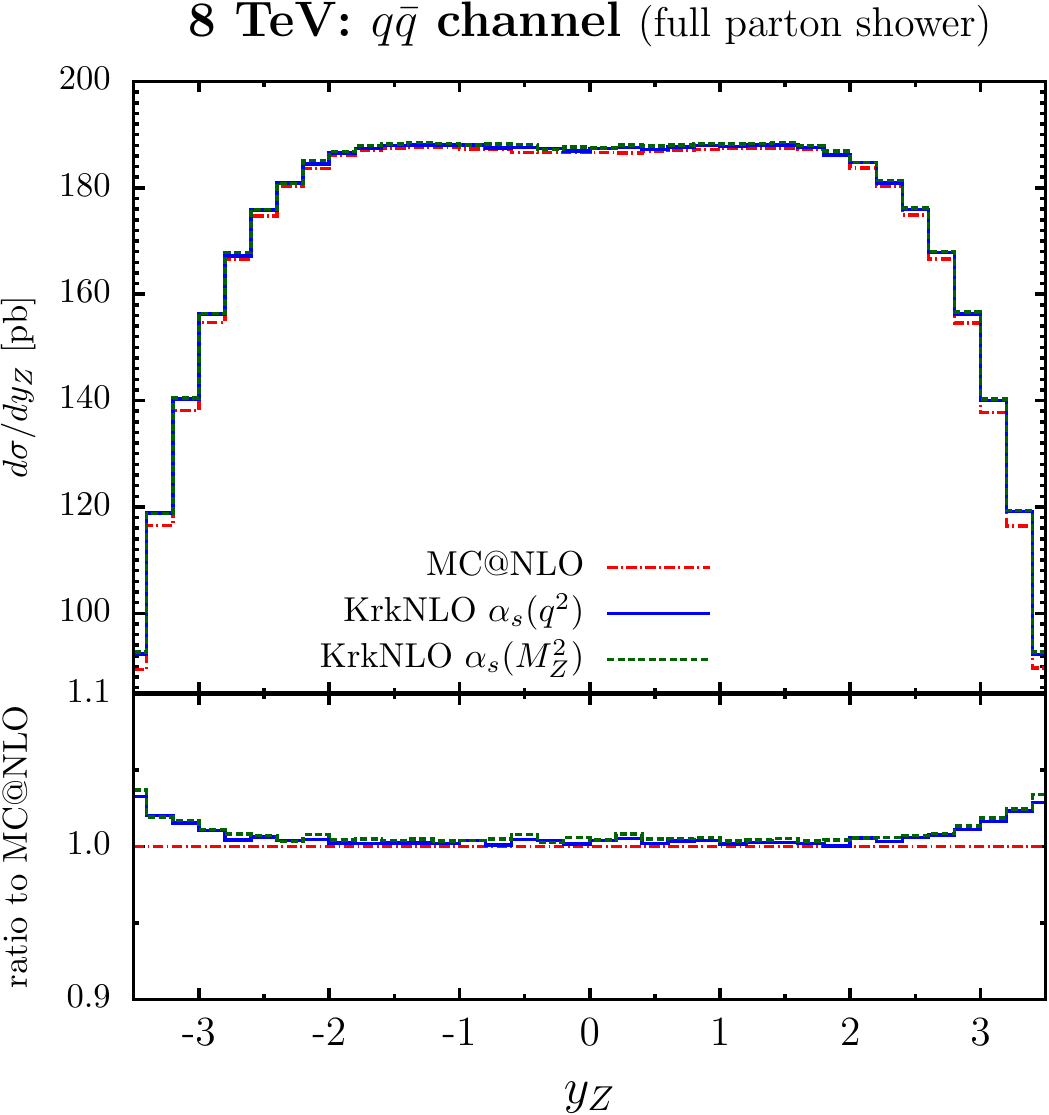}
\caption{\sf
Comparisons of the transverse momentum and rapidity distributions from
\mcatnlo{} and two versions of \krknlo{}~
for the $q\qbar$ channel only as in Fig.~\protect\ref{fig:krk-qqbar-1em},
but the parton shower  backward evolution runs to the end as normally.
\label{fig:krk-qqbar-PS}
}
\end{figure}

In the results of Fig.~\ref{fig:krk-qqbar-1em}
the parton shower was artificially stopped after the first emission,
in order to make the meaningful comparisons with \mcfm{} and limit the effects of the recoil 
due to subsequent emissions. In Fig.~\ref{fig:krk-qqbar-PS} we lift this limitation and 
the parton shower goes to the very end, as in the normal operational mode of PSMC.
As we see, switching on to the full PS influences considerably the
low $p_T$ part of the spectrum, in spite of a negligible effect
on the rapidity distribution (and hence, on the total cross section).

Again, \krknlo{} with the running $\alpha_s$ agrees very well with \mcatnlo{} at low
and moderate $p_T$, which results from the domination of the parton shower contribution in that region.
However, the two \krknlo{} results stay above \mcatnlo{} at higher $p_T$,
which we again attribute to the admixture of some NNLO terms in the former.
The difference between \krknlo{} $\as(q^2)$ and \krknlo{} $\as(M^2_Z)$ at high $p_T$
comes purely from the running of the coupling.
%
%

\subsection{All channels}

In the second round of the numerical tests, all the initial-state
parton combinations contributing at NLO, that is the $q\qbar$ and $qg$ channels
are included.
The total cross sections from \krknlo{}, \mcfm{} and \mcatnlo{}
are compared in Table~\ref{tab:krknlo-both-totalcs}.
\begin{table}[h]
\centering
\begin{tabular}{|l|c|}
\hline
           &      $\sigma_\text{tot}^{q\bar q + qg}$ [pb]  \\ \hline \hline
\mcfm{}       &      $1086.5  \pm  0.1$        \\
\mcatnlo{}     &      $1086.5  \pm  0.1$        \\
\powheg{}     &      $1084.2  \pm  0.6$           \\ \hline
\krknlo{} $\alpha_s(q^2)$    &      $1045.4  \pm  0.1$        \\ 
\krknlo{} $\alpha_s(M_Z^2)$  &      $1039.0  \pm  0.1$        \\ 
\hline
\end{tabular}
  \caption{\sf
    Values of the total cross section with statistical errors for the Drell--Yan process, both channels, from the \krknlo{} method
    compared to the fixed-order result of \mcfm{} and the results of \powheg{} and \mcatnlo{}.
  }
  \label{tab:krknlo-both-totalcs}
\end{table}
As we can see, the difference between \mcfm{}/\mcatnlo{} and \krknlo{}, which
comes from the partial inclusion of the higher-order effects in the latter,
is about twice as big as in the $q\qbar$ channel only. 
As discussed in Section~\ref{sec:mc-msbar-atNLO}, this is related to the large gluon
luminosity which leads to sizable differences between the MC and $\msbar$
PDFs, and part of that difference shows up in the total cross section.
The results in Table~\ref{tab:krknlo-both-totalcs} are generally consistent
with the differences staying below $5\%$.

\begin{figure}[!ht]
\centering
  \includegraphics[width=0.50\textwidth]{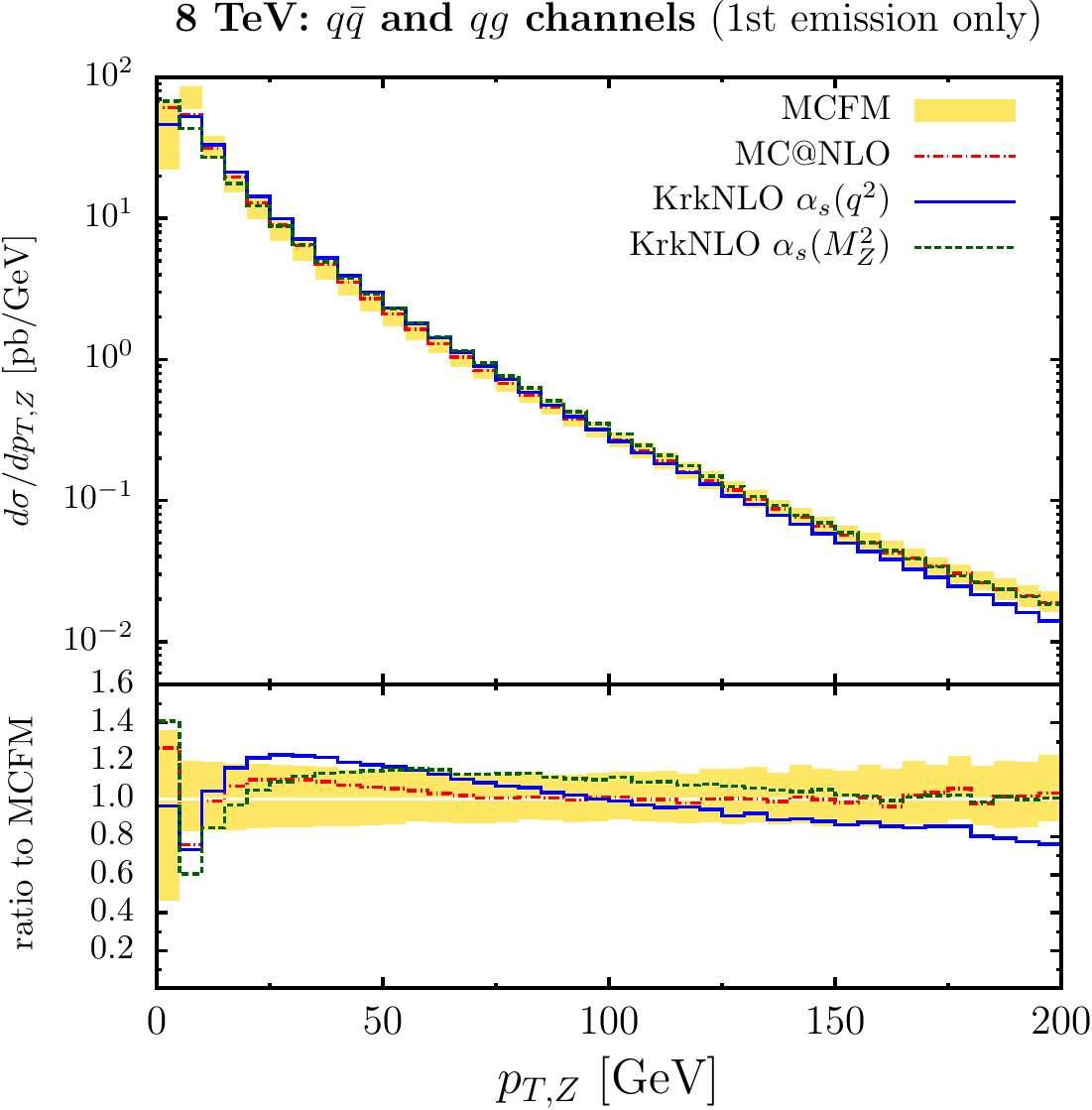}
  \includegraphics[width=0.48\textwidth]{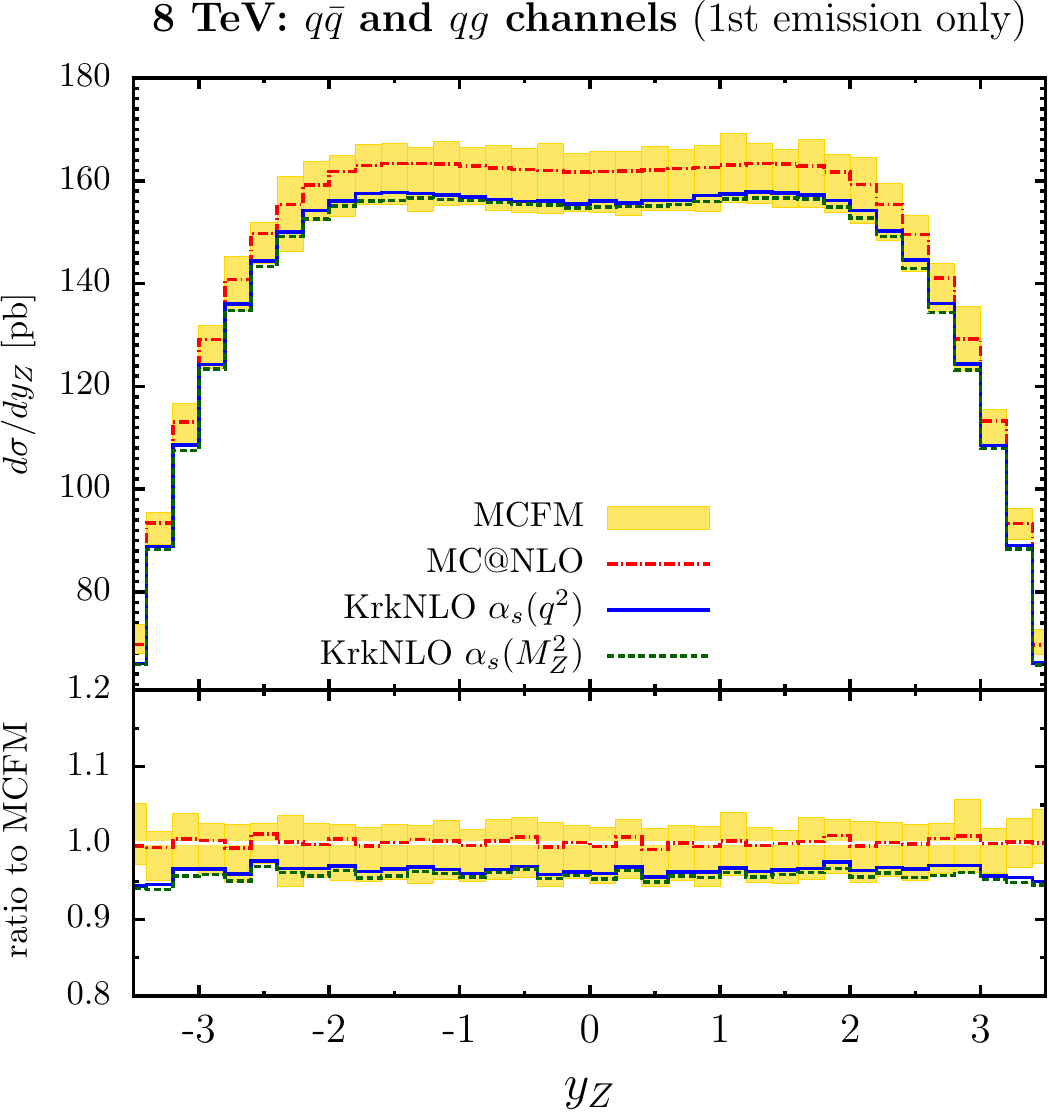}
\caption{\sf
Comparisons of the transverse momentum and rapidity distributions from
\mcfm{}, \mcatnlo{} and two versions of \krknlo{,} now
for both the $q\qbar$ and $qg$ channels, with $\alpha_s$ set as in
the previous figures and the parton shower backward evolution stopped 
after the first emission as in Fig.~\ref{fig:krk-qqbar-1em}.
}
\label{fig:krk-2ch-1emi} 
\end{figure}
In Fig.~\ref{fig:krk-2ch-1emi} we show the distributions of the transverse momentum and
rapidity of the $Z$-boson with the parton shower stopped after the first emission.
The lower panel contains the ratio with respect to \mcfm{}.
Similarly to the case of the pure $q\qbar$ channel, we see the expected agreement
between \mcatnlo{} and \mcfm{} at large $p_{T,Z}$.
The \krknlo{} results with the fixed $\alpha_s$ 
are not far from the \mcfm{} ones in that region as well. 
On the other hand, all the PS-matched results differ from those of \mcfm{} 
at smaller $p_T$, because of the lack of the Sudakov resummation in \mcfm{},
see below for more discussion on this issue.
Each curve behaves slightly differently but the variations are moderate.
The rapidity distributions shown in Fig.~\ref{fig:krk-2ch-1emi} (right) are
close to each other for all calculations.  The $\sim 5\%$ difference between
\krknlo{} and \mcfm{} is the same as for the total cross sections in
Table~\ref{tab:krknlo-both-totalcs}.
In Fig.~\ref{fig:krk-2ch-1emi} we have also
included the scale-uncertainty bands, which were produced
using \mcfm{} by means of changing (independently) both the factorization 
and the renormalization scales by the customary factors of $2$ and $1/2$.
The differences between \mcfm{} and \krknlo{} are enveloped by these bands.

\begin{figure}[!ht]
\centering
  \includegraphics[width=0.46\textwidth]{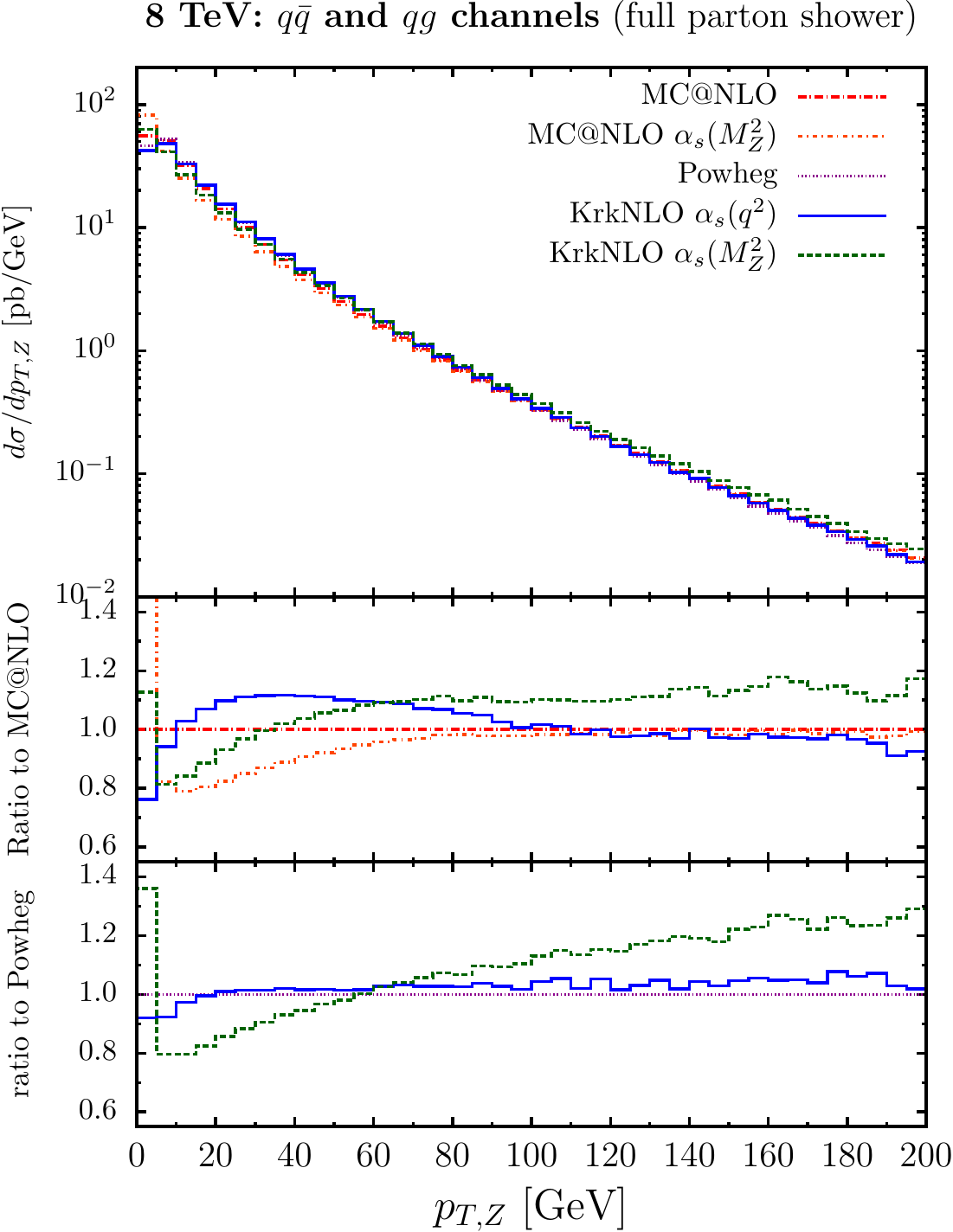}
  \includegraphics[width=0.45\textwidth]{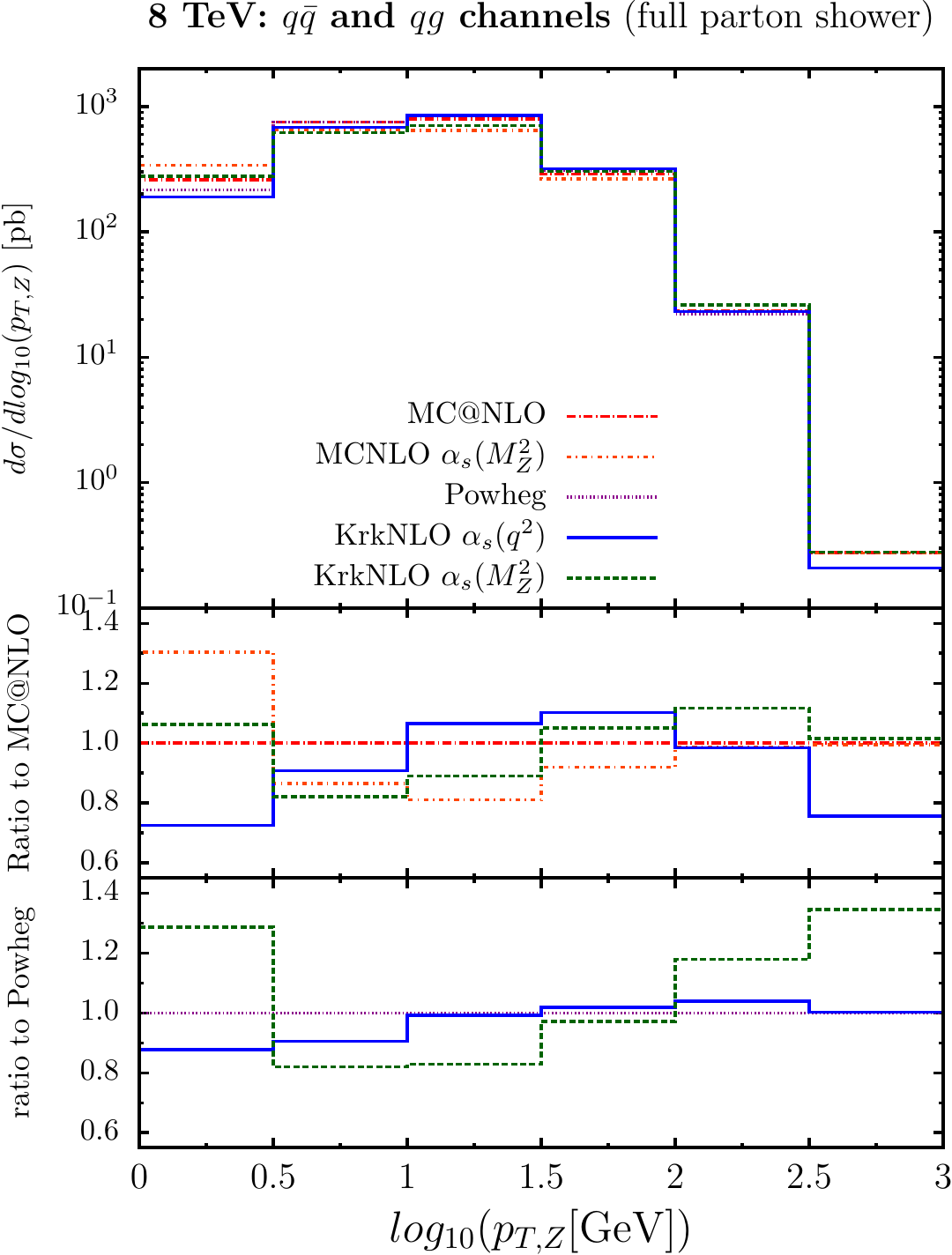}\\
\caption{\sf
Comparisons of the transverse momentum distribution from
\mcatnlo{}, \powheg{} and two versions of \krknlo{}
for both the $q\qbar$ and $qg$ channels as in Fig.~\protect\ref{fig:krk-2ch-1emi},
but the parton shower backward evolution runs to the end as normally.
}
\label{fig:krk-2ch-pt}  
\end{figure}

Fig.~\ref{fig:krk-2ch-pt} shows the similar distributions but, this time, the parton
shower is allowed for an arbitrary number of emissions. 

\begin{figure}[!h]
\centering
  \includegraphics[width=0.45\textwidth]{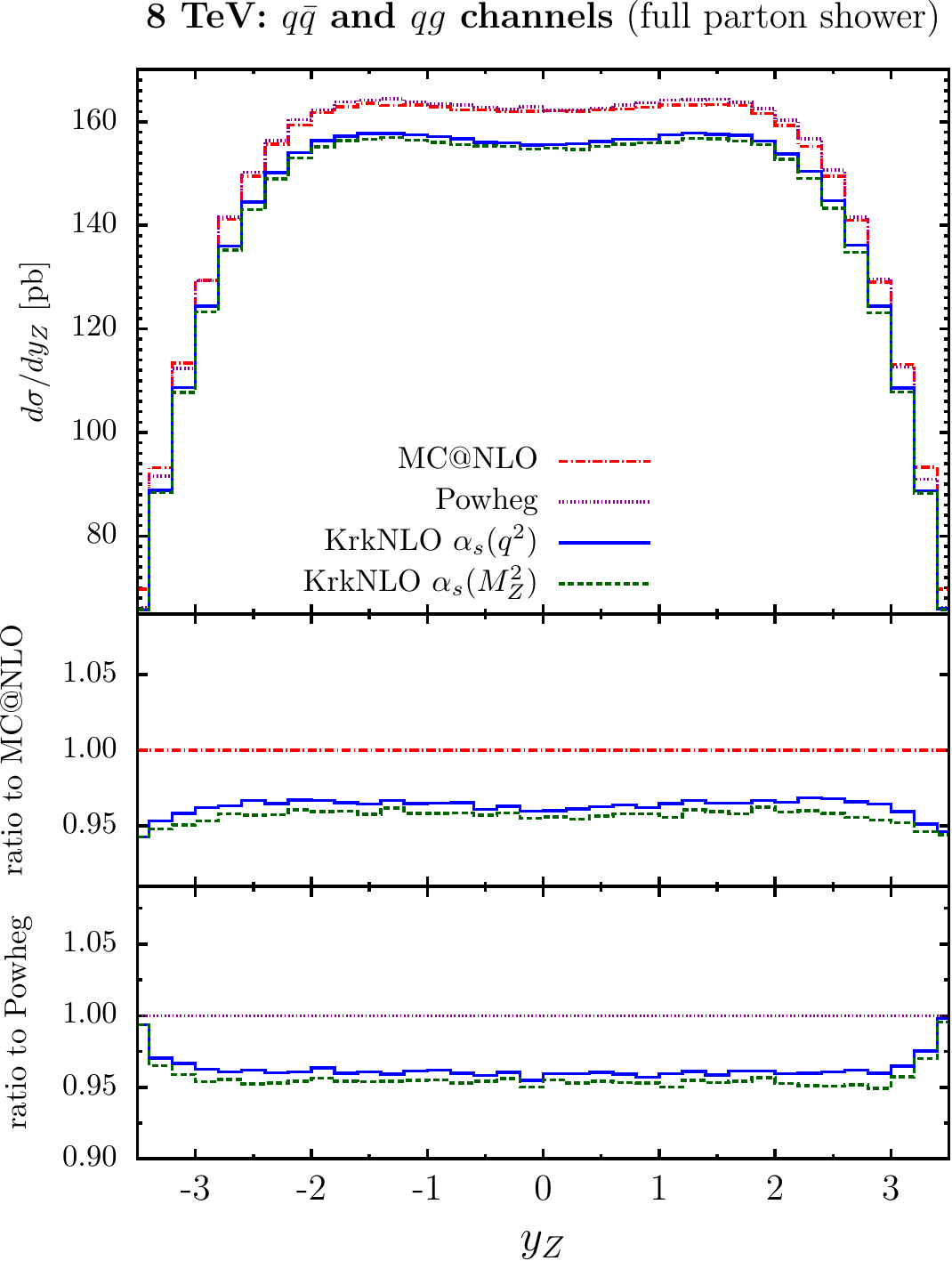}
\caption{\sf
Comparison of the rapidity distribution from
\mcfm{}, \mcatnlo{} and two versions of \krknlo{}
for $q\qbar$ and $qg$ channels as in Fig.~\protect\ref{fig:krk-2ch-1emi},
but parton shower backward evolution runs to the end as normally.
}
\label{fig:krk-2ch-rap} 
\end{figure}

\begin{figure}[!t]
\centering
  \includegraphics[width=0.45\textwidth]{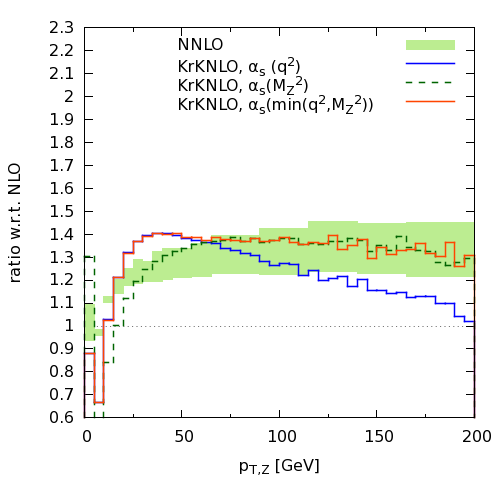}
  \includegraphics[width=0.45\textwidth]{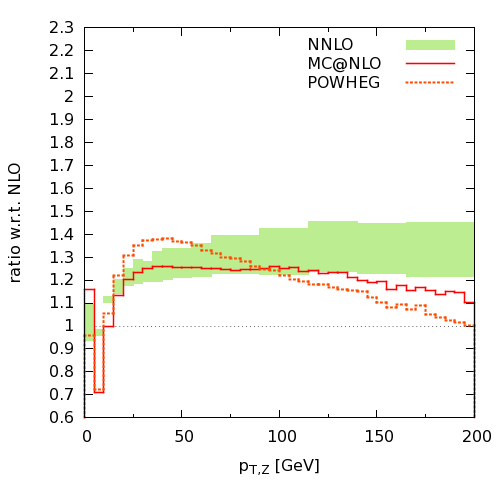}\\
\caption{\sf
The $Z$-boson transverse-momentum distributions from
\krknlo{} compared with the fixed-order NNLO result
from the {\sf DYNNLO} program~\cite{Catani:2009sm} (left).
Similar comparisons for \powheg{} and \mcatnlo{} are also shown (right).
All distributions are divided by the NLO results from \mcfm.
}
\label{fig:14}  
\end{figure}

There, we also show the results from \powheg{}, as implemented in \herwig{}. Even
though the evolution variable is different ($k_T$ rather than $q$ of
Eq.~(\ref{eq:q2vardef}) in all the other PS-matched results), 
\powheg{} agrees quite well with the \krknlo{} $\alpha_s(q^2)$ version.
This shows that the choice of the evolution variable in PS between $k_T$ 
and $q$ is not so important for numerical results. 
The distributions of the transverse momentum and rapidity
look quite similar to the previous results for the pure $q\qbar$ channel
in Fig.~\ref{fig:krk-qqbar-PS}.
The transverse momentum distribution from \krknlo{} agrees with \mcatnlo{}
at high $p_{T,Z}$ (up to a constant factor in the case with the fixed $\alpha_s$),
while larger differences are seen in the low $p_T$ region.  

The differences of the order of $10$--$20\%$ seen at low $p_T$  
between \mcfm{}, \mcatnlo{}, \powheg{} and \krknlo{} 
in Figs.~\ref{fig:krk-qqbar-PS}, \ref{fig:krk-2ch-1emi} and \ref{fig:krk-2ch-pt}
reflect mainly the parton-shower feature of the
soft-gluon resummation in the case of the matched results
and the lack of it in the case of the fixed-order NLO calculations.
More precisely, parton shower (without intrinsic $k_T$)
has a gap below the minimum $p_T$ ending  the backward evolution
and a non-physical spike $\sim\delta(p_T)$.
Normalization of the spike is governed by unitarization enforced by the shower.
Kinematics, histogramming, intrinsic $k_T$, etc.\ are smearing 
the shape of this non-physical structure.
Different implementations of the argument of $\alpha_s$ strongly
influence this structure near $p_T=0$ and may show up in the ratios
of the $p_T$ distribution as large effects there.
In particular, our results with $\alpha_s(M_Z^2)$ at low $p_T$ are very
different from the results with $\alpha_s(q^2)$. 
The latter choice better represents resummation of subclasses of higher-order
QCD contributions, relevant at low $p_T$, and it is therefore 
commonly adopted in PS MCs.
In Fig.~\ref{fig:krk-2ch-pt} another curve is added for \mcatnlo{} with $\alpha_s(M_Z^2)$,
demonstrating that the same kind of spike
and dip near $p_T=0$ is present in the corresponding ratios, 
even for the \mcatnlo{} alone.
Hence we conclude that the numerical results from \krknlo{} near $p_T=0$ are well
understood and look as expected.

Even better agreement is found in the comparison of results from
\krknlo{}~$\as(q^2)$ and  \powheg{}. 
Here, the two $p_{T,Z}$ distributions almost coincide in the whole range
shown in Fig.~\ref{fig:krk-2ch-pt} (left bottom panel). This is related to a
similar, multiplicative, way of applying the NLO correction to the parton shower
in \krknlo{} and \powheg{}.
In both cases, this corresponds to the virtual part of the NLO correction being
applied over the entire $p_{T,Z}$ range. This is different in \mcatnlo{}, where
the virtual correction is spread by the shower only up to the upper limit of 
shower's evolution variable (starting scale for BEV shower). That upper limit is
typically set at the scale of a vector boson mass. Hence, in \mcatnlo{},
the~$p_{T,Z}$ distribution  in the region $p_{T,Z} > m_Z$ is corrected only with
the real part of NLO, and therefore it recovers exactly the fixed order NLO
result in this region.
The multiplicative nature of the \krknlo{} and \powheg{} methods leads to mixed
real-virtual corrections of the order $\as^2$ at high $p_{T,Z}$. These
corrections are however part of NNLO, hence they go beyond the accuracy of
NLO+PS result.


The rapidity distributions shown in Fig.~\ref{fig:krk-2ch-rap} agree very well
in the central region between all the PS-matched results, up to the
normalization differences, the same as in Table~\ref{tab:krknlo-both-totalcs}.
Differences at forward and backward rapidities come from increasing differences
between PDFs in MC and $\msbar$ schemes as $x\to 1$, \cf Fig.~\ref{fig:mc-pdfs}.

Finally, in Fig.~\ref{fig:14}~(left) we compare the \krknlo{} results for the
$Z$-boson transverse momentum distribution, from both channels and with full PS,
with the corresponding result obtained from the fixed-order NNLO calculations
implemented in the {\sf DYNNLO} program~\cite{Catani:2009sm}.
The NNLO distributions were obtained with the  {\tt NNLO MSTW2008} PDFs 
and using $\mu_F=\mu_R=m_Z$, varied by the customary factors of $1/2$ and $2$ to
estimate the uncertainty from neglected higher orders.
Similar comparisons between the NNLO results and those of \powheg{} and
\mcatnlo{} are shown  in Fig.~\ref{fig:14}~(right). All the results are
divided by the NLO distribution from \mcfm.
  
The calculations within the \krknlo{} method were performed with two choices of the
argument of the strong coupling $\as$: $q^2$ and $M_Z^2$, as was done earlier in the
paper. These results are shown in Fig.~\ref{fig:14}~(left) as a solid blue line
and dashed green line, respectively. In addition, in solid orange, we show the
result of the \krknlo{} matching in the case where the argument of $\as$
is set to the minimum of $q^2$ and $M_Z^2$. This is a physically
motivated choice as at high $p_{T,Z}$ our matched result should be driven by
the fixed-order component, whose natural scale is $M_Z$.

As we see in Fig.~\ref{fig:14}~(left), both the \krknlo{} and the NNLO results
show the same trends, quickly raising above the NLO result at low and moderate 
$p_{T,Z}$ and then staying above it at high $p_{T,Z}$ 
(the curve for the case with $\as(q^2)$ 
falls down due to the fact that the strong coupling is running even at high
$p_{T,Z}$).
The fact that our method gives the result that is higher than the NLO one 
at high $p_{T,Z}$
is a consequence of the mixed real--virtual $\order{\as^2}$ terms, which
constitute part of the NNLO correction and arise because of the multiplicative
nature of the \krknlo{} approach.
Especially the $\as(\min(q^2, m_Z^2))$ choice does a good job, reproducing very
closely the full NNLO distribution. Therefore this argument of the strong
coupling is likely to be adopted in the future phenomenological studies.

In Fig.~\ref{fig:14}~(right) we show similar comparisons with NNLO
for \mcatnlo{} and \powheg{}. The behavior at low $p_{T,Z}$ is close to that
from \krknlo{}. At high $p_{T,Z}$, however, \mcatnlo{} and \powheg{} converge by
construction to the NLO results, departing from the NNLO predictions.

\section{Summary and outlook}

We have discussed the \krknlo{} method of matching the LO parton shower with
the fixed-order NLO QCD corrections. The method is based on two elements: 
the change of the factorization
scheme from $\msbar$ to the MC scheme, and upgrading the hardest
emission to the full NLO accuracy by reweighting with a simple, positive weight.
Details of the method and, in particular, demonstration of its NLO accuracy
have been elaborated on in Section~\ref{sec:krknlo-meth}.
 
The change of the factorization scheme allows one to eliminate troublesome
$z$-dependent terms from the coefficient function and, effectively, 
it amounts to creating the MC PDFs. In Section~\ref{sec:nlo-bench}, 
we have discussed how such PDFs can be obtained and how
they differ from the standard $\msbar$ parton distributions.
There, we have also validated the MC factorization scheme 
by studying the Drell--Yan process at the 
fixed-order NLO level and showing that the $\msbar$ and MC
scheme results are identical up to the order $\order{\as}$.

We have implemented the \krknlo{} method on top of the Catani--Seymour 
type of the parton shower in
the \sherpa{} event generator for the case of the $Z/\gamma^{\star}$-boson production process
(hence, the initial-state parton shower). In Section~\ref{sec:NLOwith-PS}, we have presented
the comparisons of the NLO--PS matched results
obtained with our technique with the fixed-order NLO results from \mcfm{} and with
other matched results, namely those of \mcatnlo{} and \powheg{}.
In particular, we have demonstrated that the KrkNLO results recover the fixed-order
NLO predictions (up to sub-percent differences for the $q\qbar$ channel only and $\sim 5\%$
for all channels, coming from the beyond-NLO terms). 
 
As for the comparisons of \krknlo{} with \mcatnlo{} and \powheg{} at the level
of differential distributions, all three methods turn out to give essentially
identical results for the $y_Z$ spectrum. 
The $p_{T, Z}$ distributions look somewhat different for each method and the
exact features depend on the initial channels and the recoil schemes.
In general, the \krknlo{} method provides similar predictions to the other two
well-established approaches. In particular, the results with both channels and
the full parton shower stay very close to those from the \powheg{} method
implemented in \herwig{}.
Residual differences come from spurious $\order{\as^2}$ terms, which are
different in each of the three matching methods.

Applying the \krknlo{} method to the Higgs production process is now under development
and will be reported in the next paper.
Applying the same \krknlo{} method to more processes is quite straightforward, 
in the sense that the modification of the PDFs (to the MC scheme)
and subtracting the hard process with the MC scheme counter-term
(instead of the $\msbar$ counter-term) can be done in the usual way for any process, 
also with more colored particles in the final state.
The interesting question is rather whether this method eliminates {\em all}
the non-physical $\sim \delta(p_T)$ singularities,
which are incompatible with the physical phase space of any PS MC?
Following the study of Ref.~\cite{Jadach:2013dfd},
we are confident that for color-singlet heavy object production in the $s$ and
$t$ channels this is true, although a general formal proof would be welcome.
For more colored particles in the final state, this is still an open question.
However, it is quite likely that the \krknlo{} method reduces significantly
the number of such pathological terms, and therefore will facilitate matching
parton shower with any NLO and/or NNLO corrected hard processes.
This is why, in our opinion, it is definitely worth to pursue 
this new development path in matching techniques.

The current methods of the type NNLO+PS 
\cite{Hamilton:2012rf,Hoeche:2014aia,Alioli:2013hqa,Hoche:2015sya}
represent a clear progress in matching of the fixed-order NNLO QCD calculations with PS MCs, but still suffer from various limitations;
for instance, they are limited to a certain class of observables only.
We are definitely thinking about extending \krknlo{} to NNLO+NLOPS, 
in which NLOPS is a parton shower MC
implementing the NLO evolution kernels in the fully exclusive form, 
thus providing the full set of the soft-collinear counter-terms for the hard process.
Ref.~\cite{Jadach:2013dfd} reviews several feasibility studies,
which show that constructing such a NLOPS is, in principle, plausible.
In our opinion, any simplifications of the NLO+PS matching, 
as in the \krknlo{} method,
will be instrumental and very useful towards more ambitious
fully exclusive NNLO+NLOPS projects.


\section*{\large Acknowledgments}

S.S. is grateful for hospitality to the IFJ PAN in Krak\'ow, where part of this
work has been done.
We acknowledge useful discussions with Simone Alioli, Stefan Hoeche, Simon
Platzer, Gavin Salam, Marek Schoenherr, Steffen Schumann and Alexandra Wilcock.
We thank A. Kusina for critical reading the manuscript.
We are also grateful to the Cloud Computing for Science and Economy 
project (CC1) at IFJ PAN (POIG 02.03.03-00-033/09-04) in Krak\'ow whose resources 
were used to carry out most of the numerical computations for this project. 
Thanks also to Mariusz Witek and Mi{\l}osz Zdyba{\l} for their help with CC1. 
This work was funded in part by 
the MCnetITN FP7 Marie Curie Initial Training Network PITN-GA-2012-315877.


\newpage
\appendix
\vspace{10mm}
\noindent
{\bf\large APPENDIX}
\section{First emission in backward evolution}
\label{sec:AppedA}

In case of two LO parton showers in the BEV algorithm the cross section
with exactly zero gluons in both showers includes
two non-emission $\Delta$-functions
\begin{equation}
\label{eq:NOglu}
\sigma^{\lo}_0 =
\int d \hat{x}_\F d\hat{x}_\B
  d\Omega\;
\frac{d\sigma}{d\Omega}(\hat{s},\hat\theta)\;
e^{ -\Delta^\F_\mc(\hat{x}_\F|s,q_s^2) -\Delta^\B_\mc(\hat{x}_\B|s,q_s^2) }\;
\Dbar^\F_\mc( \hat{s}| s, \hat{x}_\F)\;
\Dbar^\B_\mc( \hat{s}| s, \hat{x}_\B),
\end{equation}
where $\hat{s}=s \hat{x}_\F \hat{x}_\B$.
Using the identity of Eq.~(\ref{eq:DeltaBEV2MC}) twice,
the above transforms immediately into
\begin{equation}
\label{eq:FEVtwoNoemi}
\sigma^{\lo}_0 =
\int d x_\F dx_\B
  d\Omega\;
\frac{d\sigma}{d\Omega}(\hat{s},\hat\theta)\;
e^{ -S_\mc(\hat{s}| s,q_s^2) -S_\mc(\hat{s}| s,q_s^2)} 
\Dbar^\F_\mc(\hat{s}| q_s^2, x_\F),
\Dbar^\B_\mc(\hat{s}| q_s^2, x_\B),
\end{equation}
which coincides exactly with the $n=0$ result in the forward evolution,
starting from $q^2_\F =q^2_\B = q_s^2$.
NB. In the case $n=0$, we identify $\hat{x}_\F=x_\F$ and $\hat{x}_\B=x_\B$.

In the following we shall simplify the notation omitting the $\hat{s}$ argument
in $\Dbar_\mc( q^2, x) = \Dbar_\mc(\hat{s}|q^2,x)$
and $S_\mc(q_2^2,q_1^2)= S_\mc(\hat{s}|q_2^2,q_1^2) $, wherever it is unambiguous.

The distributions from the BEV algorithm with at least one gluon 
($n_\F+n_\B=1,2,3...$) reads as follows:
\begin{equation}
\label{eq:be2dipole}
\begin{split}
&\sigma^{\lo}_{1+}
=\int d \hat{x}_\F d\hat{x}_\B d\Omega
\Bigg\{
\int^{s}_{q_s^2}   \frac{d q_{1\F}^2}{ q_{1\F}^2}
\int_{\hat{x}\F}^1 \frac{dz_1}{z_1}\;
\Kbbm_\mc(\hat{x}_\F|z_1,q_{1\F}^2)
e^{ -\Delta^\F_\mc(\hat{x}_\F|s,q_{1\F}^2) -\Delta^\B_\mc( \hat{x}_\B|s,q_{1\F}^2) }
\\&~~~~~~~~~~~~~~~~~~~~~~~~
+
\int^{s}_{q_s^2}   \frac{d q_{1\B}^2}{ q_{1\B}^2}
\int_{\hat{x}\B}^1 \frac{dz_1}{z_1}\;
\Kbbm_\mc(\hat{x}_\B|z_1,q_{1\B}^2)
e^{ -\Delta^\B_\mc(\hat{x}_\B|s,q_{1\B}^2) -\Delta^\F_\mc(\hat{x}_\F|s,q_{1\B}^2) }
\Bigg\}
\\&~~~~~~~~~~~~~~~~~~~~~~~~~~~\times
\Dbar^\F_\mc(\hat{s}, \hat{x}_\F)\; 
\Dbar^\B_\mc(\hat{s}, \hat{x}_\B)\;
\frac{d\sigma}{d\Omega}(s \hat{x}_\F \hat{x}_\B,\hat\theta),
\end{split}
\end{equation}
where $z_1$ integration limits are imposed by $\hat{s}(1-z_1)^2/z_1>q_1^2$
in the $\Kbbm$ kernels.

The ``unitarity'' sum rule $\sigma^{\lo}_0 +\sigma^{\lo}_{1^+}=\sigma^{\lo}$ 
is of course automatic in the BEV algorithm.
Nevertheless, proving it provides an interesting cross-check.
The differentiation over the lower boundary in the $q^2$ integral
defining the $\Delta$-function leads to:
\begin{equation}
\label{eq:q2diff-trick}
\begin{split}
&\frac{\partial}{\partial \ln q^2} 
  e^{ -\Delta_\mc(\hat{x}_\F|s,q^2)-\Delta_\mc(\hat{x}_\B|s,q^2) }=
\\&~~~~
= \int \frac{dz}{z}
  \big[ \Kbbm_\mc(\hat{x}_\F|z,q^2) + \Kbbm_\mc(\hat{x}_\B|z,q^2) \big]\;
  e^{ -\Delta_\mc(\hat{x}_\F|s,q^2) -\Delta_\mc(\hat{x}_\B|s,q^2) }.
\end{split}
\end{equation}
After combining two integrals in Eq.~(\ref{eq:be2dipole}) we obtain
\begin{equation}
\begin{split}
&\sigma^{\lo}_{1+} 
= \int d \hat{x}_\F d\hat{x}_\B d\Omega\;
\int\limits^{\ln s}_{\ln q_s^2} d \ln(q_1^2)
\frac{\partial}{\partial \ln q_1^2}
\bigg[
e^{ -\Delta^\F_\mc( \hat{x}_\F|s,q_1^2) -\Delta^\B_\mc( \hat{x}_\B|s,q_1^2) }
\bigg]
\\&~~~~~~~~~~~~~~~~~~~~~~~~~~~~~~\times
\frac{d\sigma}{d\Omega}(s \hat{x}_\F \hat{x}_\B,\hat\theta)\;\;
\Dbar^\F_\mc(s, \hat{x}_\F)\; \Dbar^\B_\mc(s, \hat{x}_\B)
\\&
= \int d \hat{x}_\F d\hat{x}_\B d\Omega
\Big[
1- e^{ -\Delta^\F_\mc( \hat{x}_\F|s,q_s^2) -\Delta^\B_\mc( \hat{x}_\B|s,q_s^2) }
\Big]\;
\frac{d\sigma}{d\Omega}(s \hat{x}_\F \hat{x}_\B,\hat\theta)\;\;
\Dbar^\F_\mc(s, \hat{x}_\F)\;
\Dbar^\B_\mc(s, \hat{x}_\B),
\end{split}
\end{equation}
from which the sum rule 
$\sigma^{\lo}_0 +\sigma^{\lo}_{1^+}=\sigma^{\lo}$ results immediately.

Before we transform Eq.~(\ref{eq:be2dipole}) into the FEV picture, 
let us first expand $\Kbbm$-kernels%
\footnote{ We denote $\Peu(z)= \frac{C_F \alpha_s}{\pi} \frac{\bar{P}(z)}{1-z}$.
  }:
\begin{equation}
\label{eq:BEV2dipole}
\begin{split}
&\sigma^{\lo}_{1+}
=\int d \hat{x}_\F d\hat{x}_\B\; d\Omega\;
\\&\times
\Bigg\{
\int\limits^{s}_{q_s^2}   \frac{d q_{1\F}^2}{ q_{1\F}^2}
\int\limits_{x_\F}^1  \frac{dz_1}{z_1}\;\theta_{\hat{s}(1-z_1)^2/z_1>q_{1\F}^2}
\Peu(z_1)\;
e^{ -\Delta^\F_\mc( \hat{x}_\F|s,q_{1\F}^2) }
e^{ -\Delta^\B_\mc( \hat{x}_\B|s,q_{1\F}^2) }
\frac{ \Dbar^\F_\mc \Big( q_{1\F}^2, \frac{\hat{x}_\F}{z_1} \Big)}%
     { \Dbar^\F_\mc ( q_{1\F}^2, \hat{x}_\F) }
\\&~~
+
\int\limits^{s}_{q_s^2}   \frac{d q_{1\B}^2}{ q_{1\B}^2}
\int\limits_{x_\B}^1 \frac{dz_1}{z_1}\;\theta_{\hat{s}(1-z_1)^2/z_1>q_{1\B}^2}
\Peu(z_1)\;
e^{ -\Delta^\F_\mc( \hat{x}_\F|s,q_{1\B}^2) }
e^{ -\Delta^\B_\mc( \hat{x}_\B|s,q_{1\B}^2) }
\frac{ \Dbar^\B_\mc \Big( q_{1\B}^2, \frac{\hat{x}_\B}{z_1} \Big)}%
     { \Dbar^\B_\mc ( q_{1\B}^2, \hat{x}_\F) }
\Bigg\}
\\&\times
\Dbar^\F_\mc(s, \hat{x}_\F)\; \Dbar^\B_\mc(s, \hat{x}_\B)\;
\frac{d\sigma}{d\Omega}(s \hat{x}_\F \hat{x}_\B,\hat\theta).
\end{split}
\end{equation}
Now using again the identities
\[  
 e^{-\Delta^\F_\mc( \hat{x}_\F|s,q_{1\F}^2)} 
 = e^{-S^\F_\mc(s,q_{1\F}^2)} 
   \frac{ \Dbar^\F_\mc(q_{1\F}^2,\hat{x}_\F)}{ \Dbar^\F_\mc(s,\hat{x}_\F) },
\;\;
 e^{-\Delta^\B_\mc( \hat{x}_\B|s,q_{1\F}^2)} 
 = e^{-S^\B_\mc(s,q_{1\F}^2)} 
\frac{ \Dbar^\B_\mc(q_{1\F}^2,\hat{x}_\B)}{ \Dbar^\B_\mc(s,\hat{x}_\F) },
\]
we arrive at the following FEV representation
\begin{equation}
\label{eq:FEV3}
\begin{split}
&\sigma^{\lo}_{1+}
=\int d \hat{x}_\F d\hat{x}_\B\; d\Omega\;
\frac{d\sigma}{d\Omega}(s \hat{x}_\F \hat{x}_\B,\hat\theta)
\\&\times
\Bigg\{
\int\limits^{s}_{q_s^2}   \frac{d q_{1\F}^2}{ q_{1\F}^2}
\int\limits_{x_\F}^1  \frac{dz_1}{z_1}\;\theta_{\hat{s}(1-z_1)^2/z_1>q_{1\F}^2}
\Peu(z_1)\;
e^{ -S^\F_\mc(s,q_{1\F}^2) -S^\B_\mc(s,q_{1\F}^2) }
D^\F_\mc\big( q_{1\F}^2, \frac{\hat{x}_\F}{z_1} \big)
D^\B_\mc(     q_{1\F}^2, \hat{x}_\B)
\\&~
+
\int\limits^{s}_{q_s^2}   \frac{d q_{1\B}^2}{ q_{1\B}^2}
\int\limits_{x_\B}^1 \frac{dz_1}{z_1}\;\theta_{\hat{s}(1-z_1)^2/z_1>q_{1\B}^2}
\Peu(z_1)\;
e^{ -S^\F_\mc(s,q_{1\B}^2) -S^\B_\mc(s,q_{1\B}^2) }
D^\F_\mc ( q_{1\B}^2, \hat{x}_\F)
D^\B_\mc \big( q_{1\B}^2, \frac{\hat{x}_\B}{z_1} \big)
\Bigg\}.
\end{split}
\end{equation}

Finally, the longitudinal integrations are streamlined 
with the help of the substitutions to the $x$-variables before the emission,
$x_\F/z_1\to x_\F,\; x_{1\B}=x_\B$ or
$x_\B/z_1\to x_\B,\; x_{1\F}=x_\B$,
respectively in each shower,
and the convolution structure is made manifestly symmetric:
\begin{equation}
\label{eq:FEV3b}
\begin{split}
&\sigma^{\lo}_{1+}
=\int dx \int d x_\F dx_\B\; d\Omega\;
\\&\times
\Bigg\{
\int\limits^{s}_{q_s^2}   \frac{d q_{1\F}^2}{ q_{1\F}^2}
\int\limits_{x_\F}^1  \frac{dz_1}{z_1}\;\theta_{\hat{s}(1-z_1)^2/z_1>q_{1\F}^2}
\Peu(z_1)\;
e^{ -S^\F_\mc(s,q_{1\F}^2) -S^\B_\mc(s,q_{1\F}^2) }
\bar{D}^\F_\mc( q_{1\F}^2, x_\F) \bar{D}^\B_\mc( q_{1\F}^2, x_\B)
\\&~~
+
\int\limits^{s}_{q_s^2}   \frac{d q_{1\B}^2}{ q_{1\B}^2}
\int\limits_{x_\B}^1 \frac{dz_1}{z_1}\;\theta_{\hat{s}(1-z_1)^2/z_1>q_{1\B}^2}
\Peu(z_1)\;
e^{ -S^\F_\mc(s,q_{1\B}^2) -S^\B_\mc(s,q_{1\B}^2) }
\bar{D}^\F_\mc( q_{1\B}^2, x_\F) \bar{D}^\B_\mc( q_{1\B}^2, x_\B)
\Bigg\}
\\&\times
\frac{d\sigma}{d\Omega}(s x_\F x_\B,\hat\theta)\;
\delta(x-z_1x_\F x_\B)
\end{split}
\end{equation}
The above formula is for the FEV algorithm
but for one gluon next to the hard process and any number of trailing
gluons down to $q_s^2$.
The same formula in Eq.~(\ref{eq:detach2PDF})
was obtained by means of factorizing such a gluon from the fully exclusive
FEV formula of Eq.~(\ref{eq:PSMCfev2scale}).

\section{Exclusive NLO corrections in $\msbar$ scheme}
\label{append:B}
Following Ref.~\cite{Altarelli:1979ub},
the bare NLO real$\,+\,$virtual unintegrated DY
cross section for the incoming $q\qbar$ 
with the effective mass of $\sqrt{s_1}$,
in $d=4+2\epsilon $ dimensions, is quite simple:
\begin{equation}
\label{eq:NLOraw}
\begin{split}
&
\sigma^{\nlo}_{q\qbar,B}(s_1,\epsilon)=
\int d\Omega \!\!\!
\int\limits_{\alpha+\beta\leq 1}  \!\!\! d\alpha d\beta\;\;
\Big\{
    \delta(\alpha)\delta(\beta)\;\Weu_0(s_1)
   +\Weu_2(s_1,\alpha,\beta)
\Big\},
\\&
\Weu_0(s_1)=
  \big(1+V(s_1,\epsilon) \big)
  \frac{d\sigma_0}{d\Omega}(s_1,\theta),
\\&
\Weu_2(s_1,\alpha,\beta)=
 H_2(s_1,\alpha,\beta,\epsilon)\;
 \frac{d\sigma_0}{d\Omega} \big(zs_1,\theta_\F\big)
+H_2(s_1,\beta,\alpha,\epsilon)\;
 \frac{d\sigma_0}{d\Omega} \big(zs_1,\theta_\B\big)
\\&
 V(s_1,\epsilon)=
     \frac{2 C_F \alpha_s}{\pi}
     \frac{ (4\pi)^{-\epsilon} \Gamma(1+\epsilon)}{\Gamma(1+2\epsilon)} 
     \bigg( \frac{s_1}{\mu^2} \bigg)^{\epsilon}
     \bigg( 
       -\frac{1}{2\epsilon^2}  
       +\frac{3}{4}\frac{1}{\epsilon}  
       -2 
       -\frac{\pi^2}{12}
     \bigg),
\\&
 H_2(s_1,\alpha,\beta,\epsilon)=
  \frac{ 2C_F \alpha_s}{\pi}
  \frac{ (4\pi)^{-\epsilon}}{\Gamma(1+\epsilon)} 
  \bigg( \frac{s_1 \alpha\beta}{\mu^2} \bigg)^{\epsilon}
  \frac{(1-\alpha)^2 +\epsilon \alpha^2}{2\alpha\beta},
\end{split}
\end{equation}
where $z=1-\alpha-\beta$ 
and notation of Section~\ref{sec:sigNLO} is used.
The integration over the azimuthal angle is
already done in order to improve readability.
The integration decay lepton angles and $\alpha$ and $\beta$ 
keeping $z=1-\alpha-\beta$ fixed, leads to a divergent (bare) result
\begin{equation}
\label{eq:rhoAEMbare}
\begin{split}
&\sigma^{\nlo}_{q\qbar,B}(s_1,\epsilon)=
 \int_0^1 dz\; \sigma_0(s_1 z)\; 
[ \delta(1-z)+\rho_B(z,s_1,\epsilon)],
\\&
\rho_B(z,s_1,\epsilon) =
 \frac{2\alpha_s}{\pi} C_F\, 
 \frac{ (4\pi)^{-\epsilon}}{\Gamma(1+\epsilon)} 
 \left( \frac{s_1}{\mu^2} \right)^{\epsilon}
 \bigg\{ 
   \left(\frac{\pi^2}{6}-2 \right) \delta(1-z) 
\\&~~~~~~~~~~~~~~~~~~~~~~~~~~~~
  -\frac{1}{\epsilon}
   \left(\frac{1+z^2}{2(1-z)} \right)_+
   + (1+z^2)\left(\frac{\ln(1-z)}{1-z}\right)_+  
 \bigg\}.
\end{split}
\end{equation}
The $\msbar$ recipe tells us to subtract from the
bare radiator function $\rho_B$ its pole part
\begin{equation}
 2\Gamma^{[1]}_{\msbar}(\epsilon)=
  {\rm PP}\; \rho_B(z,s_1,\epsilon) 
  = \frac{2\alpha_s}{\pi} C_F\,
    \frac{1}{\epsilon} \frac{ (4\pi)^{-\epsilon}}{\Gamma(1+\epsilon)}
    \left(\frac{1+z^2}{2(1-z)} \right)_+ .
\end{equation}
After including PDFs and integrating over $s_1$ which leads to the replacement 
$s_1\to sx_\F x_\B$, we obtain 
the finite hadron-level cross section
\begin{equation}
\label{eq:rhoAEMbare2}
\begin{split}
&\sigma^{\nlo}(s)=
\int dx_\F dx_\B dz\;
\Big\{ \delta(1-z)
\\&~~~~~~~~~~~~
+\big[(1 - {\rm PP} ) \rho_B(z,sx_\F x_\B,\epsilon)\big]
\Big\}
 \sigma_0(s z x_F x_\B)
 D^{\msbar}_q(   \mu^2,x_F)
 D^{\msbar}_\qbar(\mu^2,x_B).
\end{split}
\end{equation}
The contribution from the $q\qbar$ channel to the
inclusive coefficient function of the Drell--Yan process is%
\footnote{ The assignment $ \mu^2=zs_1 $ induces
  the presence of the term $\sim \ln(z) (1+z^2)/(1-z)$ in $C^{\msbar}_{2q}$.  }
$
  C^{\msbar}_{2q}(z)= (1 - {\rm PP} ) \rho_B(z,\epsilon) \big|_{\mu^2=zs_1}
$
of Eq.~\ref{eq:C2q-MSbar}.
The above formula can be used to define the PDFs and dig them out from
experimental DY data.
Alternatively, it can be used to predict the DY cross section taking the PDFs 
obtained from other processes, e.g. deep inelastic lepton--proton scattering 
(DIS).

The above formulation is suitable for the total cross section.
In the case of any NLO-class exclusive observable
defined by the function $J= J_{\nlo}(x_\F,x_\B,z,k^2_{1T})$
of Eq.~(\ref{eq:jetfun}),
the above formulas are reorganized using the technique of the Catani--Seymour
soft-collinear counter-terms (SCC).
In the present case SCC coincides (for $\veps=0$) with
the single gluon LO distribution of the parton shower MC:
\begin{equation}
 \Weu_{ct}(s_1,\alpha,\beta,\epsilon)=
  \frac{ 2C_F \alpha_s}{\pi}
  \frac{ (4\pi)^{-\epsilon}}{\Gamma(1+\epsilon)} 
  \bigg( \frac{s_1 \alpha\beta}{\mu^2} \bigg)^{\epsilon}
  \frac{1+z^2+\veps(1-z)^2}{2\alpha\beta}\;
  \frac{d\sigma_0}{d\Omega}(zs_1,\theta),
\end{equation}
where $z=1-\alpha-\beta$.
The above SCC is subtracted and added:
\begin{equation}
\begin{split}
& \sigma^{\nlo}(s)[J]=
\int dx_\F dx_\B
D^{\msbar}_q(   \mu^2,x_F)
D^{\msbar}_\qbar(\mu^2,x_B)
\\&\times
\Bigg\{
(1-{\rm PP })
\bigg[
  \int d\Omega\;  \Weu_0(s_1) \;
  +\!\!\! \int\limits_{\alpha+\beta\leq 1}  \!\!\! d\alpha d\beta\;
   \int d\Omega\;   \Weu_{ct}(s_1,\alpha,\beta,\epsilon)
\bigg] 
J(x_\F,x_\B,1,0)
\\&
+\int d\Omega \!\!\!\!\!
\int\limits_{\alpha+\beta\leq 1}  \!\!\!\!\! d\alpha d\beta
\big[
      \Weu_2(s_1,\alpha,\beta)
      J(x_\F,x_\B,z,k^2_{1T})
     -\Weu_{ct}(s_1,\alpha,\beta)
      J(x_\F,x_\B,1,0)
\big]_{\epsilon=0} 
\bigg\},
\end{split}
\end{equation}
where $s_1 \equiv sx_\F x_\B$.
In the second real emission integral
$\Weu_{ct}$ eliminates completely the soft/collinear singularities,
so it can be evaluated in the $\veps\to 0$ limit.
The key point here is that $\Weu_{ct}$
is multiplied by $J_{\lo}=J(x_\F,x_\B,1,0)$,
hence the partial integration 
$\int d\alpha d\beta \delta(z-\alpha-\beta)$
can be performed for the fixed $z$ in order
to isolate and subtract the $1/\epsilon$ pole:
\begin{equation}
\label{eq:DYinCS}
\begin{split}
& \sigma^{\nlo}(s)[J]=
\int dx_\F dx_\B d\Omega \int dz\;
D^{\msbar}_q(   \mu^2,x_F)
D^{\msbar}_\qbar(\mu^2,x_B)
\\&~~~~~~~~~~~~\times
\Big\{
 \big( 1+\Delta_{VS}^{q\qbar} \big)\; \delta(1-z)
 +2\Sigma_q(z)
\Big\}
\frac{d\sigma_0}{d\Omega}(zs_1,\theta)
J(x_\F,x_\B,1,0)
\\&
+
\int dx_\F dx_\B d\Omega \;\;
D^{\msbar}_q(   \mu^2,x_F)
D^{\msbar}_\qbar(\mu^2,x_B)
\!\!\!
\int\limits_{\alpha+\beta\leq 1}  \!\!\!\! d\alpha d\beta\;
\Delta \Weu_{real}[J]\,,
\end{split}
\end{equation}
with
\begin{equation}
\begin{split}
2\Sigma_q(z)= \frac{2C_F \alpha_s}{\pi}
 \bigg\{
    \frac{1+z^2}{2(1-z)} \ln\frac{(1-z)^2}{z}
   +\frac{1+z^2}{2(1-z)} \ln\frac{\hat{s}}{\mu^2}
   +\frac{1-z}{2}
 \bigg\}_+,
\\
\Delta \Weu_{real}[J]=
\big[
      \Weu_2(s_1,\alpha,\beta)
      J(x_\F,x_\B,z,k^2_{1T})
     -\Weu_{ct}(s_1,\alpha,\beta)
      J(x_\F,x_\B,1,0)
\big]_{\epsilon=0},
\end{split}
\end{equation}
where
$\Delta_{VS}^{q\qbar}=\frac{2C_F \alpha_s}{\pi}
 \big(\frac{\pi^2}{3} -\frac{5}{8} \big)
$
is that of Eq.~(\ref{eq:wt-vs}),
$\Sigma_q(z)|_{\hat{s}=\mu^2} = \Delta C_{2q}$,
see Eq.~(\ref{eq:DeltaC2q}),
and  $\hat{s}=szx_\F x_\B$ as usual.

In the MC scheme the entire $\sim \delta(k_{1T}^2) \Sigma(z)$ 
part gets eliminated
(modulo ${\cal O}(\alpha_s^2) $ terms)
thanks to the assignment $\hat{s}=\mu^2$
and redefinition of the PDFs
\begin{equation}
\label{eq:MSnar2MC}
 D^{\mc}_{q,\qbar}(   \mu^2,x) =  
 \int dz dx' \delta(x-zx')
 [ \delta(1-z) + \Sigma_q(z) ]_{\hat{s}=\mu^2}\;
 D^{\msbar}_{q,\qbar}(   \mu^2,x'),
\end{equation}
see Eq.~(\ref{eq:PDFs-MC-quark}).
The only freedom%
\footnote{
 The $z \neq 0$ part of $\Sigma$ is unique --
 it has to be eliminate completely
 from the NLO distribution,
 because the singular non-positive $\sim \delta(k_{1T}^2) \Sigma $ term
 cannot be included in the multiplicative MC weight.}
is in the coefficient of $\delta(1-z)$.
However, the convention $\int dz \Sigma(z)=0$
adopted in Ref.~\cite{Jadach:2011cr}
removes this freedom and fixes the content of $\Delta_{VS}$.

Inserting Eq.~(\ref{eq:MSnar2MC}) into Eq.~(\ref{eq:DYinCS}),
we obtain the formula for the NLO cross section in the MC scheme
\begin{equation}
\label{eq:DYinCSMC}
\begin{split}
& \sigma^{\nlo}_{\mc}(s)[J]=
\int dx_\F dx_\B d\Omega \int dz\;
D^{\mc}_q(   \hat{s},x_F)
D^{\mc}_\qbar(\hat{s},x_B)
\\&~~~~~~~~~~~~~~~\times
 \big( 1+\Delta_{VS}^{q\qbar} \big)
\frac{d\sigma_0}{d\Omega}(s_1,\theta)
J(x_\F,x_\B,1,0)
\\&
~~~~~~~~~~~+
\int dx_\F dx_\B d\Omega \;\;
\!\!\!
\int\limits_{\alpha+\beta\leq 1}  \!\!\!\! d\alpha d\beta\;
\Delta \Weu_{real}[J]\;
D^{\mc}_q(   \hat{s},x_F)
D^{\mc}_\qbar(\hat{s},x_B),
\end{split}
\end{equation}
where we have replaced the $\msbar$ PDFs by the $\mc$ PDFs also
in the {\em real} part. The cross sections in Eqs.~(\ref{eq:DYinCSMC}) and
(\ref{eq:DYinCS}) are equal up to ${\cal O}(\alpha_s^2)$.

Finally, let us remark, following Ref.~\cite{Jadach:2011cr},
that formally the transition from the $\msbar$ to MC scheme 
can be done by means of subtracting from the bare NLO distribution the term
\begin{equation}
\label{eq:GammaMC}
  \Gamma^{[1]}_{\mc}(\epsilon)
  = \Gamma^{[1]}_{\msbar}(\epsilon)
   +\Sigma_q(z)|_{\hat{s}=\mu^2},
\end{equation}
instead of the pure pole $\Gamma^{[1]}_{\msbar}(\epsilon)$.
It is appealing, that $\Gamma^{[1]}_{\mc}(\epsilon)$ concides with the integrated
single-gluon emission distribution of PS MC
extrapolated to $d=4+2\epsilon$ dimensions.
This justifies the name of the MC factorization scheme.


\providecommand{\href}[2]{#2}

\end{document}